\PassOptionsToPackage{hyphens}{url}

\documentclass[pdflatex,sn-mathphys-num]{sn-jnl}

\usepackage{graphicx}%
\usepackage{multirow}%
\usepackage{amsmath,amssymb,amsfonts}%
\usepackage{amsthm}%
\usepackage{mathrsfs}%
\usepackage[title]{appendix}%
\usepackage{xcolor}%
\usepackage{textcomp}%
\usepackage{manyfoot}%
\usepackage{booktabs}%
\usepackage{algorithm}%
\usepackage{algorithmicx}%
\usepackage{algpseudocode}%
\usepackage{listings}%
\usepackage{pifont}
\usepackage{subcaption}
\newcommand{\cmark}{\ding{51}}%
\newcommand{\xmark}{\ding{55}}%
\usepackage[T1]{fontenc}
\usepackage[utf8]{inputenc}
\usepackage[numbers]{natbib} 

\usepackage{longtable, booktabs, tabularx, multirow} 
\usepackage{rotating}
\usepackage{pdflscape} 

\theoremstyle{thmstyleone}%
\theoremstyle{thmstyletwo}%

\theoremstyle{thmstylethree}%

\begin{document}

\title[Article Title]{Understanding the Performance Plateau in Text-to-Video Retrieval: A Comprehensive Empirical and Linguistic Analysis}


\author[1,2]{\fnm{Maria-Eirini} \sur{Pegia}}\email{mpegia@iti.gr}

\author[1]{\fnm{Dimitrios} \sur{Stefanopoulos}}\email{distef@iti.gr}

\author[2]{\fnm{Bj\"{o}rn {\TH}\'{o}r   J\'{o}nsson}}\email{bjorn@ru.is}

\author[1]{\fnm{Anastasia} \sur{Moumtzidou}}\email{moumtzid@iti.gr}

\author[1]{\fnm{Ilias} \sur{Gialampoukidis}}\email{heliasgj@iti.gr}

\author[1]{\fnm{Stefanos} \sur{Vrochidis}}\email{stefanos@iti.gr}

\author[1]{\fnm{Ioannis} \sur{Kompatsiaris}}\email{ikom@iti.gr}

\affil[1]{\orgdiv{Information Technologies Institute}, \orgname{CERTH-ITI}, \orgaddress{\city{Thessaloniki}, \country{Greece}}}

\affil[2]{\orgdiv{School of Computer Science}, \orgname{Reykjavik University}, \orgaddress{\city{Reykjavík}, \country{Iceland}}}


\abstract{Text-to-video retrieval enables users to find relevant video content using natural language queries, a task growing rapidly in importance. Over the past six years (2020–2025), research has produced numerous methods, such as dual encoders, attention-driven models and multimodal fusion approaches. Fundamental questions remain, however, about model behavior, dataset influence and query difficulty. To address this, we evaluate 14 state-of-the-art retrieval methods across 3 widely used benchmark datasets under a unified preprocessing and evaluation framework. To better understand model behavior, we dig deep into the benchmarks and analyze caption characteristics, including length, clarity, semantic category and action vs. scene balance, and link these to model performance. \textcolor{black}{Empirically, models achieve higher recall on short, clear and simple captions (e.g., single actions or color attributes), whereas complex events and multi-step activities lower accuracy across all evaluated systems. Furthermore, attention-driven architectures excel on temporally dependent or multi-step queries, whereas dual-encoder models perform better on simpler or single-category captions. Cross-dataset generalization scales with larger, more diverse caption sets, though generative captions show variable impact on retrieval accuracy.} Overall, our findings highlight key dataset factors, benchmark challenges and the interplay between 
query content and model architecture, providing guidance for developing more effective text-to-video retrieval systems.
}


\maketitle

\section{Introduction}
The rapid growth and complexity of online video collections have made effective video retrieval a major challenge for modern systems.~\cite{gharahsouflou2022efficient, liu2021activity, zhu2025motionrag, liu2022activity, xu2020proposal, qiu2022challenges, vadicamo2024evaluating, dong2022partially, chavate2021comparative, hu2023adaclip, falcon2022feature, zhang2021personalized, jiang2021learning, shvetsova2022everything, aumuller2020ann, pegia2024time, manohar2024parlayann, Pegia2022, subramanian2022metrics, kalinin2025versatile, xu2021videoclip, wang2024videoclip, nguyen2024videoclip, bain2022clip, wang2023unified, ventura2024covr}. Because users naturally rely on text-based search as a direct extension of how they navigate the web, text-to-video retrieval has emerged as the primary interface for search, browsing and exploration~\cite{dong2022reading, tian2024holistic, zhang2025quantifying, zhao2025continual, yang2024dgl, song2023relation, han2021fine, jiang2023dual, lokoc2020w2vv++, wu2023cap4video, wang2021t2vlad, gorti2022x, duarte2022sign, yuan2020central, ibrahimi2023audio, song2021spatial, wang2022align, li2020sea, dong2021dual, jin2023text, zhao2022centerclip, ji2022cret, yakovlev2023sinkhorn}. This paradigm powers a wide variety of practical applications, including semantic video search~\cite{wray2021semantic}, digital assistants~\cite{lei2022assistsr}, video recommendations~\cite{dong2024musechat}, and interactive multimedia exploration. Over the past six years, substantial progress in retrieval performance and model scalability has been driven by improvements in large-scale datasets~\cite{afzal2023visualization} and the availability of powerful pretrained vision–language models. To tackle these challenges, recent research explores a diverse architectural landscape, encompassing dual encoders~\cite{li2020sea, lokoc2020w2vv++, croitoru2021teachtext, jiang2021learning}, multimodal fusion methods~\cite{dong2021dual, cheng2021improving, yuan2020central, luo2022clip4clip, bain2021frozen, ma2022query}, transformer-based networks~\cite{akbari2021vatt, dzabraev2021mdmmt, yan2021video}, and attention-driven systems~\cite{galanopoulos2020attention, han2021visual, tan2021logan}.

\begin{figure*}[t!]
    \centering
    \includegraphics[width=1.0\linewidth]{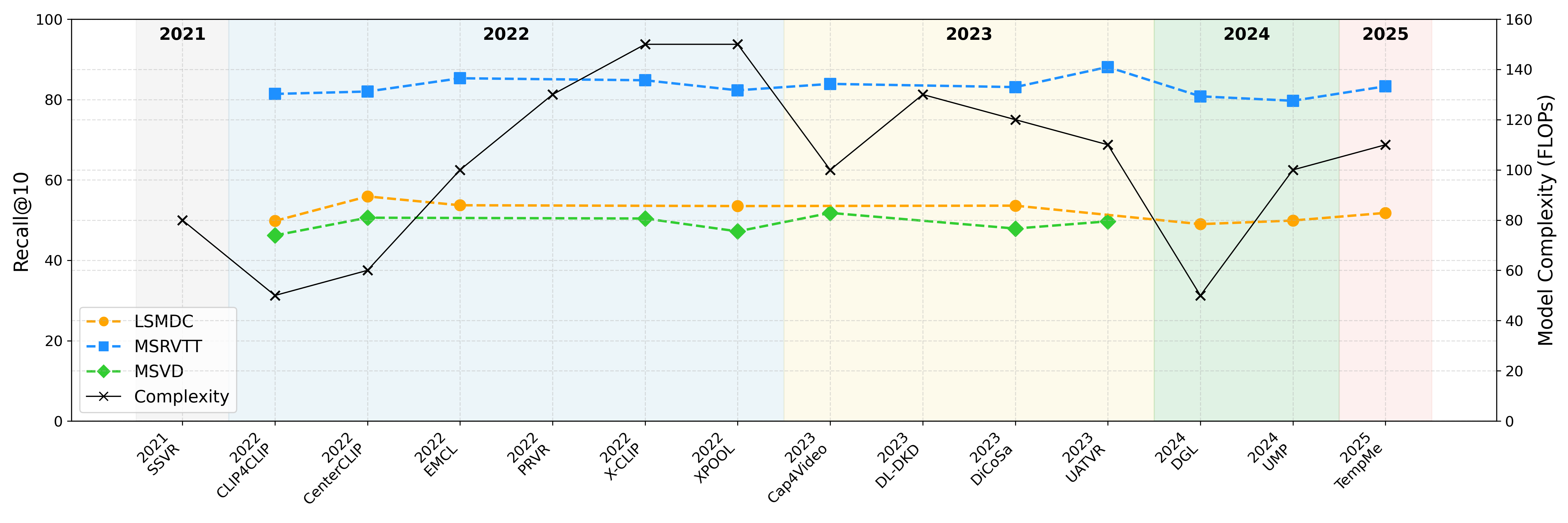}
    \caption{Recall@10 (left $y$-axis) and model complexity in FLOPs (right $y$-axis, black line) from 2021 to 2025 for various methods and datasets from the literature. Methods are sorted chronologically by year and alphabetically within the same year ($x$-axis). Datasets are shown as LSMDC (orange), MSRVTT (blue), MSVD (green). Year blocks are highlighted with soft background colors: 2021 (light gray), 2022 (blue), 2023 (yellow), 2024 (soft green), 2025 (red). The figure shows that annual performance gains have flatlined to marginal improvements.}
    \label{fig:literature_trend}
\end{figure*}

A further challenge in the field is the lack of consistent and principled evaluation practices. Many recent methods are compared primarily against a small set of baselines, often CLIP4CLIP, while direct comparisons among newer approaches are limited. Evaluation protocols, preprocessing choices and reporting standards vary across studies, making it difficult to draw reliable conclusions about relative performance. Moreover, most works focus exclusively on aggregate metrics such as Recall@k across whole benchmark collections, offering little insight into why certain methods succeed or fail.
Figure~\ref{fig:literature_trend} presents the Recall@10 performance from 2020 to 2025 \textcolor{black}{alongside model complexity in FLOPs. Specifically, it shows a rapid initial improvement, followed by a clear performance plateau, even as model complexity has continued to fluctuate and scale significantly. To formally characterize this trend rather than relying on visual inspection alone, our analysis quantifies the year-over-year diminishing marginal returns: annual performance gains have flatlined to marginal improvements of less than 1\% across benchmarks.}

In several cases, performance of highly complex models (such as X-CLIP~\cite{ma2022x} and XPOOL~\cite{gorti2022x}) is comparable to lighter, more efficient approaches (such as DGL~\cite{yang2024dgl} and TempMe~\cite{shen2024tempme}), proving that higher complexity does not translate into proportional performance gains. This empirical evidence reveals a structural saturation point in current text-to-video retrieval paradigms, underscoring the need to investigate query semantics and dataset properties rather than relying solely on brute-force architectural scaling.

Moreover, performance also varies significantly across datasets and the field lacks principled explanations for why certain methods succeed or fail. Some queries are consistently easy for all models, while others remain challenging regardless of the architecture. Most studies report only aggregate metrics, rarely analyzing temporal trends, dataset-specific differences, or qualitative behavior. Few examine the nature of positive and negative matches or identify challenging query types~\cite{mothe2022analytics}. Despite diverse architectures (discussed in detail in Section~\ref{sec:related_work}), a comprehensive understanding of linguistic properties, semantic categories, and dataset composition, and other factors driving performance differences across systems and datasets is still lacking.

To address these gaps, we present a comprehensive analysis of text-to-video retrieval that focuses on query characteristics, dataset properties, and model behavior rather than proposing a new architecture. We evaluate 14 representative state-of-the-art retrieval methods across three widely used benchmark datasets under a unified preprocessing and evaluation framework. A central focus is the role of dataset annotations: datasets with multiple captions per video capture richer semantic variability and reveal strengths and weaknesses that single-caption datasets fail to expose. We identify which retrieval tasks are consistently easy or difficult, how model behaviors align or diverge, and which semantic categories pose challenges across all systems. Our findings indicate that simple descriptions, such as single actions or color attributes, are generally handled reliably, while tasks involving complex events, multi-step activities, or fine-grained scene understanding remain challenging for all existing models.

In addition to standard metrics~\cite{rafiq2021video}, we analyze the linguistic structure of captions using semantic similarity embeddings~\cite{nozza2022state,cooper2024perplexed}, syntactic complexity measures~\cite{mohammed2022analysing,
solnyshkina2017evaluating}, and category-based clustering. This approach allows us to quantify how textual properties influence retrieval behavior and whether certain models respond consistently to specific linguistic patterns. The resulting clusters show that performance differences arise not only from architectural choices, but also from how models interpret and prioritize different forms of linguistic information. 
Attention-driven models such as TempMe and UATVR~\cite{fang2023uatvr} are better suited to temporally dependent or multi-step queries, while dual-encoder and multimodal fusion approaches perform well primarily on simpler or single-category descriptions. We further show that datasets with multiple captions per video provide more informative and challenging evaluation settings, leading to improved cross-dataset generalization, whereas single-caption datasets can overestimate performance.

Finally, we investigate the influence of caption quantity and quality across datasets. We show that datasets with many captions per video provide more informative and challenging evaluation settings, while single caption datasets can artificially inflate performance by reducing semantic diversity. We also test the use of generative captions, finding that automatically generated descriptions do not consistently improve retrieval accuracy and often introduce noise. These results show that the number and quality of captions are key for benchmarking and should guide dataset design.

\textcolor{black}{Our work bridges systematic literature consolidation with rigorous empirical investigation. Specifically, our contributions are organized into two distinct parts:}

\begin{description}
    \item[\bf 1. Synthesis and Consolidation of Prior Work:] \hfill
    \begin{itemize}
    \item \textcolor{black}{\textbf{Protocol Harmonization:} We conduct a large-scale, unified evaluation of 14 representative state-of-the-art text-to-video retrieval methods across three widely used benchmark datasets, eliminating inconsistencies caused by differing preprocessing and evaluation protocols found in prior literature.}
    \item \textcolor{black}{\textbf{Architectural Taxonomical Review:} We synthesize how different architectural paradigms (dual-encoder, attention-based and multimodal fusion) interact with established query types and dataset characteristics.}
\end{itemize}

\item[\bf 2. Novel Empirical and Methodological Contributions:] \hfill
\begin{itemize}
    \item \textcolor{black}{\textbf{The Robustness Evaluation Framework:} We introduce a semantic categorization of textual queries and a buffered difficulty-tier stratification framework (using an empirical query difficulty measure based on average ground-truth rank across models) to isolate true retrieval failures from boundary noise and identify consistently easy or hard query types.}
    \item \textcolor{black}{\textbf{Comprehensive Stress-Testing \& Generalization Analysis:} We provide new empirical findings across MSVD, MSRVTT and LSMDC showing how models degrade under linguistic and structural query shifts, while proving that single-caption datasets can overestimate retrieval effectiveness.}
    \item \textcolor{black}{\textbf{Practical Deployment Insights:} We assess the impact of operational factors such as frame rate, compression and training cost, providing actionable guidance for robust evaluation and future benchmark design.}
\end{itemize}
\end{description}

The remainder of this paper is organized as follows. Section~\ref{sec:related_work} surveys related work in video retrieval, with a focus on text-to-video approaches. Section~\ref{sec:methodology} outlines the research methodology, including the experimental setup, datasets, evaluation metrics, and implementation details, and introduces the key definitions used in this study. Section~\ref{sec:experimental_outline} presents three sets of experimental analyses: traditional experiments, which include analysis by task category, clustering of performance, task difficulty, and training times over epochs (Section~\ref{sec:main_experiment}); cross-dataset learning, examining the effects of training and testing on different datasets by reducing or increasing the number of captions per video (Section~\ref{sec:cross_dataset_analysis}); and sensitivity analysis, investigating the impact of frame rate and video compression on retrieval performance (Section~\ref{sec:sensitivity_analysis}). Finally,  Section~\ref{sec:conclusion} summarizes the study and outlines future research directions.

\section{Related Work}
\label{sec:related_work}

This section provides a survey of prior work in text-to-video retrieval, covering both the overall trends (Section~\ref{sec:overall_trends}) and key architectures (Section~\ref{sec:archs_gaps}) in the field  as well as a comparison of existing survey studies, highlighting their focus, coverage, and limitations (Section~\ref{sec:surveys}). 

\subsection{Overall Trends in Video Retrieval}
\label{sec:overall_trends}

\begin{figure*}[t!]
    \centering
    \includegraphics[width=0.6\linewidth]{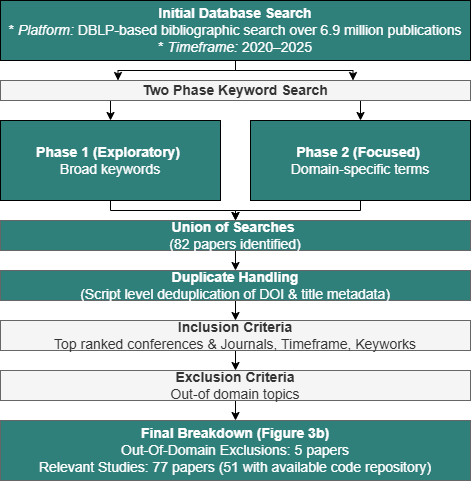} 
    \caption{Overview of the process used for literature exploration, screening and filtering for the video retrieval survey.}
    \label{fig:search_flowchart}
\end{figure*}

Recent trends in video retrieval research are highlighted by surveying representative methods published between 2020 and 2025, following the systematic workflow illustrated in Figure~\ref{fig:search_flowchart}. We conducted a large-scale literature search using the DBLP Survey Python script\footnote{\url{https://github.com/JanZahalka/dblp\_survey}} by Zahálka~\cite{zahalka2014towards}, which enables querying over 6.9 million DBLP-indexed publications. \textcolor{black}{Script-level deduplication was performed on DOI and title metadata to merge identical records or cross-conference listings prior to screening.} Our analysis focuses on papers published in the following top-ranked conferences and journals across computer vision, machine learning, multimedia retrieval, and information retrieval, as defined by the CORE conference and journal rankings.\footnote{\url{https://portal.core.edu.au/conf-ranks/}
 and \url{https://portal.core.edu.au/jnl-ranks/}, respectively.}

\begin{description}
      \item[\textbf{A*-rank conferences}]
      AAAI, NeurIPS, ICML, CVPR, ICCV, ECCV, ACM Multimedia, IJCAI, ICLR, SIGIR, ICDM, WWW
      \item[\textbf{A-rank conferences}]
      CIKM, WSDM, RecSys, ECAI
      \item[\textbf{B-rank conferences}]
      MMM, ICMR
      \item[\textbf{A*-rank journals}]
      IEEE TPAMI, IEEE TMM, JMLR, IJCV, Artificial Intelligence, Cognitive Science, Pattern Recognition, IEEE TKDE
      \item[\textbf{A-rank journals}]
      Machine Learning, CVIU, JAIR
      \item[\textbf{B-rank journals}]
      Conn. Sci.
\end{description}

Our search followed two main steps. First, we conducted an exploratory search using the following broad keywords to capture general trends in multimodal learning: \textit{multimodality}~\cite{christel2004exploiting,spolaor2021video,khan2021impact,kumar2022metadata}, \textit{multimodal retrieval}, \textit{foundation models}, \textit{language-visual models}, and \textit{large language models}. Second, we performed a more focused search with the following domain-specific terms directly related to video retrieval: \textit{video retrieval}, \textit{video search}, \textit{text-to-video retrieval}, \textit{video clip retrieval}, \textit{composed image retrieval}, and \textit{composed video retrieval}. The union of these two search phases yielded an initial pool of papers prior to duplicate handling.

\begin{figure*}[!t]
    \centering
    \setlength{\abovecaptionskip}{2pt}  
    \setlength{\belowcaptionskip}{1pt}  
    \begin{subfigure}{0.48\textwidth}
        \centering
        \includegraphics[width=1.0\linewidth]{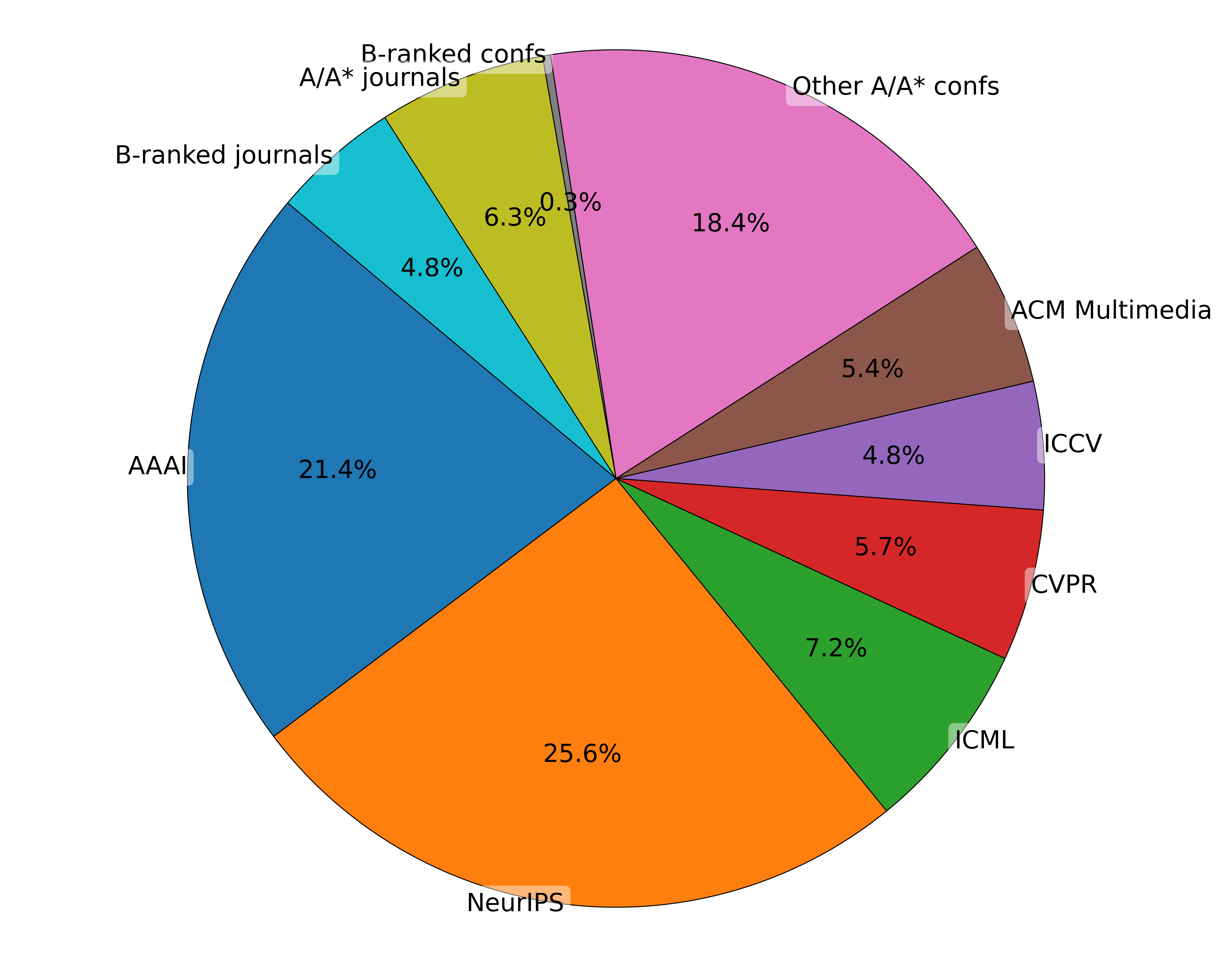}
        \caption{\small Papers per venue\label{fig:papers_per_venue}}
    \end{subfigure}\hspace{0.02\textwidth}
    \begin{subfigure}{0.44\textwidth}
        \centering
        \includegraphics[width=1.0\linewidth]{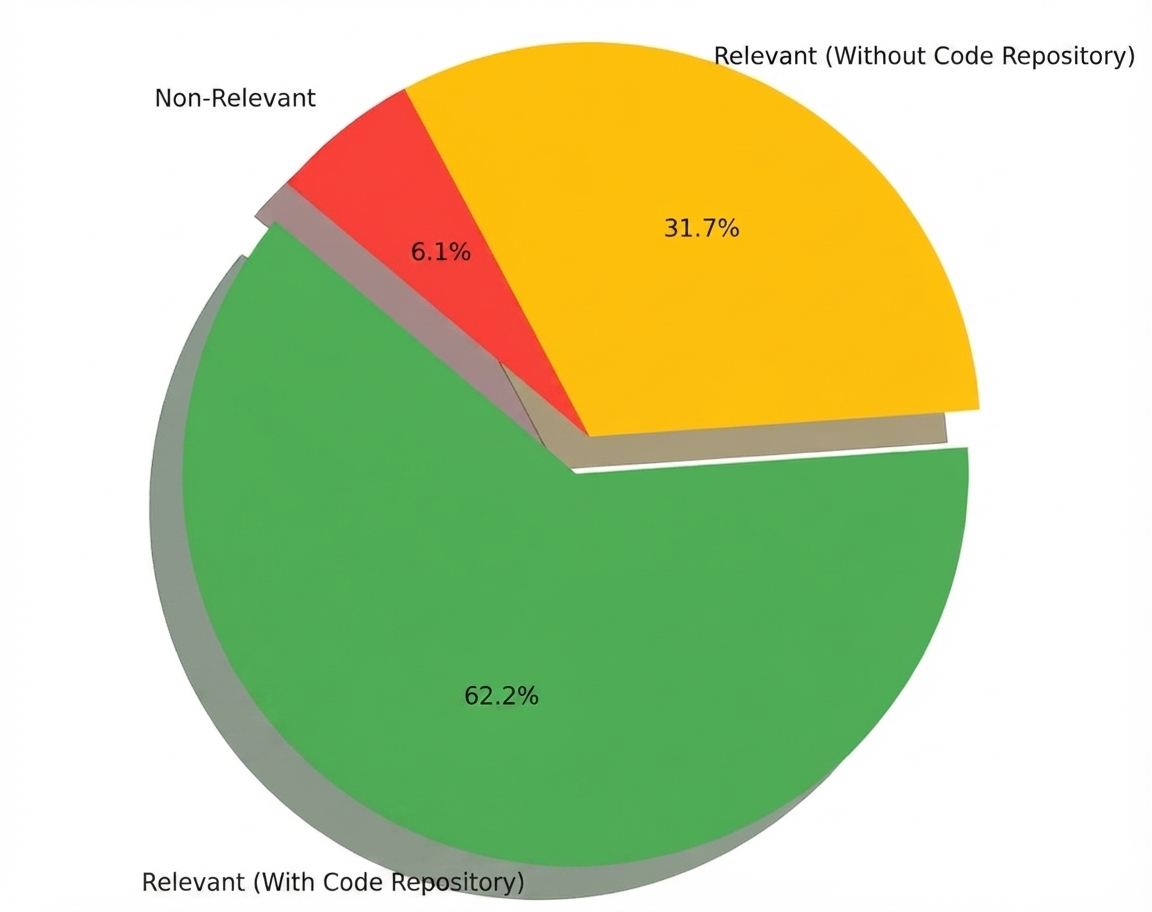}
        \caption{\small Relevance of paper\label{fig:relevant_papers}}
    \end{subfigure}
    
    \caption{\label{fig:exploration_papers} Distribution of papers and their relevance: (a) Number of papers published across different conferences and journals, highlighting the contribution per venue; (b) Breakdown of papers by relevance, distinguishing between relevant, non-relevant, and those associated with Code Repositories.}
\end{figure*}

Using this process, we identified 82 papers on video retrieval \textcolor{black}{after script-level deduplication of DOI and title metadata}. Among them, 5 papers were excluded as they addressed different domains (e.g., face retrieval or incident retrieval) \cite{dong2022partially,yin2024exploiting,liang2023simple,xiao2025videoqa,Lin2017}. The remaining 77 papers were relevant, of which 51 provided publicly available code repository implementations. 
Figure~\ref{fig:exploration_papers} breaks down the characteristics of the surveyed literature. As shown in Figure~\ref{fig:papers_per_venue}, the vast majority of the collected papers are published in top-tier A*-rank conferences  (such as NeurIPS, AAAI, ICML and CVPR), which collectively account for nearly 90$\%$ of all venues. Building on this baseline, Figure~\ref{fig:relevant_papers} illustrates that more than half (62.2\%) of the relevant studies release open-source code, while a notable fraction (31.7\%) does not. Because almost all papers originate from A*-rank venues, the concentration of code availability in these top-tier forums  primarily appears to be a reflection of the field's overall publication distribution rather than a venue-specific reproducibility mandate.

\begin{figure}[t!]
    \centering
    
     \begin{subfigure}[b]{0.45\textwidth}
        \centering
        \includegraphics[width=\linewidth]{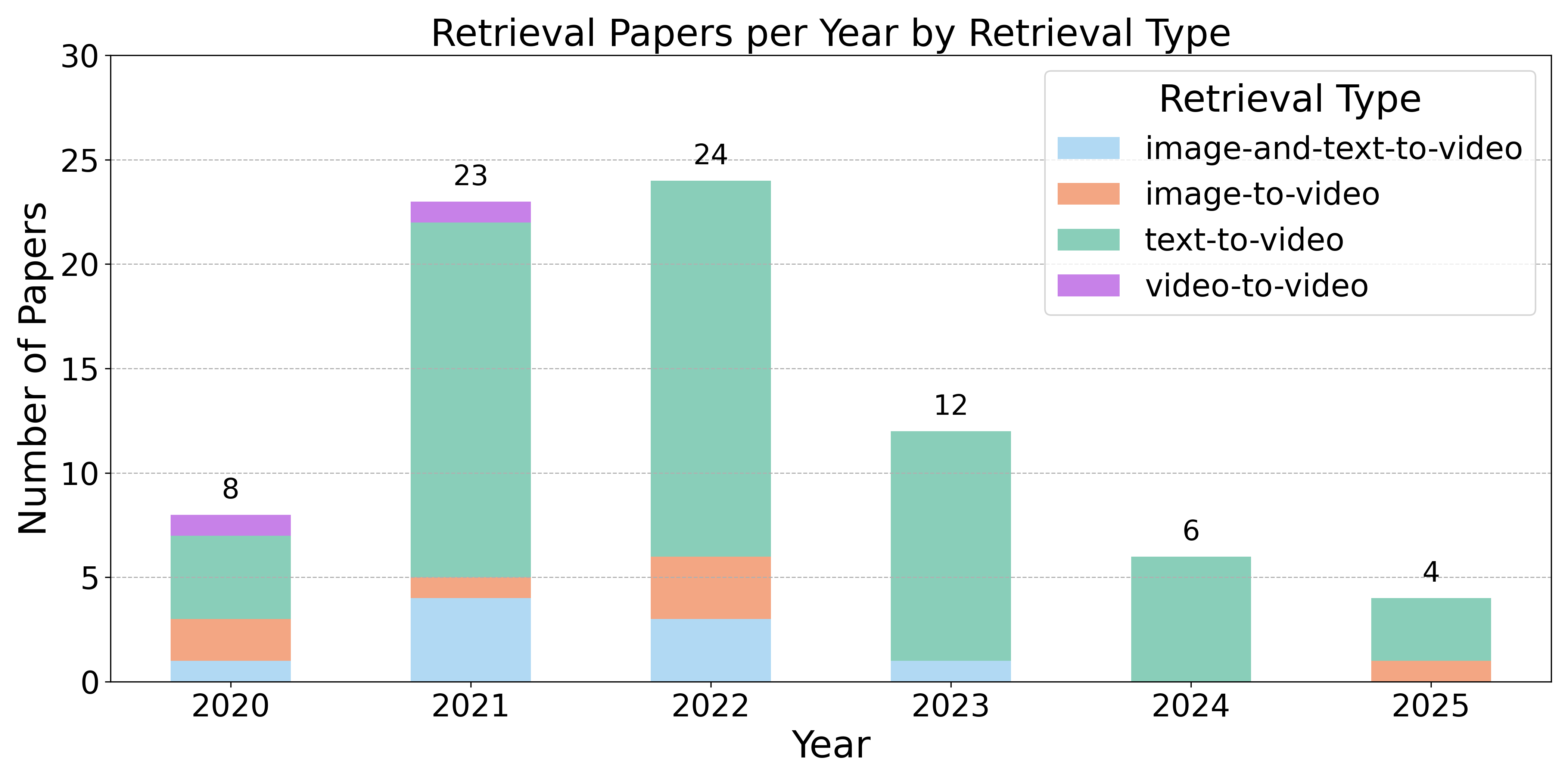}
        \caption{Retrieval Task}
        \label{fig:retrieval_task}
    \end{subfigure}
    \hfill
   \begin{subfigure}[b]{0.45\textwidth}
        \centering
        \includegraphics[width=\linewidth]{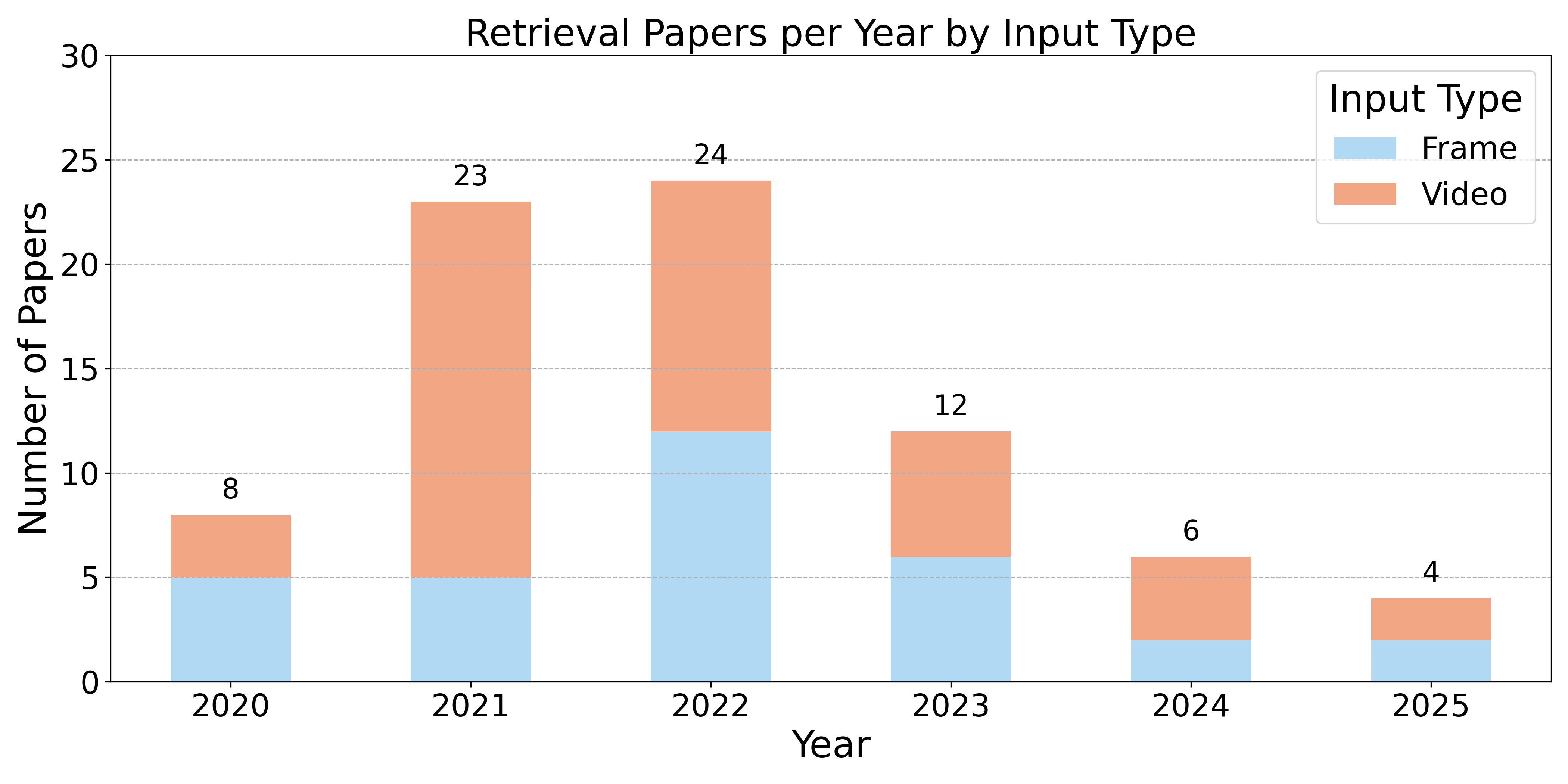}
        \caption{Input Type}
        \label{fig:input_type}
    \end{subfigure}

    \vspace{0.2em}
    
    \begin{subfigure}[b]{0.45\textwidth}
        \centering
        \includegraphics[width=\linewidth]{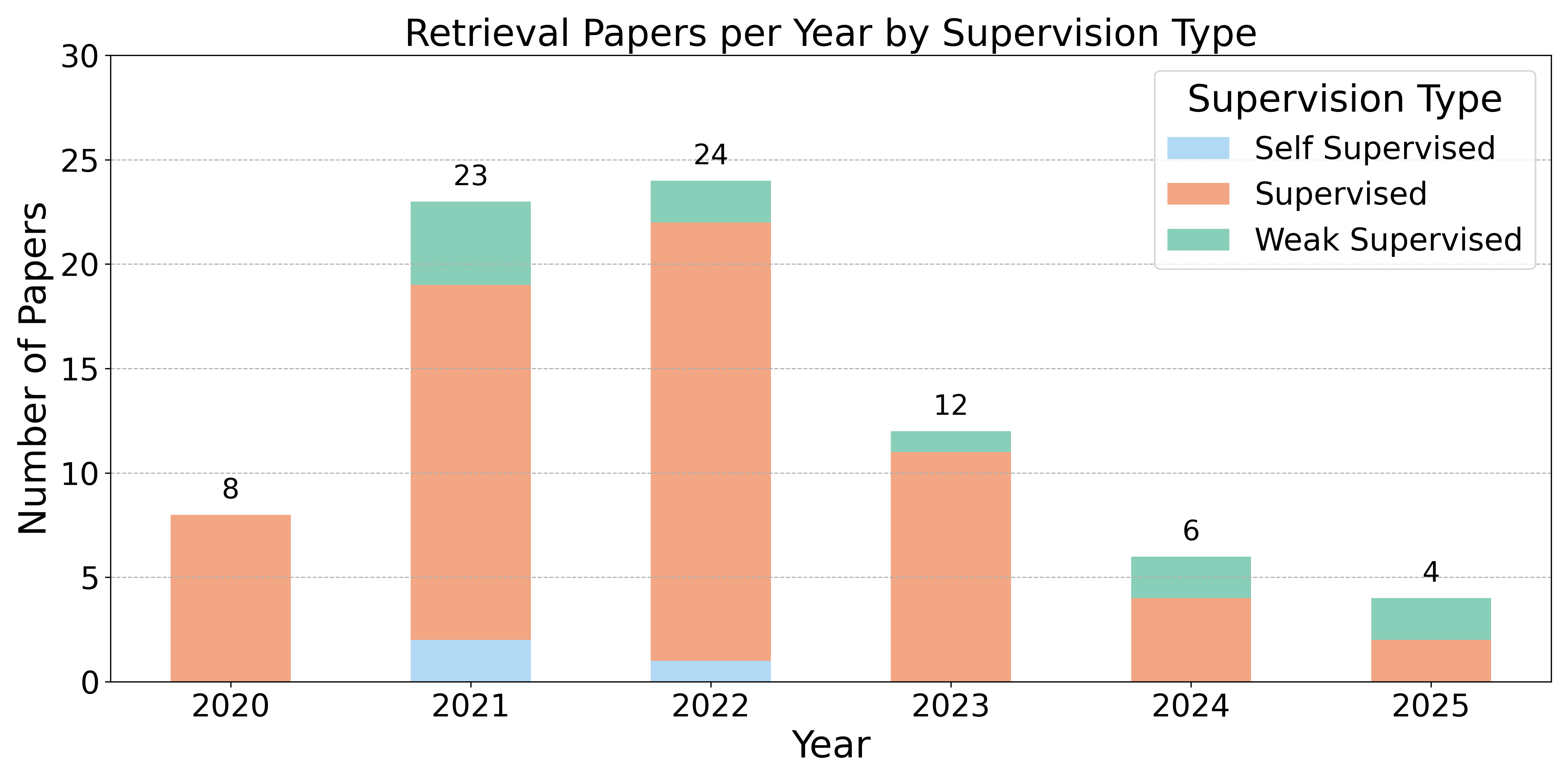}
        \caption{Supervision}
        \label{fig:supervision}
    \end{subfigure} 
    \hfill
    \begin{subfigure}[b]{0.45\textwidth}
        \centering
        \includegraphics[width=\linewidth]{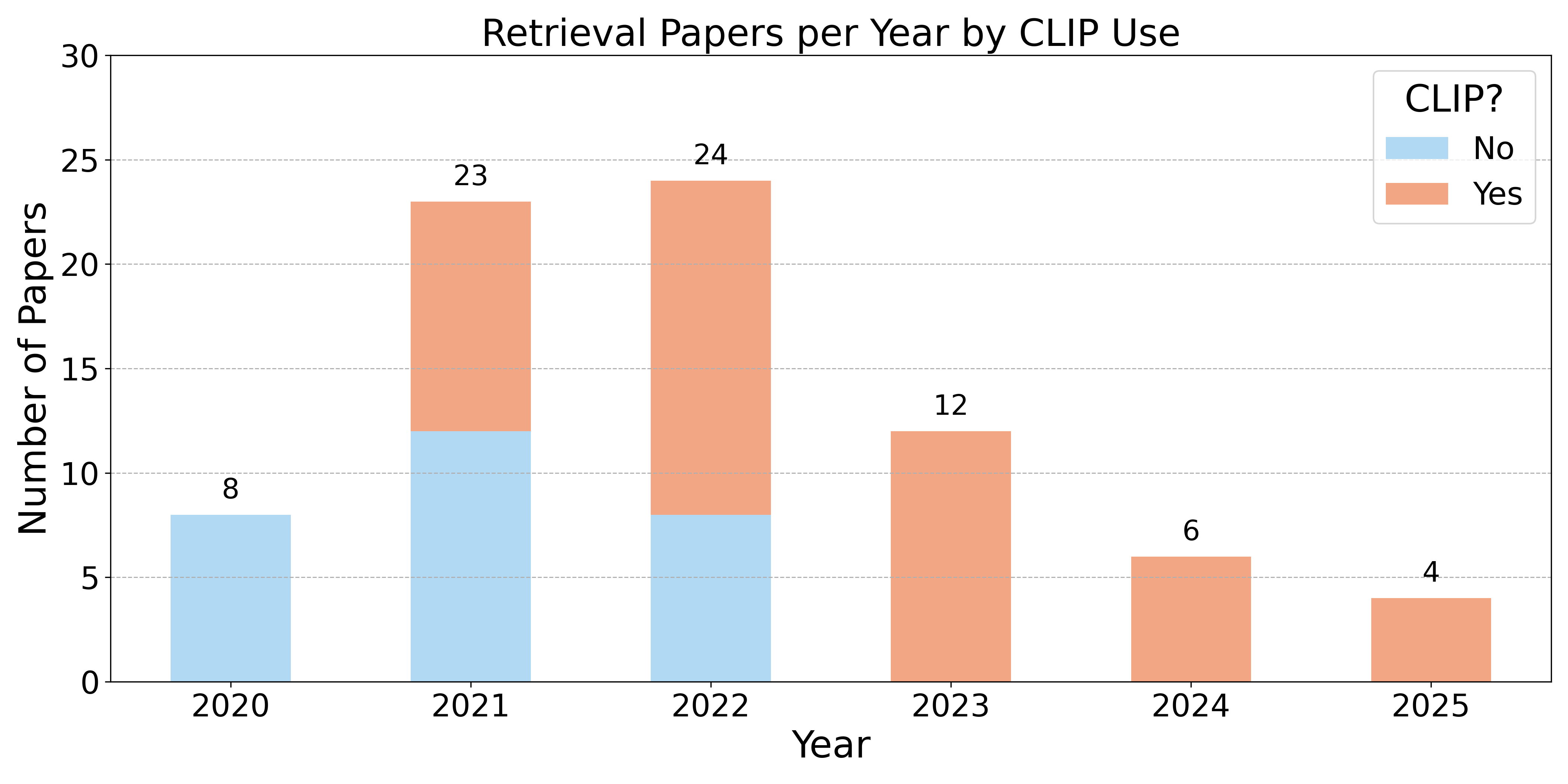}
        \caption{CLIP Type}
        \label{fig:clip_type}
    \end{subfigure}

    \vspace{0.2em}
    
    \begin{subfigure}[b]{0.45\textwidth}
        \centering
        \includegraphics[width=\linewidth]{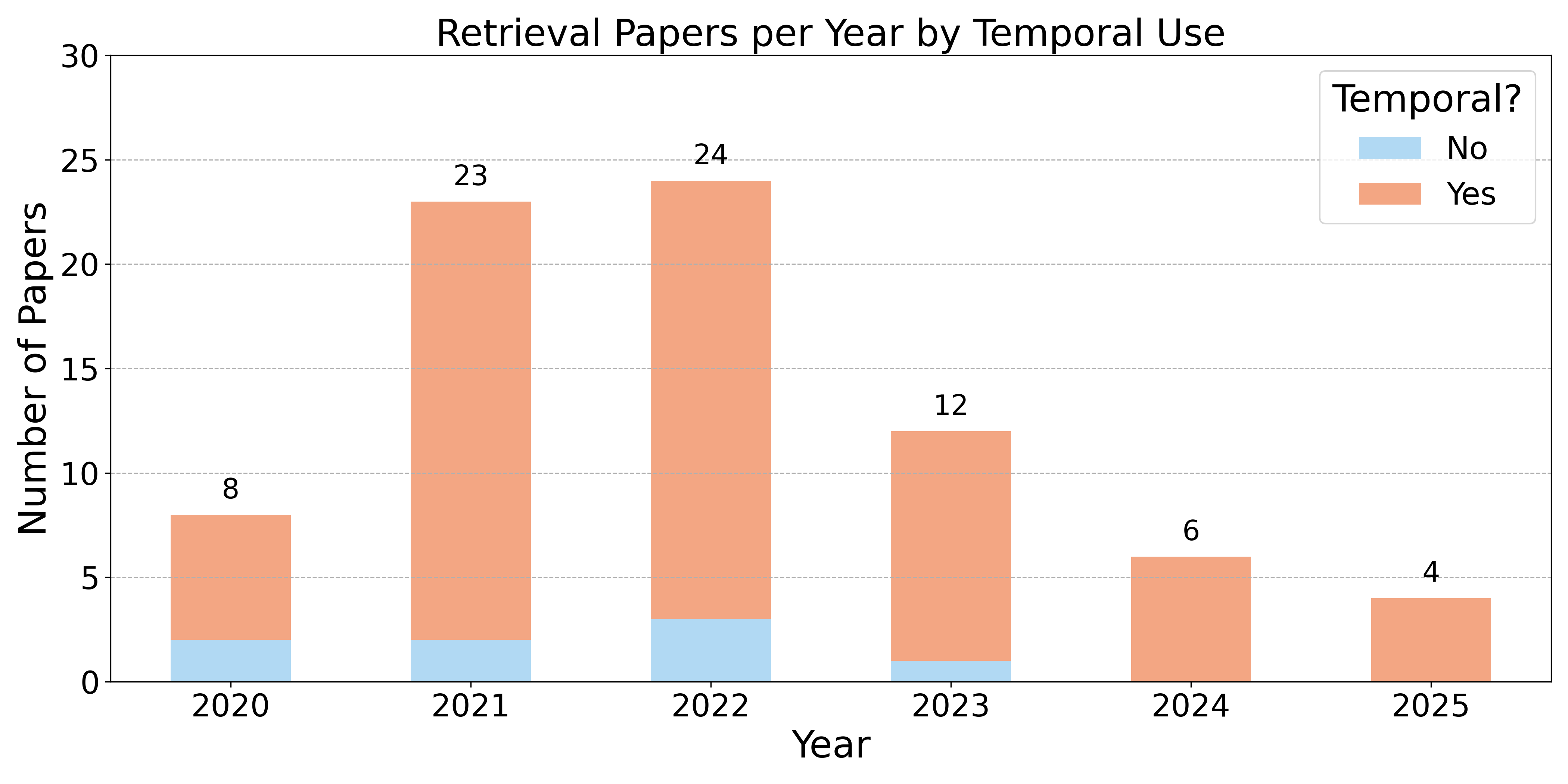}
        \caption{Temporal}
        \label{fig:temporal}
    \end{subfigure}
   \hfill 
    \begin{subfigure}[b]{0.45\textwidth}
        \centering
        \includegraphics[width=\linewidth]{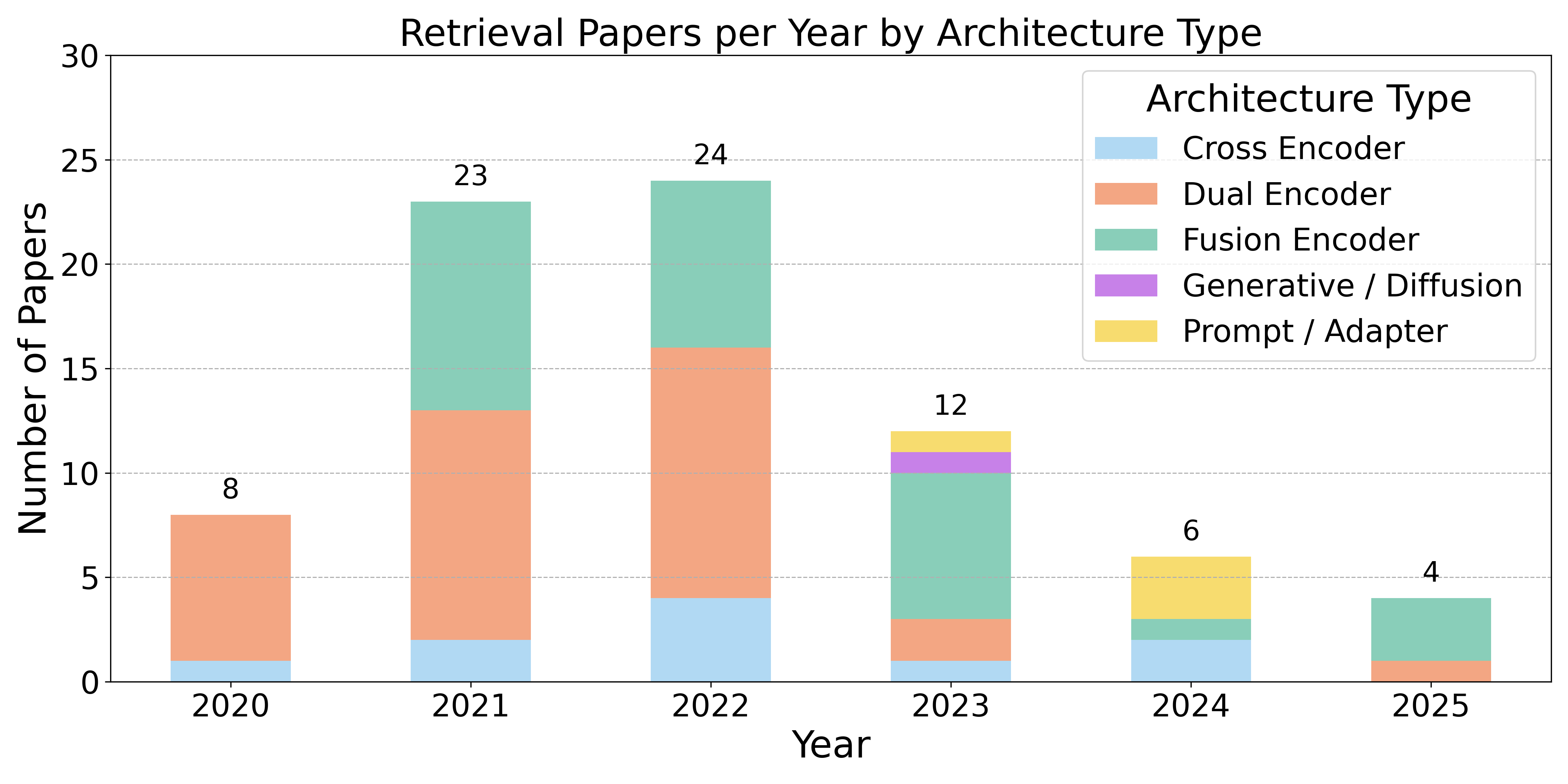}
        \caption{Architecture Type}
        \label{fig:architecture_type}
    \end{subfigure}
    
    \caption{Trends of video retrieval papers per year.}
    \label{fig:papers_per_year_3x3}
\end{figure}

Figure \ref{fig:papers_per_year_3x3} illustrates detailed yearly trends in video retrieval research between 2020 and 2025 across 6 aspects: (a)~retrieval task (b) input type, (c) supervision, (d) CLIP usage, (e) temporal modeling, and (f) architecture type. The figure shows a peak in publications in 2021-2022, followed by a decrease in the next years to the same level as in 2020. In short summary, the figure furthermore shows that (a)~text-to-video retrieval clearly dominates among tasks; (b)~frame-based inputs are slightly more common than full-video inputs due to their efficiency and easier preprocessing; most works use (c)~fully supervised and (d)~CLIP-based methods, though weakly supervised approaches are growing; (e)~temporal modeling is increasingly used to capture motion; and (f)~dual-encoder or prompt-based architectures are now the most common, supported by large pre-trained vision–language models.

Exploring the results in more detail, Figure~\ref{fig:retrieval_task} shows the distribution of retrieval tasks.
Among the 77 papers, text-to-video retrieval was the most common category, with 59 papers in total. Most methods (48 papers)~\cite{wray2021semantic,wang2021t2vlad,luo2022clip4clip,wang2022learn,dong2022partially,duarte2022sign,gorti2022x,wu2023cap4video,jiang2023dual,kang2023meme,li2023progressive,fang2023uatvr,yang2024dgl,zhang2024ump,galanopoulos2020attention,li2020sea,lokoc2020w2vv++,dong2021dual,croitoru2021teachtext,jin2022expectation,ma2022x,zhao2022centerclip,shvetsova2022everything,falcon2022feature,falcon2022relevance,jin2023text,jin2023diffusionret,dong2023dual,deng2023prompt,amato2023visione,tian2024holistic,shen2024tempme,portillo2021straightforward,yin2024exploiting,gao2021clip2tv,han2021visual,fang2021clip2video,yan2021video,wang2022disentangled,liu2021hit,dzabraev2021mdmmt,patricksupport,wang2022hybrid,lei2021less,ge2022miles,cheng2021improving,dong2022reading,ali2022video,yang2020tree} addressed the text-to-video retrieval task in a single-shot scenario using pre-trimmed videos~\cite{qing2022learning,wang2023protege,luo2022exploring}. The second most common category is the related image-and-text-to-video, with 9 papers~\cite{hu2023adaclip,falcon2022feature,zhang2021personalized,jiang2021learning,shvetsova2022everything,han2021visual,dzabraev2021mdmmt,wang2022hybrid,tang2025muse} while image-to-video retrieval has 7 papers \cite{gharahsouflou2022efficient,liu2021activity,zhu2025motionrag,liu2022activity,xu2020proposal,qiu2022challenges,yuan2020central}. The final category is video-to-video retrieval, with only 2 papers~\cite{jiang2021learning,nie2019joint}. This distribution shows the emphasis on text-to-video retrieval, for the reasons we outlined in the introduction.


Regarding input type, shown in Figure~\ref{fig:input_type}, some methods~\cite{wray2021semantic,wang2021t2vlad,luo2022clip4clip,wang2022learn,dong2022partially,duarte2022sign,gorti2022x,wu2023cap4video,jiang2023dual,kang2023meme,li2023progressive,fang2023uatvr,yang2024dgl,zhang2024ump,yin2024exploiting,gao2021clip2tv,han2021visual,wang2022disentangled,liu2021hit,ge2022miles,cheng2021improving} preprocess videos into frames internally, offering flexibility and a “black-box” setup for researchers, though this can lead to variable evaluation results. In contrast, others~\cite{galanopoulos2020attention,li2020sea,lokoc2020w2vv++,dong2021dual,croitoru2021teachtext,jin2022expectation,ma2022x,zhao2022centerclip,shvetsova2022everything,falcon2022feature,falcon2022relevance,jin2023text,jin2023diffusionret,dong2023dual,deng2023prompt,amato2023visione,tian2024holistic,shen2024tempme,tang2025muse} use pre-extracted frames as input, allowing flexible frame-splitting strategies that influence performance. Some methods~\cite{wang2022learn,dong2023dual} use precomputed visual and textual features, typically I3D (1024-D)~\cite{carreira2017quo} with or without ResNet152 (2048-D)~\cite{shafiq2022deep,panda2023modified,zhang2022infrared} and RoBERTa (1024-D)~\cite{lei2020tvr,ozkurt2024comparative,cheruku2024sentiment}.

Turning to Figure~\ref{fig:supervision}, methods can be categorized based on the type of supervision they employ: fully supervised~\cite{wray2021semantic,zhao2022centerclip,luo2022clip4clip,jin2022expectation,dong2022partially,ma2022x,gorti2022x,wu2023cap4video,jin2023text,dong2023dual,fang2023uatvr,yang2024dgl,shen2024tempme,zhang2024ump,wang2022learn,wang2021t2vlad,duarte2022sign,jiang2023dual,kang2023meme,li2023progressive,galanopoulos2020attention,li2020sea,lokoc2020w2vv++,dong2021dual,croitoru2021teachtext,shvetsova2022everything,falcon2022feature,falcon2022relevance,jin2023diffusionret,deng2023prompt,amato2023visione,tian2024holistic,tang2025muse,bain2021frozen}, weakly supervised~\cite{portillo2021straightforward,yin2024exploiting,yoon2021weakly,lu2024exploiting,wang2021weakly,fang2025multi,lv2025variational,yoon2023scanet,zheng2022weakly,zhang2022weakly,tan2021logan}, and self-supervised~\cite{akbari2021vatt,he2021self,wang2022align}. Fully supervised approaches dominate the field, mainly due to the availability of large-scale captioned datasets.

Figure~\ref{fig:clip_type} shows that many recent methods leverage CLIP-based~\cite{jin2022expectation,ma2022x,zhao2022centerclip,luo2022clip4clip,wang2022learn,shvetsova2022everything,gorti2022x,jin2023text,jin2023diffusionret,dong2023dual,jiang2023dual,kang2023meme,li2023progressive,deng2023prompt,fang2023uatvr,amato2023visione,yang2024dgl,tian2024holistic,shen2024tempme,tang2025muse,zhang2024ump,yin2024exploiting,fang2021clip2video,lei2021less} or other vision-language architectures such as Frozen~\cite{bain2021frozen} and VATT~\cite{akbari2021vatt}, often incorporating temporal information~\cite{galanopoulos2020attention,li2020sea,lokoc2020w2vv++,dong2021dual,wang2021t2vlad,zhao2022centerclip,luo2022clip4clip,wang2022learn,shvetsova2022everything,falcon2022feature,dong2022partially,duarte2022sign,falcon2022relevance,jin2022expectation,ma2022x,gorti2022x,wu2023cap4video,jin2023text,jin2023diffusionret,jiang2023dual,kang2023meme,li2023progressive,deng2023prompt,fang2023uatvr,amato2023visione,yang2024dgl,tian2024holistic,shen2024tempme,tang2025muse,zhang2024ump,bain2021frozen,yin2024exploiting,fang2021clip2video,yan2021video,wang2022disentangled,liu2021hit,ge2022miles,lei2021less}. Figure~\ref{fig:temporal} then shows that most approaches exploit temporal structure in video frames, highlighting its importance for text-to-video performance.

Finally, Figure~\ref{fig:architecture_type} distinguishes the methods by architectural design. Some methods focus on dual or multi-level encoding. Galanopoulos et al.~\cite{galanopoulos2020attention} use dual encoding with attention and fusion, Dong et al.~\cite{dong2021dual} employ multi-level dual encoding with hybrid space learning, and Li et al.~\cite{li2020sea} match videos and text using multiple sentence encoders (SEA). Loko\v{c} et al.~\cite{lokoc2020w2vv++} show that user involvement improves retrieval. Other works emphasize fine-grained and global-local alignment. Wang et al.~\cite{wang2021t2vlad} use global-local alignment, Wang et al.~\cite{wang2022learn} handle negation with soft negatives, Jiang et al.~\cite{jiang2023dual} mine hard negatives and model fine-grained similarity, Li et al.~\cite{li2023progressive} align objects and temporal events, and Kang et al.~\cite{kang2023meme} combine coarse- and fine-grained features. Several methods target representation enhancement and multi-modal fusion. Croitoru et al.~\cite{croitoru2021teachtext} reduce dataset noise with TEACHTEXT, Shvetsova et al.~\cite{shvetsova2022everything} fuse video, audio, and text, Falkon et al.~\cite{falcon2022feature} apply feature-space augmentation and Deng et al.~\cite{deng2023prompt} use a Prompt Cube with captioning.

Overall, text-to-video retrieval dominates recent research, with architectures ranging from CLIP-based to feature- or transformer-based models. These methods differ in input type, use of temporal information, and feature representation. We  discuss our choice of 14 representative methods and provide a detailed comparative analysis in Section~\ref{sec:architectures}.

\subsection{Architectures and Research Gaps}
\label{sec:archs_gaps}

\textcolor{black}{Before the dominance of large-scale transformer and contrastive vision-language models (2020–2025), the evolution of video and multimedia retrieval was rooted in classical computer vision and structured database indexing techniques. Early video retrieval systems relied heavily on handcrafted visual features (such as color histograms~\cite{swain1991color}, texture patterns~\cite{datta2008image}, scale-invariant feature transforms~\cite{lowe2004distinctive} and motion descriptors~\cite{el2017content}) coupled with Bag-of-Visual-Words (BoVW)~\cite{shekhar2012word} frameworks. Throughout the 2000s and early 2010s, temporal modeling depended on rigid mathematical structures such as Hidden Markov Models (HMMs)~\cite{li2002applying} and trajectory analysis to bridge the gap between static frames and dynamic video content. As deep learning matured between 2015 and 2019, the field transitioned toward hybrid architectures that combined Convolutional Neural Networks (CNNs)~\cite{yan2007review} for spatial frame feature extraction with Recurrent Neural Networks (RNNs)~\cite{chen2007interactive} or LSTMs~\cite{reiter2006combined} for temporal sequence modeling. While these classical and early deep learning paradigms laid essential engineering foundations for real-world visual understanding, contemporary research has largely shifted away from pipeline-heavy feature engineering toward joint multimodal embedding spaces to optimize semantic query alignment. Building upon the general video retrieval foundations discussed in Section~\ref{sec:overall_trends}, recent literature (particularly over the last five years) has focused intensely on text-to-video retrieval.} These approaches differ in how they model video–text interactions, temporal dependencies, and semantic alignment. Table~\ref{tab:arch_gaps} provides an overview of the main architectures, their limitations, and open research gaps, highlighting areas where improvements are needed, such as efficient temporal modeling, robust cross-modal alignment, and better generalization.

As Table~\ref{tab:arch_gaps}  lays out, video retrieval architectures can be broadly categorized as follows. CLIP-based methods~\cite{luo2022clip4clip,jin2022expectation,jiang2023dual,fang2021clip2video,gao2021clip2tv,bain2021frozen,ma2022x,gorti2022x,zhao2022centerclip,dong2022partially,shen2024tempme,yang2024dgl,yin2024exploiting,jin2023text,duarte2022sign,wang2022learn,li2023progressive,fang2023uatvr,deng2023prompt,zhang2024ump,tang2025muse,wu2023cap4video} leverage pretrained CLIP embeddings and include variants such as fine-tuning, structural enhancement, alignment, prompting, and generative assistance. These approaches excel at semantic alignment and transfer learning but often struggle with temporal modeling, domain shifts, and computational cost. Dual encoder and joint embedding methods~\cite{li2020sea,lokoc2020w2vv++,croitoru2021teachtext,jiang2021learning,ali2022video,nie2019joint,liu2021activity,liu2022activity,falcon2022relevance,wang2022disentangled,falcon2022feature,wang2022align,wang2022hybrid} learn a shared embedding space for videos and text, enabling efficient retrieval through similarity search. While scalable and efficient, these models typically exhibit limited temporal reasoning and may overfit on fine-grained tasks. Multimodal fusion approaches~\cite{dong2021dual,cheng2021improving,yuan2020central,falcon2022feature,jiang2021learning,zhang2021personalized,ge2022miles,patricksupport,luo2022clip4clip,bain2021frozen,ma2022query} \textcolor{black}{generally divide into early, late and intermediate integration approaches. Early fusion combines representations at the input or feature-extraction stage, allowing low-level interactions but often struggling with modality-specific noise and differing sampling rates. Conversely, late fusion processes each modality independently through dedicated encoders before merging their final predictions, which simplifies training yet misses fine-grained, cross-modal correlations. Recent advancements favor intermediate fusion paradigms (utilizing cross-attention mechanisms to dynamically exchange information across modalities at multiple network depths) combining information from multiple modalities through multi-branch networks or memory-based mechanisms. They capture complementary signals from video and text but tend to be computationally intensive and sensitive to modality imbalance.} Transformer-based methods~\cite{akbari2021vatt,dzabraev2021mdmmt,yan2021video,liu2021hit,lei2021less} model spatio-temporal dependencies through attention mechanisms. They achieve strong temporal reasoning and flexible cross-modal interactions, yet their high computational cost and slow inference limit scalability. Structured, region, and graph-based methods~\cite{yang2020tree,tan2021logan,han2021visual,yoon2023scanet,dong2022partially,xu2020proposal,he2021self} leverage object relations, local regions, or temporal proposals to improve fine-grained reasoning. While effective at capturing local interactions, they are sensitive to noise and often require complex implementations. Generative approaches~\cite{jin2023diffusionret,lv2025variational,zhu2025motionrag,wu2023cap4video} generate intermediate representations such as captions or embeddings to aid retrieval. These methods can improve semantic coverage but are slow during inference and computationally expensive to train. Finally, attention-based architectures~\cite{galanopoulos2020attention,han2021visual,tan2021logan} use token-level cross-attention to align video and text features. They capture fine-grained interactions effectively but are sensitive to noisy frames and limited in scalability to long sequences.

Finally, we note that some of the works cited above focus on the efficiency or system-level aspects of retrieval. Amato et al.~\cite{amato2023visione} introduce VISIONE for large-scale retrieval, Tian et al.~\cite{tian2024holistic} transfer fine-grained knowledge with TeachCLIP, Tang et al.~\cite{tang2025muse} use multi-scale features in MUSE, Duarte et al.~\cite{duarte2022sign} tackle sign language retrieval with SPOT-ALIGN, Falcon et al.~\cite{falcon2022relevance} improve ranking with a relevance-based margin, and Jin et al.~\cite{jin2023diffusionret} adopt a generative-discriminative approach in DiffusionRet.

\subsection{Comparison to Existing Surveys}
\label{sec:surveys}

Previous surveys~\cite{rossetto2020interactive,zhu2023deep,liu2023survey,perez2022comprehensive,lei2024survey,lan2023survey,    schiappa2023self,madan2024foundation,xing2024survey,tang2025video,wan2025composed} focused on benchmark metrics, rarely considering the impact of query characteristics. Rosetto et al.~\cite{rossetto2020interactive} analyzed interactive video retrieval via the 8th Video Browser Showdown, summarizing team systems, user behavior, and feature trends. Zhu et al.~\cite{zhu2023deep} reviewed 100+ video-to-text methods, highlighting deep learning advances and cross-modal challenges. Perez et al.~\cite{perez2022comprehensive} focused on video-to-text retrieval and captioning, emphasizing spatiotemporal complexity. Liu et al.~\cite{liu2023survey} surveyed video moment localization, summarizing methods, datasets, and future directions. Lei et al.~\cite{lei2024survey} reviewed video-language models, focusing on feature extraction and embeddings. Lan et al.~\cite{lan2023survey} categorized temporal sentence grounding methods and datasets. Madan et al.~\cite{madan2024foundation} examined 200+ Video Foundation Models, highlighting universal multimodal and self-supervised approaches. Xing et al.~\cite{xing2024survey} covered video diffusion models, including generation, editing, and understanding. Tang et al.~\cite{tang2025video} surveyed Vid-LLMs for video understanding, including reasoning, tasks, and datasets. Wan et al.~\cite{wan2025composed} reviewed composed image retrieval, summarizing features, alignment, fusion, datasets, and metrics. Schiappa et al.~\cite{schiappa2023self} surveyed self-supervised video representation learning, providing core representations relevant for retrieval.

\noindent
\begin{turn}{90} 
    \begin{minipage}{\textheight} 
        \small
        \centering
        \captionof{table}{Video retrieval architectures, limitations, and research gaps (highlighting missing capabilities).}
        \begin{tabular}{p{4.7cm} p{3.5cm} p{3.5cm} p{4cm}}
            \toprule
            \textbf{Architecture} & \textbf{Subcategory} & \textbf{Limitations} & \textbf{Research Gaps} \\
            \midrule
            CLIP-based
             & Fine-Tuning / Adaptation~\cite{luo2022clip4clip, jin2022expectation, jiang2023dual, fang2021clip2video, gao2021clip2tv, bain2021frozen}
             & Limited temporal modeling, domain shift sensitivity
             & Temporal reasoning across datasets is underexplored \\
            \cmidrule{2-4}
             & Structural Enhancement~\cite{ma2022x, gorti2022x, zhao2022centerclip, dong2022partially, shen2024tempme, yang2024dgl, yin2024exploiting}
             & High computation and memory
             & Efficient multi-scale structural enhancements are lacking \\
             & Alignment~\cite{jin2023text, duarte2022sign, wang2022learn, li2023progressive}
            & Limited local and fine-grained alignment
            & Robust local and temporal alignment methods are scarce \\
            \cmidrule{2-4}
            
            & Prompting~\cite{fang2023uatvr, deng2023prompt, zhang2024ump, tang2025muse}
            & Sensitive to prompt design
            & Adaptive and generalizable prompt strategies are underdeveloped \\
            \cmidrule{2-4}
            
            & Generative Assistance~\cite{wu2023cap4video}
            & Slow inference, complex training
            & Scalable generative retrieval for real-time applications is missing \\
            
            \midrule
            Dual Encoder / Joint Embedding
            & Classical Embedding~\cite{li2020sea, lokoc2020w2vv++, croitoru2021teachtext, jiang2021learning, ali2022video, nie2019joint, liu2021activity, liu2022activity}
            & Weak generalization, limited temporal reasoning
            & Zero-shot and cross-domain retrieval capabilities are limited \\
            \cmidrule{2-4}
            
            & Metric Learning~\cite{falcon2022relevance, wang2022disentangled, falcon2022feature, wang2022align, wang2022hybrid}
            & Mostly global embeddings, risk of overfitting
            & Local and temporal metric learning methods are underexplored \\
            
            \midrule
            Multimodal Fusion
            & Multi-Branch~\cite{dong2021dual, cheng2021improving, yuan2020central, falcon2022feature, jiang2021learning, zhang2021personalized,luo2022clip4clip,bain2021frozen,ma2022query}
            & Computationally heavy, modality imbalance
            & Balanced and efficient multi-branch fusion techniques are lacking \\
            \cmidrule{2-4}
            
            & Memory-Based~\cite{ge2022miles, patricksupport}
            & Memory-intensive, slow inference
            & Scalable memory-efficient fusion approaches are missing \\
            \midrule
            \multicolumn{4}{r}{\emph{Continued on next page}} \\
            \bottomrule
        \end{tabular}
        \label{tab:arch_gaps}
    \end{minipage}
\end{turn}

\addtocounter{table}{-1} 

\noindent
\begin{turn}{90} 
    \begin{minipage}{\textheight} 
        \small
        \centering
        \captionof{table}{Continued from previous page}
        \begin{tabular}{p{4.7cm} p{3.5cm} p{3.5cm} p{4cm}}
        \toprule
          \textbf{Architecture} & \textbf{Subcategory} & \textbf{Limitations} & \textbf{Research Gaps} \\
            \midrule
            \multirow{1}{*}{Transformer-based}
            & Full Transformer~\cite{akbari2021vatt, dzabraev2021mdmmt, yan2021video, liu2021hit, lei2021less}
            & High computation, slow inference
            & Efficient long-range temporal modeling is underexplored \\
            
            \midrule
            \multirow{3}{*}{Structured / Region / Graph}
            & Graph~\cite{yang2020tree, tan2021logan, han2021visual}
            & Sensitive to noise, complex implementation
            & Robust graph-based video representations are scarce \\
            \cmidrule{2-4}
            
            & Region / Part / Scale~\cite{yoon2023scanet, dong2022partially}
            & Depends on region proposals, high computation
            & Adaptive region and scale selection methods are limited \\
            \cmidrule{2-4}
            
            & Temporal / Proposal~\cite{xu2020proposal, he2021self}
            & Heavy computation, inconsistent proposals
            & Efficient and robust temporal proposal strategies are underexplored \\
            
            \midrule
            \multirow{3}{*}{Generative}
            & Diffusion / Variational~\cite{jin2023diffusionret, lv2025variational}
            & Expensive training, slow inference
            & Fast generative retrieval approaches are underexplored \\
            \cmidrule{2-4}
            
            & Retrieval-Augmented~\cite{zhu2025motionrag}
            & Resource-heavy, slow inference
            & Efficient RAG strategies for large-scale video retrieval are lacking \\
            \cmidrule{2-4}
            
            & Caption-Guided~\cite{wu2023cap4video}
            & Caption mismatch, slow inference
            & Accurate caption-guided retrieval in real-time remains a challenge \\
            
            \midrule
            \multirow{1}{*}{Attention}
            & Token-level cross-attention~\cite{galanopoulos2020attention, han2021visual, tan2021logan}
            & Sensitive to noisy frames, high computation
            & Noise-robust, scalable cross-attention methods are underexplored \\
            
            \bottomrule
        \end{tabular}
\end{minipage}
\end{turn}

\begin{table*}[t!]
    \centering
    \caption{Comparison of prior surveys and this study in text-video retrieval (2020–2025).}
    \footnotesize
    \setlength{\tabcolsep}{4pt} 
    \begin{tabular}{p{3cm}p{4cm}p{5cm}}
        \toprule
        \multicolumn{1}{l}{\bf Aspect} & \multicolumn{1}{l}{\bf Prior Surveys} & \multicolumn{1}{l}{\bf This Study} \\
          & \multicolumn{1}{l}{\bf (2020–2025)} &  \\
        \midrule
        Focus & \raggedright Reviewed architectures, benchmarks, datasets~\cite{rossetto2020interactive,zhu2023deep,perez2022comprehensive,liu2023survey,lei2024survey,lan2023survey,madan2024foundation,xing2024survey,tang2025video,wan2025composed,schiappa2023self} & 14 methods, 3 datasets evaluated (Section~\ref{sec:traditional_experiments}) \\
        Query Analysis & \raggedright Rarely examined~\cite{rossetto2020interactive,zhu2023deep,perez2022comprehensive,liu2023survey} & Difficulty and category analyzed (Section~\ref{sec:task_category}-\ref{sec:task_difficulty}) \\
        Dataset Influence & \raggedright Broad overview~\cite{zhu2023deep,perez2022comprehensive,lei2024survey,lan2023survey} & Composition, size, bias quantified (Section~\ref{sec:main_experiment}) \\
        Captions & \raggedright Mostly ignored~\cite{rossetto2020interactive,zhu2023deep,schiappa2023self} & Clarity, simplicity, length examined (Section~\ref{sec:task_difficulty}) \\
        Performance & \raggedright Overall SOTA reported~\cite{rossetto2020interactive,zhu2023deep,perez2022comprehensive,liu2023survey} & Strengths/weaknesses highlighted (Section~\ref{sec:metrics}) \\
        Reproducibility & \raggedright Limited discussion~\cite{zhu2023deep,perez2022comprehensive,schiappa2023self} & Preprocessing, frame rate, compression assessed (Section~\ref{sec:sensitivity_analysis}) \\
        Efficiency & \raggedright Often ignored~\cite{madan2024foundation,xing2024survey} & Training vs performance trade-offs shown (Section~\ref{sec:training_time_vs_epochs}) \\
        Cross-Dataset Transfer & \raggedright Not deeply studied~\cite{lei2024survey,lan2023survey} & Generalization and diversity effects examined (Section~\ref{sec:cross_dataset_analysis}) \\
        Recommendations & \raggedright Few~\cite{rossetto2020interactive,zhu2023deep,perez2022comprehensive} & Guidance on captions, frame sampling, query balance (Section~\ref{sec:clustering_by_task_category}) \\
        \bottomrule
    \end{tabular}
    \label{tab:ours_vs_surveys}
\end{table*}

Although previous surveys provide overviews of video-to-text methods and benchmarks, they rarely consider dataset or query characteristics. Table~\ref{tab:ours_vs_surveys} summarizes these surveys and shows how this study addresses their gaps. In this work, we analyze how task category, query difficulty, number of captions, and video properties affect retrieval performance, focusing on dataset features rather than new architectures.

\section{Methodology}
\label{sec:methodology}

This section presents our methodology for analyzing the relationship between textual query characteristics and video retrieval performance across multiple datasets and methods. Crucially, rather than aiming to propose a new state-of-the-art model or crown a single ``best'' method, our primary goal is diagnostic and analytical: to understand how and why query properties influence retrieval behavior across different architectures. To this end, we first describe the video retrieval architectures considered (Section~\ref{sec:architectures}), followed by a description of the datasets used in our evaluation (Section~\ref{sec:datasets}).
Next, we define evaluation metrics (Section~\ref{sec:metrics}), quantify query difficulty  to systematically identify what makes specific queries challenging for current models, and to investigate whether models with comparable overall performance share underlying failure modes by succeeding or stumbling on the same specific queries. (Section~\ref{sec:quantifying_query_difficulty}), and introduce a semantic categorization of textual queries along with the notion of  \textit{single}  semantic queries (Section~\ref{sec:semantic_query_categorization}) 
to study the impact of query composition on retrieval performance.

\subsection{Architectures}
\label{sec:architectures}

From the broad range of video retrieval approaches summarized in Table~\ref{tab:arch_gaps}, we select 14 representative methods for detailed evaluation, chosen for strong benchmark performance, public code availability, and coverage of diverse paradigms including CLIP-based adaptation, dual encoders, transformers, prompt tuning and temporal modeling.\footnote{Note that not all categories from Table~\ref{tab:arch_gaps} are represented, as for some categories no methods had publicly available code and their authors did not respond to our requests for access to their code.}

Table~\ref{tab:methods} gives details of these 14 selected implementations. They span four distinct focus areas across broader architectural categories: (1) six CLIP-based methods (CLIP4CLIP~\cite{luo2022clip4clip},  Cap4Video~\cite{wu2023cap4video} (also features generative aspects), X-CLIP~\cite{ma2022x}, DGL~\cite{yang2024dgl}, UMP~\cite{zhang2024ump}, CenterCLIP~\cite{zhao2022centerclip}); (2) two dual-encoder-based methods (EMCL~\cite{jin2022expectation}, TempMe~\cite{shen2024tempme}); (3) three structured-based (PRVR~\cite{dong2022partially}, DL-DKD~\cite{dong2023dual}, UATVR~\cite{fang2023uatvr}) and (4) three attention-based (XPOOL~\cite{gorti2022x}, DiCoSA~\cite{jin2023text}, SSVR~\cite{wray2021semantic}).

\begin{table}[t!]
    \centering
    \caption{State-of-the-art methods from the past 6 years with available GitHub repositories and key features used in our evaluation.}
    \footnotesize
    \setlength{\tabcolsep}{3pt} 
    \renewcommand{\arraystretch}{0.9} 
    \begin{tabular}{lccccl}
        \toprule
        \multicolumn{1}{c}{\bf Method} 
        & \multicolumn{1}{c}{\bf Input}
        & \multicolumn{1}{c}{\bf Temporal}
        & \multicolumn{1}{c}{\bf CLIP?} 
        & \multicolumn{1}{c}{\bf Year}
        & \multicolumn{1}{c}{\bf GitHub}\\ 
        \multicolumn{1}{c}{\bf } 
        & \multicolumn{1}{c}{\bf (V / F)}
        & \multicolumn{1}{c}{\bf (\cmark /\ \xmark)}
        & \multicolumn{1}{c}{\bf (\cmark /\ \xmark)}
        & \multicolumn{1}{c}{\bf }
        & \multicolumn{1}{c}{\bf Link}\\
        \midrule 
        SSVR~\cite{wray2021semantic} & V & \xmark & \xmark & 2021 & \url{https://mwray.github.io/SSVR/}\\
        CenterCLIP~\cite{zhao2022centerclip} & F & \cmark & \cmark & 2022 & \url{https://github.com/mzhaoshuai/CenterCLIP}\\
        CLIP4CLIP~\cite{luo2022clip4clip} & V & \cmark & \cmark & 2022 & \url{https://github.com/ArrowLuo/CLIP4Clip}\\
        EMCL~\cite{jin2022expectation} & F & \cmark & \cmark & 2022 & \url{https://github.com/jpthu17/EMCL}\\
        PRVR~\cite{dong2022partially} & V & \cmark & \xmark & 2022 & \url{https://github.com/HuiGuanLab/ms-sl}\\
        X-CLIP~\cite{ma2022x} & F & \cmark & \cmark & 2022 & \url{https://github.com/xuguohai/X-CLIP}\\
        XPOOL~\cite{gorti2022x} & V & \cmark & \cmark & 2022 & \url{https://github.com/layer6ai-labs/xpool}\\
        Cap4Video~\cite{wu2023cap4video} & V & \cmark & \cmark & 2023 & \url{https://github.com/whwu95/Cap4Video}\\
        DiCoSA~\cite{jin2023text} & F & \cmark & \cmark & 2023 & \url{https://github.com/jpthu17/DiCoSA}\\
        DL-DKD~\cite{dong2023dual} & F & \xmark & \cmark & 2023 & \url{https://github.com/HuiGuanLab/DL-DKD} \\
        UATVR~\cite{fang2023uatvr} & V & \cmark & \cmark & 2023 & \url{https://github.com/bofang98/UATVR}\\
        DGL~\cite{yang2024dgl} & V & \cmark & \cmark & 2024 & \url{https://github.com/knightyxp/DGL}\\
        UMP~\cite{zhang2024ump} & V & \cmark & \cmark & 2024 & \url{https://github.com/zchoi/UMP_TVR}\\
        TempMe~\cite{shen2024tempme} & F & \cmark & \cmark & 2025 & \url{https://github.com/LunarShen/TempMe}\\
        \bottomrule
    \end{tabular}
    \label{tab:methods}
\end{table}

These methods highlight key advances across architectures: SSVR and CenterCLIP target semantic alignment and efficiency; CLIP4CLIP, EMCL, and Cap4Video utilize pretrained CLIP features; PRVR, X-CLIP, XPOOL, DiCoSA, DL-DKD, and UATVR focus on fine-grained retrieval; and DGL, UMP and TempMe improve prompt design and temporal modeling. Together, these methods provide a diverse and representative collection for evaluating retrieval performance.

\textcolor{black}{To contextualize model complexity alongside retrieval performance, practical deployment also requires examining efficiency, inference latency, and memory consumption. Because these evaluated models rely on existing, off-the-shelf codebases rather than newly proposed network architectures, exact runtimes and VRAM footprints vary by implementation. Nonetheless, their operational scalability is associated with three key principles:}

\begin{enumerate}
    \item \textcolor{black}{\textbf{Asymptotic Bounds:} Feature extraction scales linearly ($\mathcal{O}(N)$) with the number of input frames or tokens, ensuring predictable computational scaling.}
    \item \textcolor{black}{\textbf{Operational Decoupling:} Heavy workloads—such as raw video frame decoding and deep backbone feature extraction—are decoupled from online queries and pre-computed offline.}
    \item \textcolor{black}{\textbf{Lightweight Query Execution:} Online retrieval relies exclusively on text encoding and rapid vector similarity lookups against pre-cached indices.}
\end{enumerate}

\label{sec:model_encoders}

All 14 encoders are trained and tested under the same setup for fair comparison; the code is made available to the community for reproducibility.\footnote{\textcolor{black}{\url{https://github.com/mpegia/survey-and-reproducibility-scripts-for-text-to-video-retrieval-methods.git}}}  Implementations use \texttt{Python 3.10} and \texttt{PyTorch} with default parameters on a workstation with an \texttt{Intel i9-10920X} CPU, \texttt{NVIDIA RTX 4090} GPU, and \texttt{Ubuntu 20.04.5 LTS}. Models are trained for 1–5 epochs \textcolor{black}{using a standardized batch size of 32} on videos compressed at 3 FPS; some extract frames as preprocessing, others handle it internally. 
Results are grouped to reflect these differences, with Section~\ref{sec:sensitivity_analysis} analyzing the impact of FPS and compression.

\subsection{Datasets}
\label{sec:datasets}

We test our methods on several popular benchmark datasets. Table~\ref{tab:datasets} shows details for 3 commonly used datasets: MSRVTT\footnote{\url{https://ms-multimedia-challenge.com/2017/dataset}}~\cite{xu2016msr}, MSVD\footnote{\url{https://www.cs.utexas.edu/\textasciitilde ml/clamp/videoDescription/}}~\cite{chen2011collecting}\, and LSMDC\footnote{\url{https://sites.google.com/site/describingmovies/download}}~\cite{lsmdc}. It lists the number of videos, the average number of frames per video, the average duration (in seconds), the number of captions per video, and the train/test splits. The default parameters provided by the authors are used for all datasets across the experiments.\footnote{The exception is that the LSMDC dataset contains only 6,209 out of 7,408 videos for training, because the remaining 799 videos are corrupted.} MSVD (2012) is the earliest dataset, followed by LSMDC (2015) and MSRVTT (2016). MSVD and MSRVTT contain short YouTube videos with captions written by human annotators via crowdsourcing. LSMDC, instead, is built from movies (films from the 1990s–2010s) and uses professionally written audio descriptions aligned with video clips.

\begin{table}[t!]
    \centering
    \caption{Benchmark Datasets.}
    \begin{tabular}{lcccccc}
        \toprule
        \multicolumn{1}{l}{\bf Dataset} 
        & \multicolumn{1}{c}{\bf Videos}
        & \multicolumn{1}{c}{\bf Frames/Vid}
        & \multicolumn{1}{c}{\bf Duration (s)} 
        & \multicolumn{1}{c}{\bf Caps/Vid} & \multicolumn{1}{c}{\bf Train} & \multicolumn{1}{c}{\bf Test}\\ 
        \midrule 
        LSMDC~\cite{lsmdc} & 118,081 & 14 & \phantom{1}2-30 & ~ \phantom{0}1 & 6,209 & 1,000\\
        MSRVTT~\cite{xu2016msr} & \phantom{1}10,000 & 46 & 10-32 & ~20 & 7,000 & 1,000\\
        MSVD~\cite{chen2011collecting} & \phantom{18}1,970 & 30 & \phantom{1}1-62 & ~40 & 1,200 & \phantom{1,}670\\
        \bottomrule
    \end{tabular}
    \label{tab:datasets}
\end{table}

While LSMDC is primarily action-oriented, whereas MSRVTT and MSVD are more scene-oriented, we expect that this inherent dataset bias significantly affects semantic-category performance and generalization. Because LSMDC is derived from professional audio descriptions intended for the visually impaired, which are professionally authored rather than crowd-sourced by general users, models trained on it tend to excel at recognizing fine-grained human actions and narrative contexts. However, they may struggle with the diverse, open-domain web video scenes found in MSRVTT and MSVD. Conversely, models optimized for crowd-sourced, scene-oriented datasets might underperform on professionally narrated action clips; an expectation that our results subsequently confirm.

\subsection{Evaluation Metrics}
\label{sec:metrics}

We evaluate models using standard video–text retrieval metrics and linguistic quality measures for captions. During retrieval, we retain the 1000 top-ranked videos for each query for analysis. \textcolor{black}{Our performance evaluation focuses on \textit{Recall@k} with $k \in \{1, 10, 50\}$ alongside training time per epoch. While prior academic literature reports $k \in \{1, 5, 10\}$, we explicitly incorporate \textit{Recall@50} because it represents a large class of real-world interactive video retrieval applications (such as systems from Video Browser Showdown (VBS) competition~\cite{vadicamo2024evaluating}. Omitting \textit{Recall@50} would overlook long-tail robustness and retrieval depth essential for operational frameworks. Meanwhile, we deliberately omit \textit{Recall@5}, because it tracks closely with \textit{Recall@10} within a narrow margin. Using \textit{Recall@5}  would unnecessarily expand tables and figures, and reduce overall paper readability. To ensure that performance differences are statistically robust rather than artifacts of random variance, we evaluate significance using paired Wilcoxon signed-rank tests~\cite{woolson2007wilcoxon}.}

\subsection{Query Difficulty}
\label{sec:quantifying_query_difficulty}

To understand how challenging each textual query is for retrieval models, we define the \textit{difficulty} of a query based on its \textit{average ground-truth rank} across all evaluated methods. The ground-truth rank represents the rank position at which the correct video is retrieved \textcolor{black}{within a candidate pool of 1000 items}. Lower values indicate easier queries, while higher values indicate harder ones.
To formalize this observation, we empirically classify queries into three difficulty levels based on their average ground-truth rank.  \textcolor{black}{Rather than utilizing contiguous partitions, we introduce strategic gaps and buffers between our threshold limits to ensure clear class separability and prevent noisy intermediate data points from skewing our findings:}

\begin{description}
    \item[\textbf{Easy}] Average ground truth rank < 200
    \item[\textbf{Medium}] Average ground truth rank 401--600
    \item[\textbf{Hard}] Average ground truth rank > 800
\end{description}

\textcolor{black}{Queries falling into the intermediate unassigned intervals (i.e., ranks 201--400 and 601--800) are omitted exclusively from the comparative difficulty-tier analysis to preserve strict categorical integrity and prevent ambiguity. Crucially, these excluded queries are not discarded from the overall system evaluation, and remain fully accounted for in aggregate metrics.} This categorization allows us to analyze retrieval performance not only by aggregate metrics but also by how well methods handle queries of varying difficulty.

\subsection{Query Categories for Semantic Analysis}
\label{sec:semantic_query_categorization}

\subsubsection{Semantic Categories}
\label{sec:semantic_categories}

To analyze model performance across different types of textual queries, we first categorize queries by their semantic characteristics. 
Based on~\cite{bird2009natural,honnibal2020spacy}, we apply 10 primary semantic categories to classify the content of textual queries:

\vfill
~
\begin{description}
    \item[\textbf{Action (Act)}] run, jump, hit, throw, ride, eat, chop, etc.
    \item[\textbf{Scene-related/Static (Scn)}] sit, stand, be, exist, etc.
    \item[\textbf{Cognitive (Cog)}] think, know, believe, etc.
    \item[\textbf{Speech (Spch)}] say, talk, ask, etc.
    \item[\textbf{Perception (Perc)}] see, hear, watch, etc.
    \item[\textbf{Motion (Mot)}] go, come, arrive, leave, etc.
    \item[\textbf{Change of state (StChg)}] become, grow, fade, etc.
    \item[\textbf{Temporal (Temp)}] before, after, while, during, then, once, meanwhile, etc.
    \item[\textbf{Color (Col)}] red, blue, green, yellow, etc.
    \item[\textbf{Scene Objects (Obj)}] beach, park, room, office, kitchen, etc.
\end{description}
Each category was automatically expanded with relevant words using pre-trained embeddings via NLP Python libraries~\cite{miller1995wordnet,reimers2019sentence}, such as spaCy, NLTK and sentence-transformers. \textcolor{black}{Specifically, we leveraged the \texttt{en\_core\_web\_lg} model from spaCy and \texttt{sentence-transformers/all-MiniLM-L6-v2} to compute} Transformer-based semantic similarity and WordNet synonyms to enrich each category. Next, we computed the frequency of each category and analyzed the ground-truth rank of each query per method to \textcolor{black}{identify performance trends across categories for each approach (Section~\ref{sec:task_category}).}

\subsubsection{Types of Queries}
\label{sec:query_types}

While a query can ideally belong to a single semantic category, most queries span multiple categories. We therefore classify queries into 3 main types:

\begin{description}
    \item[\textcolor{black}{\textbf{Zero}}] Queries that cannot be confidently assigned to any known category. For example, \emph{``a man in a losing attempt to hitchhike''}.\footnote{Note that the system could identify the query as a Single query if it were rephrased as \emph{``a man attempting to hitchhike''} or \emph{``a man hitchhiking''}, but we have chosen not to manually edit queries.}
    \item[\textbf{Single}] Queries dominated by a single semantic category, e.g., \emph{``a girl riding a scooter''}.
    \item[\textbf{Multi}] Queries containing multiple semantic categories without a clear dominant one, e.g., \emph{``a woman chopping a red bell pepper into small pieces''}.
\end{description}
A query is \textit{Single} if that belongs exclusively to a single semantic category. In practice, however, most queries span multiple categories. To address this, we introduce the notion of a \textit{relaxed Single query}, in which one category clearly dominates the others, defined by a \textit{dominant rate} of at least 1.5. For multi-category queries, this same threshold ensures that the primary category outweighs secondary concepts by this margin. The impact of this threshold choice is evaluated in Section~\ref{sec:dominant_rate_threshold}. Throughout this paper, mentions of Single queries refer to this relaxed version, allowing us to examine both strictly Single and partially Multi queries.

\subsubsection{Distribution of Queries Across Datasets}

Table~\ref{tab:pure_queries} summarizes the distribution of Single category queries across datasets. Specifically, 227 for LSMDC (1,000 total), 113 for MSRVTT (1,000 total), and 89 for MSVD (670 total). LSMDC is dominated by Action queries, followed by Scene, reflecting dynamic content, whereas MSRVTT and MSVD are primarily Scene-focused, emphasizing static or environment-centered descriptions. Action and Scene queries are the most frequent, suggesting that retrieval methods may perform differently depending on dataset characteristics.  

Despite using predefined semantic categories, some textual queries could not be confidently assigned to any known class: 108 from LSMDC, 46 from MSRVTT, and 28 from MSVD remained unrecognized after the initial classification. To address this, we explored a language-model-based partial tagging strategy that assigns each query to the most probable category based on lexical and semantic similarity \cite{reimers2019sentence}. While this approach did produce assignments for previously unrecognized queries, many were noisy and unreliable. Consequently, we retained these queries with zero recognized semantic categories to avoid introducing errors into the analysis.

\subsubsection{Linguistic and Structural Complexity} \label{sec:linguistic_complexity}

To understand how text formulation influences retrieval behavior, caption quality and linguistic properties are assessed using \textit{Flesch Reading Ease} (FRE)~\cite{mohammed2022analysing,hutchings2022evaluation} and \textit{Flesch–Kincaid Grade Level} (FKG)~\cite{solnyshkina2017evaluating,counihan2021open} for readability. Although traditionally designed for human educational readability rather than machine retrieval semantics, their core computational components (specifically average sentence length and syllable-per-word counts) directly reflect syntactic and structural complexity. This is crucial because longer and structurally intricate sentences impose a heavier load on text encoders in cross-modal retrieval.

To ensure a robust evaluation that goes beyond simple readability, we combine FRE and FKG with the number of unique words, total word count, and average word length (capturing lexical diversity and complexity), a binary profanity indicator~\cite{nozza2022state,lafreniere2022power}, and \textit{perplexity}~\cite{cooper2024perplexed,fang2024wrong} computed via a pretrained language model (GPT-2) to measure fluency. Lower perplexity and higher FRE indicate smoother, more coherent captions; for instance, minimalist descriptions such as ``Person running fast'' exhibit higher perplexity and lower readability than complete, fluent sentences such as ``A person is running quickly on a track.''

\begin{table}[t!]
    \centering
    \caption{Categories of all queries per dataset, including Zero, Single and Multi queries.}
    \setlength{\tabcolsep}{3pt}      
    \renewcommand{\arraystretch}{0.9} 
    \footnotesize
    \begin{tabular}{lccccccccccccc}
        \toprule
        \multicolumn{1}{l}{} 
        & \multicolumn{11}{c}{\textbf{Single queries}} &  & \\
        \cmidrule{3-13}
        \textbf{Dataset} & \textbf{Zero} & \textbf{Act} & \textbf{Scn} & \textbf{Cog} & \textbf{Spch} & \textbf{Perc}
        & \textbf{Mot} & \textbf{StChg} & \textbf{Temp} & \textbf{Col} & \textbf{Obj} & \textbf{Total} & \textbf{Multi}\\
        \midrule
        LSMDC~\cite{lsmdc} & 108 & 82 & 34 & 6 & \phantom{1}7 & 30 & 32 & - & 17 & 11 & 8  & 227 & 665 \\
        MSRVTT~\cite{xu2016msr} & \phantom{1}46 & 29 & 48 & - & 12 & \phantom{1}4 & \phantom{1}6 & 1 & \phantom{1}1 & 12 & 3  & 113 & 841\\
        MSVD~\cite{chen2011collecting} & \phantom{1}28 & 17 & 61 & - & \phantom{1}3 & \phantom{1}- & \phantom{1}4 & 1 & \phantom{1}- & \phantom{1}4 &  - & \phantom{1}89 & 553\\
        \bottomrule
    \end{tabular}
    \label{tab:pure_queries}
\end{table}

\section{Experimental Evaluation}
\label{sec:experimental_outline}

In this section, we analyze video–text retrieval models from three angles. First, we compare 14 methods across 3 datasets to study performance differences, task effects, and training efficiency (Section~\ref{sec:main_experiment}). Second, we focus on the top 3 models (EMCL, TempMe and X-CLIP) to examine cross-dataset generalization and the impact of varying the number of captions per video (Section~\ref{sec:cross_dataset_analysis}). Third, we study how these top models are affected by frame rate and compression  level (Section~\ref{sec:sensitivity_analysis}). Together, these experiments reveal how video, query characteristics and architectural choices influence retrieval results.

\subsection{Main Experiment}
\label{sec:main_experiment}

This section presents the main experiment, aimed at understanding differences between models based on data characteristics. We compare our results with those reported in the literature for 14 text-to-video retrieval methods across 3 datasets. The analysis covers recall differences (Section~\ref{sec:traditional_experiments}), the impact of query difficulty (Section~\ref{sec:query_difficulty}), task categories (Section~\ref{sec:task_category}), 
task difficulty vs semantic categories (Section~\ref{sec:task_difficulty}), method grouping (Section~\ref{sec:clustering_by_task_category}), and training efficiency (Section~\ref{sec:training_time_vs_epochs}).

\subsubsection{Traditional Experiment}
\label{sec:traditional_experiments}

Our first experiment aims to make a uniform comparison of the 14 state-of-the-art video–text retrieval methods on 3 benchmark datasets under consistent settings. For completeness, Appendix~\ref{sec:app_supplementary} provides the full, granular performance breakdown across all evaluation metrics (Recall@1, Recall@10, Recall@50, and Average) for the 14 baseline methods on the LSMDC, MSRVTT, and MSVD datasets. In this section, however, we focus on Recall@10 for a simplified comparison, as we observe similar results for Recall@1 and Recall@50. To confirm that these performance advantages are statistically reliable rather than due to random variance, we conduct paired Wilcoxon signed-rank tests across the output distributions. The resulting $p$-values ($p < 0.05$) and confidence intervals confirm that the performance advantages of leading architectures are consistent across test splits.

\begin{figure*}[t!]
    \centering

    \begin{subfigure}[b]{1.0\linewidth}
        \centering
        \includegraphics[width=\linewidth]{Images/Papers/literature_trend_with_complexity_recall_at_10.png}
        \caption{\label{fig:literature_results} Literature: Recall@10 and model complexity (2020–2025).}
    \end{subfigure}

    \vspace{0.1em} 

    \begin{subfigure}[b]{1.0\linewidth}
        \centering
        \includegraphics[width=\linewidth]{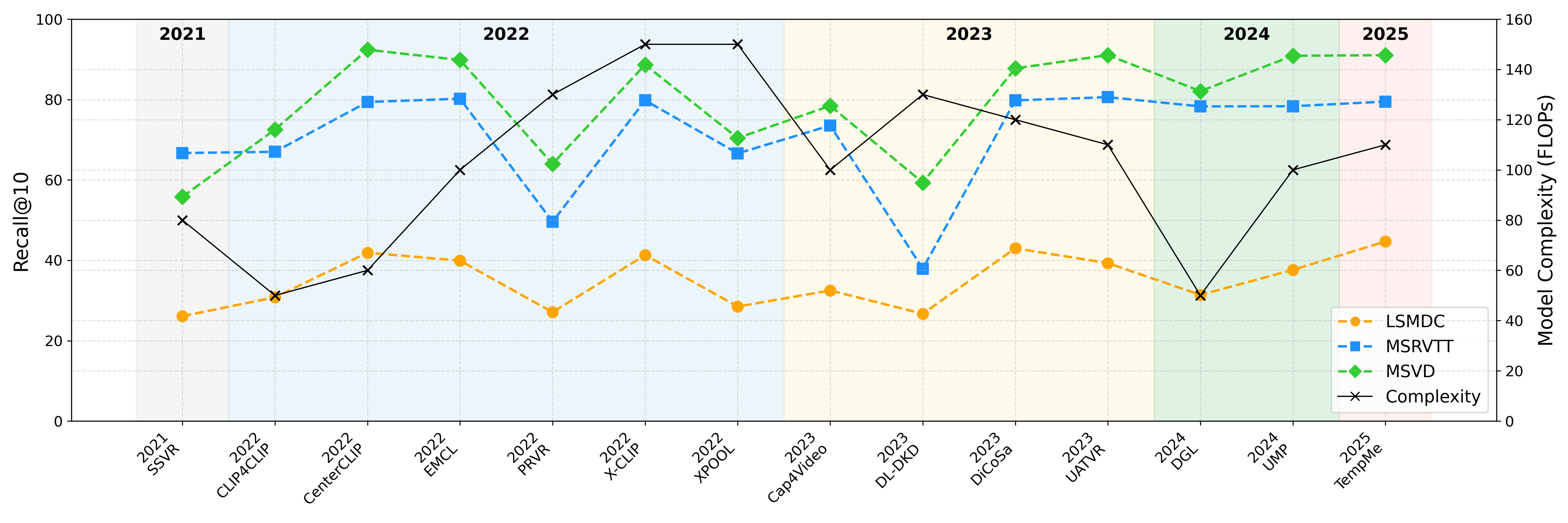}
        \caption{\label{fig:our_evaluation} Our evaluation: Recall@10 and model complexity (2020–2025).}
    \end{subfigure}

    \vspace{0.1em} 

    \begin{subfigure}[b]{1.0\linewidth}
        \centering
        \includegraphics[width=\linewidth]{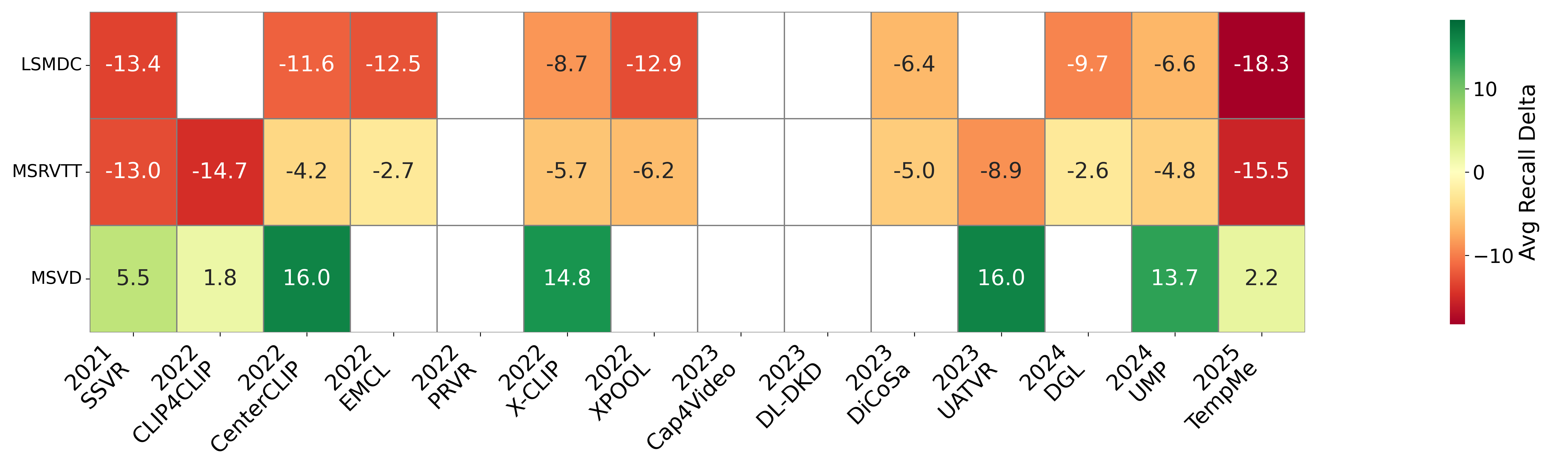}
        \caption{\label{fig:ours_vs_literature} Average recall difference (Our vs Literature) across datasets.}
    \end{subfigure}

    \caption{Recall@10 across datasets: (Top) literature results, (Middle) our results, and (Bottom) average recall difference comparison (Ours vs. Literature).}
    \label{fig:combined_evaluation}
    
\end{figure*}

Figure~\ref{fig:combined_evaluation} provides a visual comparison between our evaluation and literature-reported results. First, Figure~\ref{fig:literature_results} shows Recall@10 of the 14 methods across the 3 datasets as reported by their authors (this is the same figure as Figure~\ref{fig:literature_trend}). Figure~\ref{fig:our_evaluation} then shows Recall@10 of the methods across datasets as measured in our experiment, while Figure~\ref{fig:ours_vs_literature} presents a heatmap of mean recall differences between the two results. In Figure~\ref{fig:ours_vs_literature}, rows correspond to datasets and columns to methods, and green indicates higher performance under our evaluation, red lower performance, and near-white negligible differences.

As Figure~\ref{fig:ours_vs_literature} shows, our results differ from the literature when evaluated under a common setup. Specifically, most methods show negative deltas (red/orange) on LSMDC and MSRVTT, indicating our results are slightly lower than reported, likely due to differences in preprocessing settings. On MSVD, several methods (e.g., CenterCLIP, UATVR, DiCoSa) exhibit positive deltas (green), suggesting improved or comparable performance under our unified evaluation setup. When we used the setups from the papers, our results matched the literature more closely, with only at most 0.1$\%$ difference. This shows that these parameters can affect how the methods perform. We therefore conduct a sensitivity analysis (Section~\ref{sec:sensitivity_analysis}) on the top 3 methods, as the other methods produce analogous results.

While our results differ from the original results reported by the authors of the 14 architectures, due to differences in parameter settings, our results nevertheless indicate that many of the architectures perform at a very similar level, indicating that the progress in text-to-video retrieval has indeed reached a plateau as outlined in the introduction. To further understand which tasks present the most difficulties to the architectures, we perform  a number of different experimental analyses in the remainder of this section. 
In some of our analysis below, we wish to focus on a smaller subset of architectures to (a)~allow us to run a wider range of experiments with our limited computational resources, and (b)~to simplify the presentation.  We have selected EMCL, TempMe and X-CLIP, as their performance is at the top, both in our experiments and in the experiments reported by their authors.

\subsubsection{Analysis: Query Difficulty}
\label{sec:query_difficulty}

As mentioned previously in Section~\ref{sec:quantifying_query_difficulty}, we introduce the term \textit{query difficulty} to describe how challenging a text query is for retrieval models. Intuitively, easy queries retrieve the correct video early, Medium queries appear in the middle of the ranks, and Hard queries either retrieve the correct video at very late positions or fail to retrieve it at all.

\begin{figure*}[t!]
    \centering
    \begin{subfigure}[b]{1.0\linewidth}
        \centering
        \includegraphics[width=0.76\linewidth]{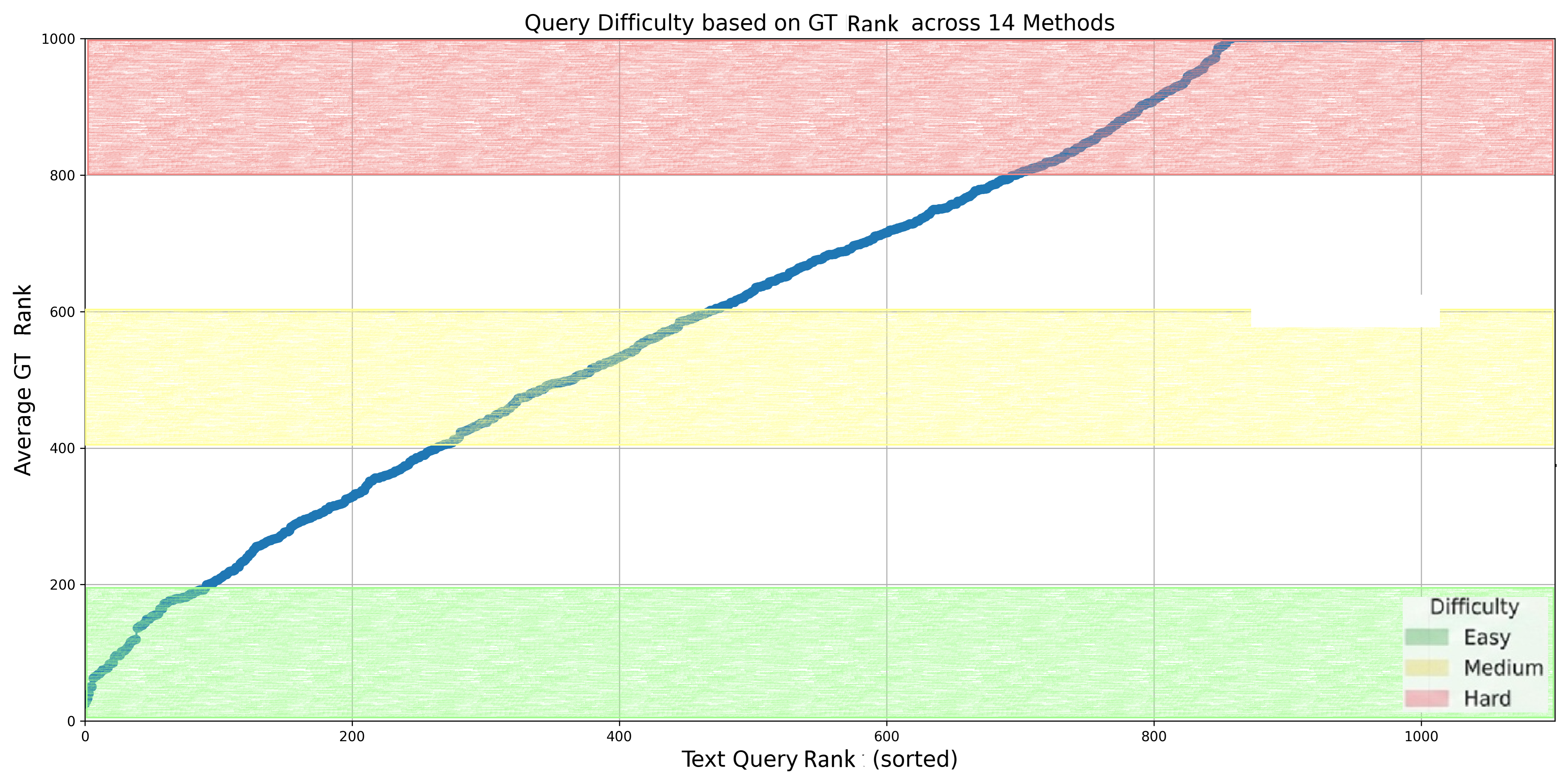}
        \caption{LSMDC}
        \label{fig:query_difficulty_plot_14_methods_pure_queries_lsmdc}
    \end{subfigure}

    \vspace{0.2em}

    \begin{subfigure}[b]{1.0\linewidth}
        \centering
        \includegraphics[width=0.76\linewidth]{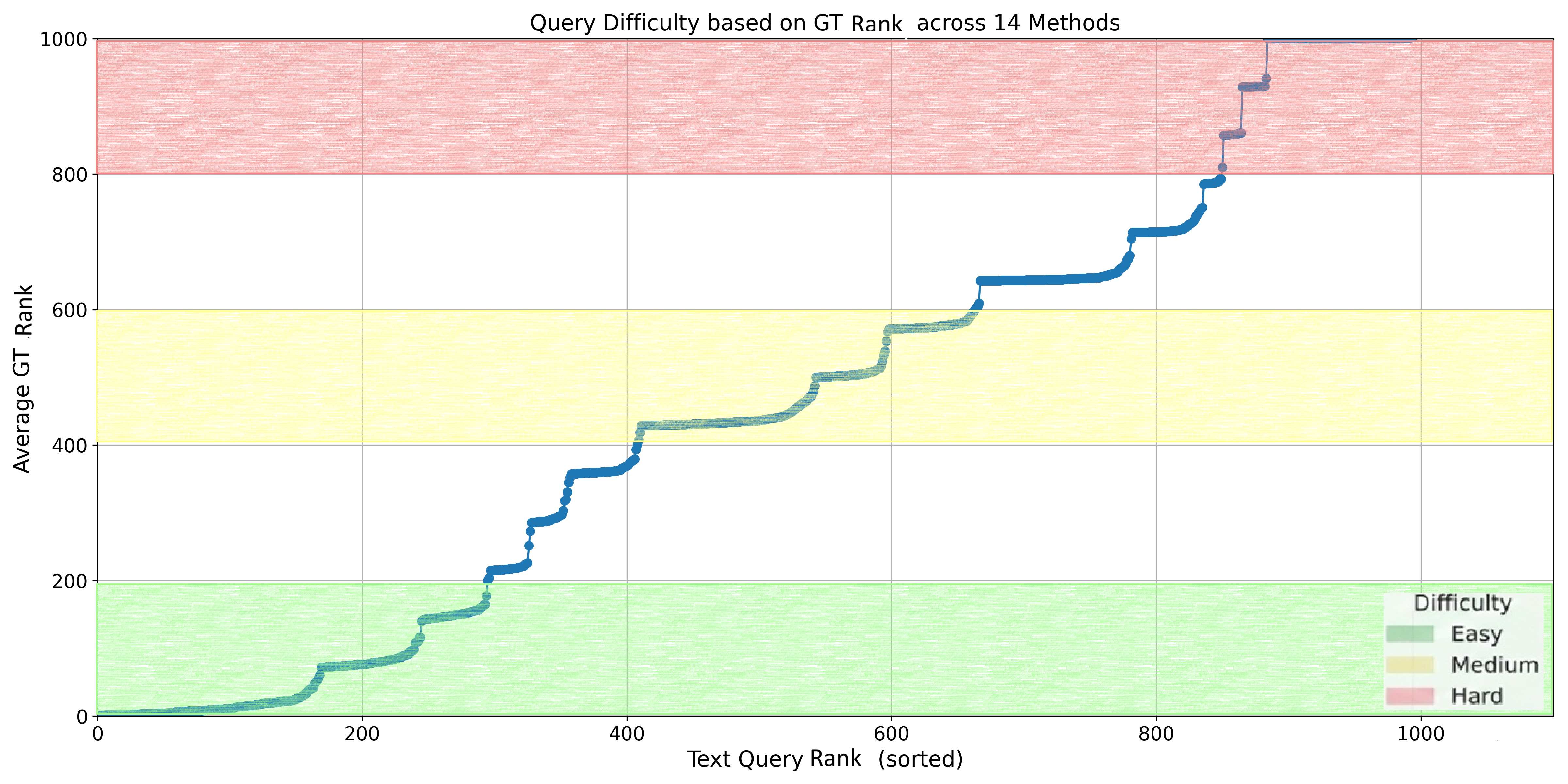}
        \caption{MSRVTT}
        \label{fig:query_difficulty_plot_14_methods_pure_queries_msrvtt}
    \end{subfigure}

    \vspace{0.2em}
    
    \begin{subfigure}[b]{1.0\linewidth}
        \centering
        \includegraphics[width=0.76\linewidth]{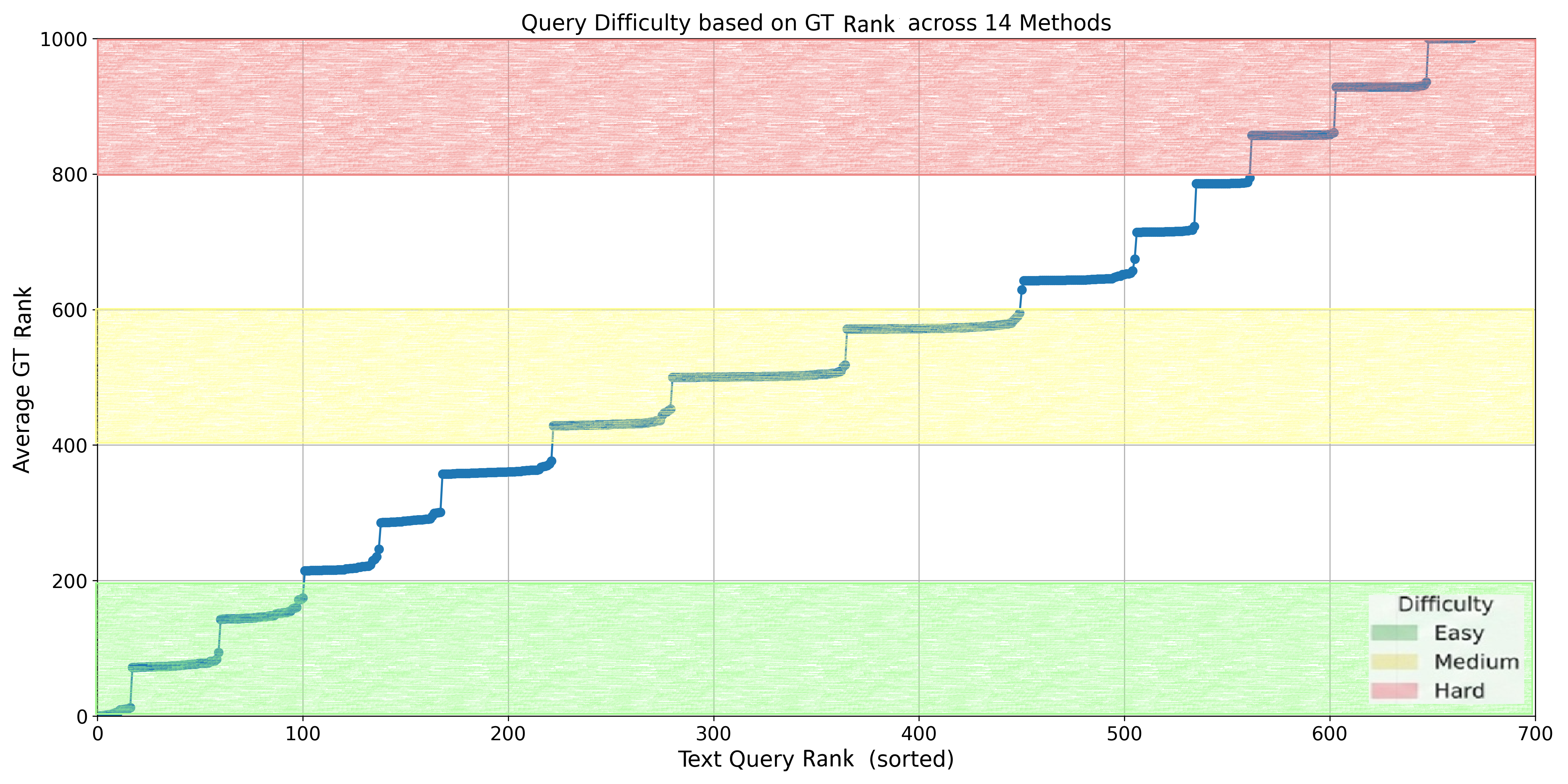}
        \caption{MSVD}
        \label{fig:query_difficulty_plot_14_methods_pure_queries_msvd}
    \end{subfigure}   
    
    \caption{Query difficulty (average GT rank) for 14 methods on LSMDC, MSRVTT and MSVD.}
    
\end{figure*}

Figure~\ref{fig:query_difficulty_plot_14_methods_pure_queries_lsmdc}, Figure~\ref{fig:query_difficulty_plot_14_methods_pure_queries_msrvtt} and Figure~\ref{fig:query_difficulty_plot_14_methods_pure_queries_msvd} present the query difficulty based on the average ground truth rank across 14 methods, on the LSMDC~\cite{lsmdc}, MSRVTT~\cite{xu2016msr}, and MSVD~\cite{chen2011collecting} benchmarks, respectively. The average ground-truth rank shows the mean rank at which the correct video is retrieved across all methods. The $x$-axis represents text queries sorted by their average ground-truth rank, while the $y$-axis indicates the corresponding average ground-truth position. Lower values are better, showing the correct video retrieved earlier; higher values (up to 1,000) indicate later retrieval. Metrics are computed considering up to the top 1,000 returned videos.

Figure~\ref{fig:query_difficulty_plot_14_methods_pure_queries_lsmdc} aligns with our expectations, showing that some text queries are very Easy, others are very Hard, while most fall into the Medium difficulty range. However, Figure~\ref{fig:query_difficulty_plot_14_methods_pure_queries_msrvtt} and Figure~\ref{fig:query_difficulty_plot_14_methods_pure_queries_msvd} reveal many clusters of difficulty levels, appearing as plateaus (15 in total) in the plots. The stepwise pattern forms a staircase shape, indicating that many queries share the same average GT rank. Several flat sections suggest groups of queries with identical or very similar difficulty levels across models, while jumps between plateaus represent significant increases in difficulty for a small number of queries. For MSRVTT, the lower portion of the plots is dense and smooth, showing that many queries are relatively Easy, with difficulty gradually increasing, and the top end shows very steep increases, reflecting a number of queries that are much harder. For MSVD, the slope of difficulty is much more uniform.

To ground these distributions in concrete data, examples from the MSVD dataset illustrate how query semantics shift across difficulty tiers. In the Easy difficulty level, straightforward queries such as ``a man is playing a guitar'' or ``a person cutting a potato'' are successfully retrieved by all evaluated models. Queries assigned to the Medium difficulty level include ``a dog is spun on the ground by a man'' or ``a siamese kitten is playing with a cat toy'' which are correctly handled by roughly half of the systems, capturing moderate complexity such as multi-object interactions. Finally, Hard queries include ``a puppy is barking at himself in a mirror'' or ``a man walks up to a dead deer he apparently shot'' which fail across all 14 models, typically due to fine-grained action nuances, occlusions, or unusual semantic contexts that current text-to-video architectures struggle to align.

\begin{figure*}[t!]
    \centering
    \begin{subfigure}[b]{0.32\linewidth}
        \centering
        \includegraphics[width=\linewidth]{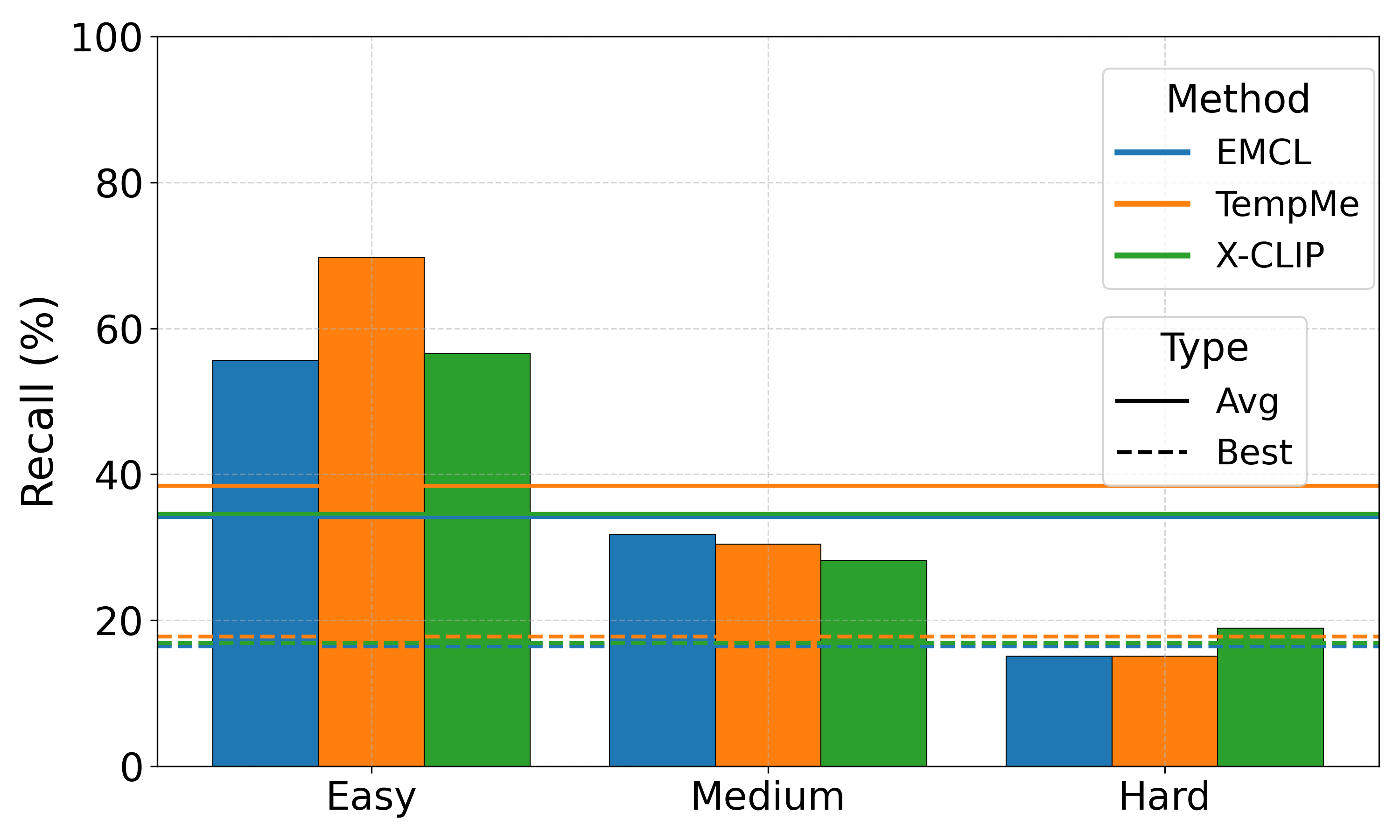}
        \caption{LSMDC: Recall@1}
        \label{fig:lsmdc_k1_per_difficulty}
    \end{subfigure}
    \hfill
    \begin{subfigure}[b]{0.32\linewidth}
        \centering
        \includegraphics[width=\linewidth]{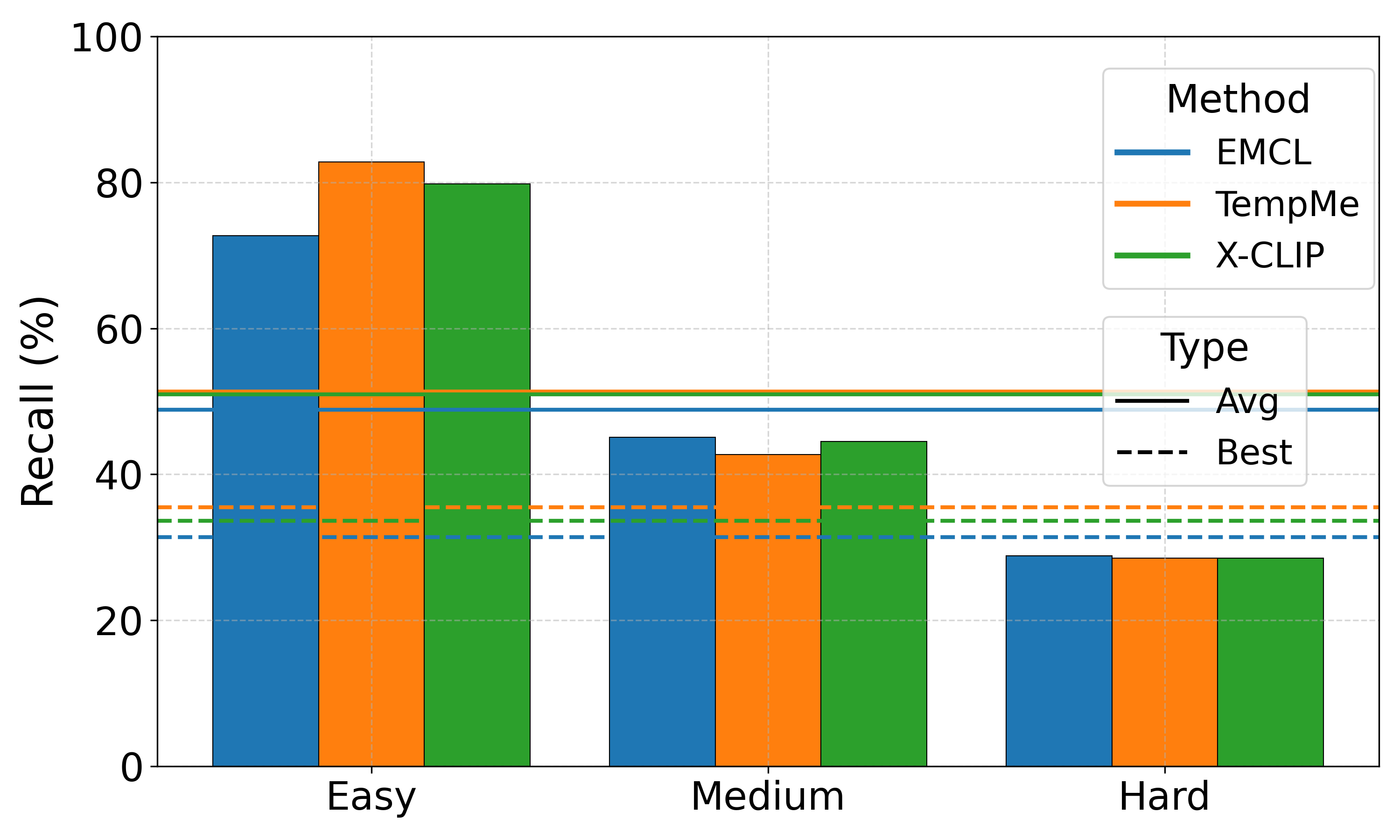}
        \caption{LSMDC: Recall@10}
        \label{fig:lsmdc_k5_per_difficulty}
    \end{subfigure}
    \hfill
    \begin{subfigure}[b]{0.32\linewidth}
        \centering
        \includegraphics[width=\linewidth]{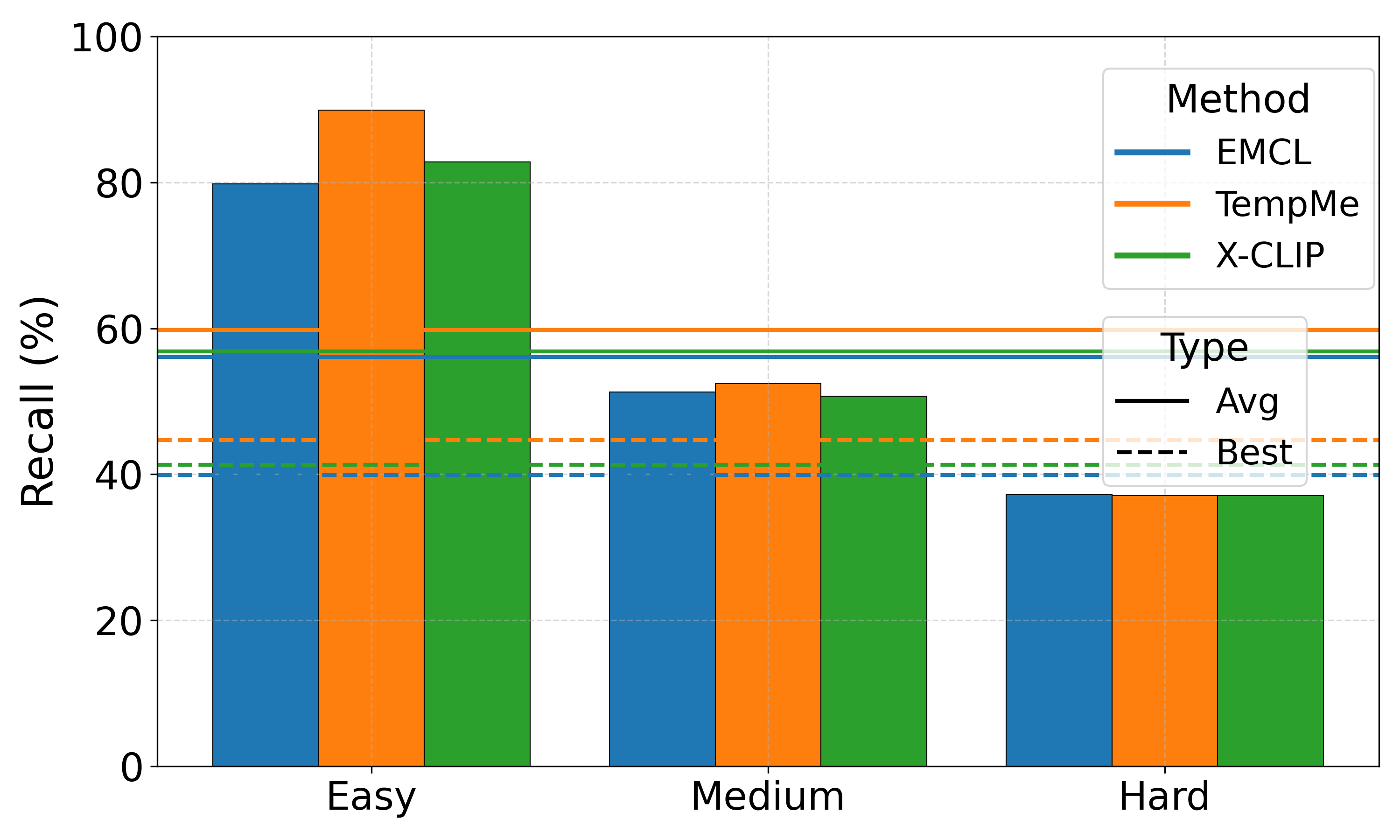}
        \caption{LSMDC: Recall@50}
        \label{fig:lsmdc_k10_per_difficulty}
    \end{subfigure}

    \begin{subfigure}[b]{0.32\linewidth}
        \centering
        \includegraphics[width=\linewidth]{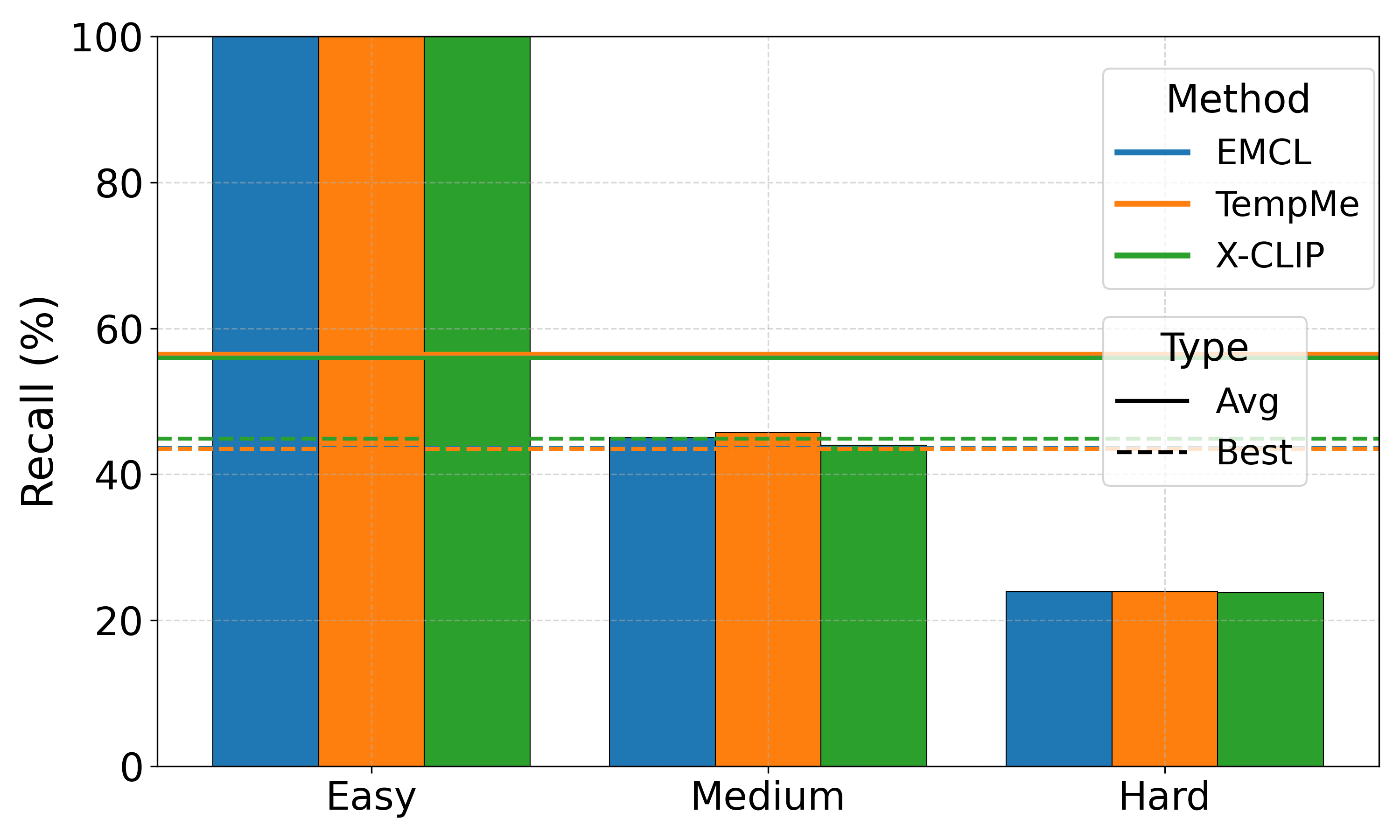}
        \caption{MSRVTT: Recall@1}
        \label{fig:msrvtt_k1_per_difficulty}
    \end{subfigure}
    \hfill
    \begin{subfigure}[b]{0.32\linewidth}
        \centering
        \includegraphics[width=\linewidth]{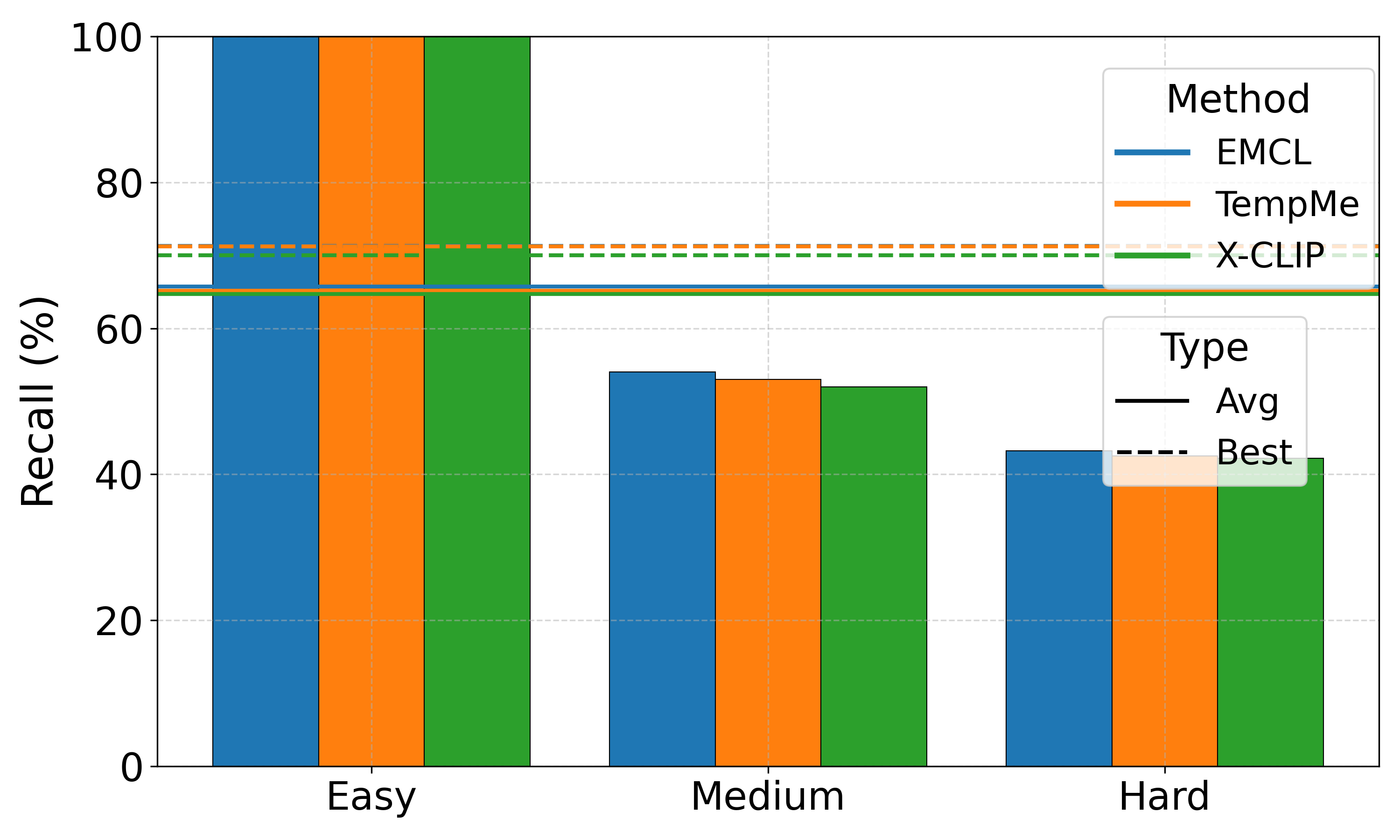}
        \caption{MSRVTT: Recall@10}
        \label{fig:msrvtt_k5_per_difficulty}
    \end{subfigure}
    \hfill
    \begin{subfigure}[b]{0.32\linewidth}
        \centering
        \includegraphics[width=\linewidth]{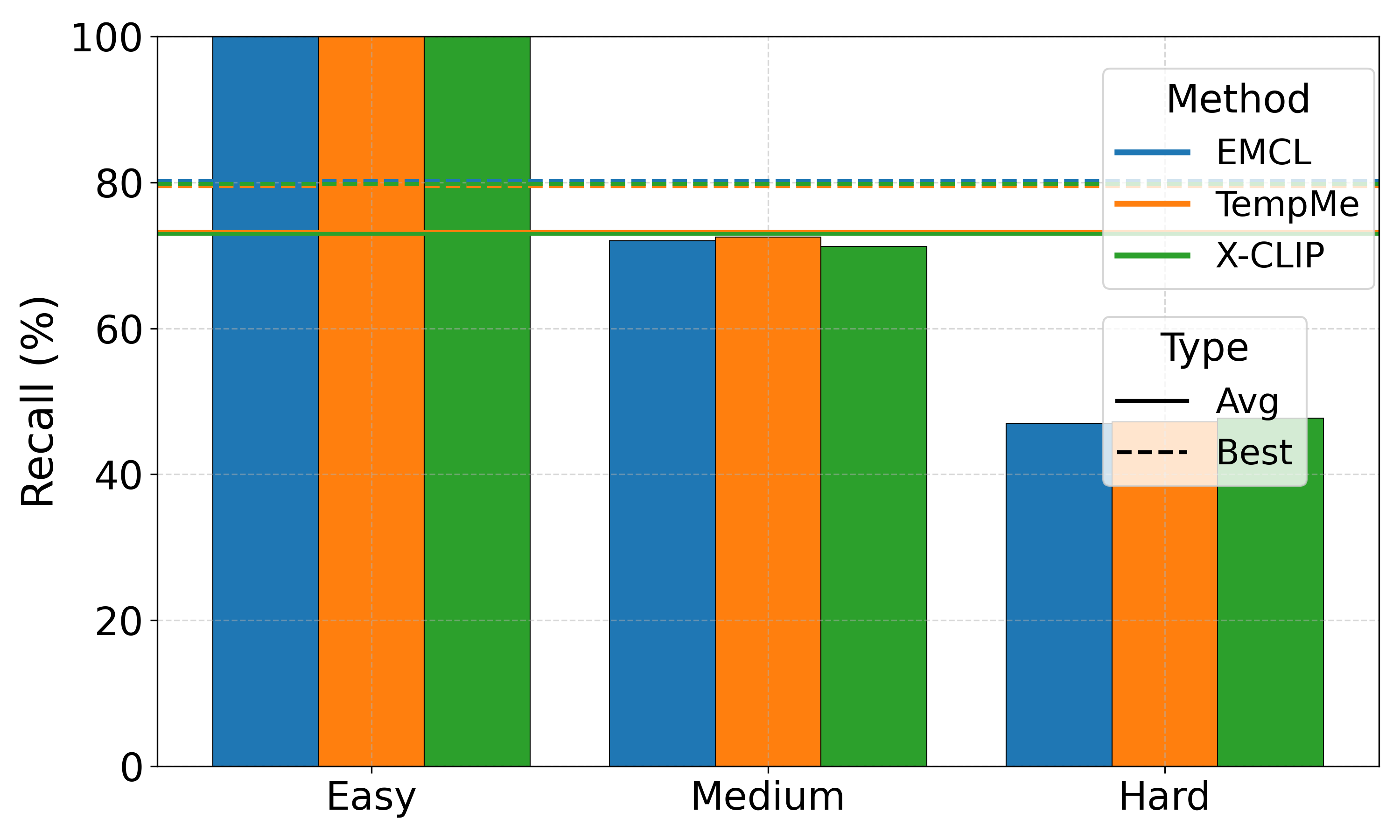}
        \caption{MSRVTT: Recall@50}
        \label{fig:msrvtt_k10_per_difficulty}
    \end{subfigure}

    \begin{subfigure}[b]{0.32\linewidth}
        \centering
        \includegraphics[width=\linewidth]{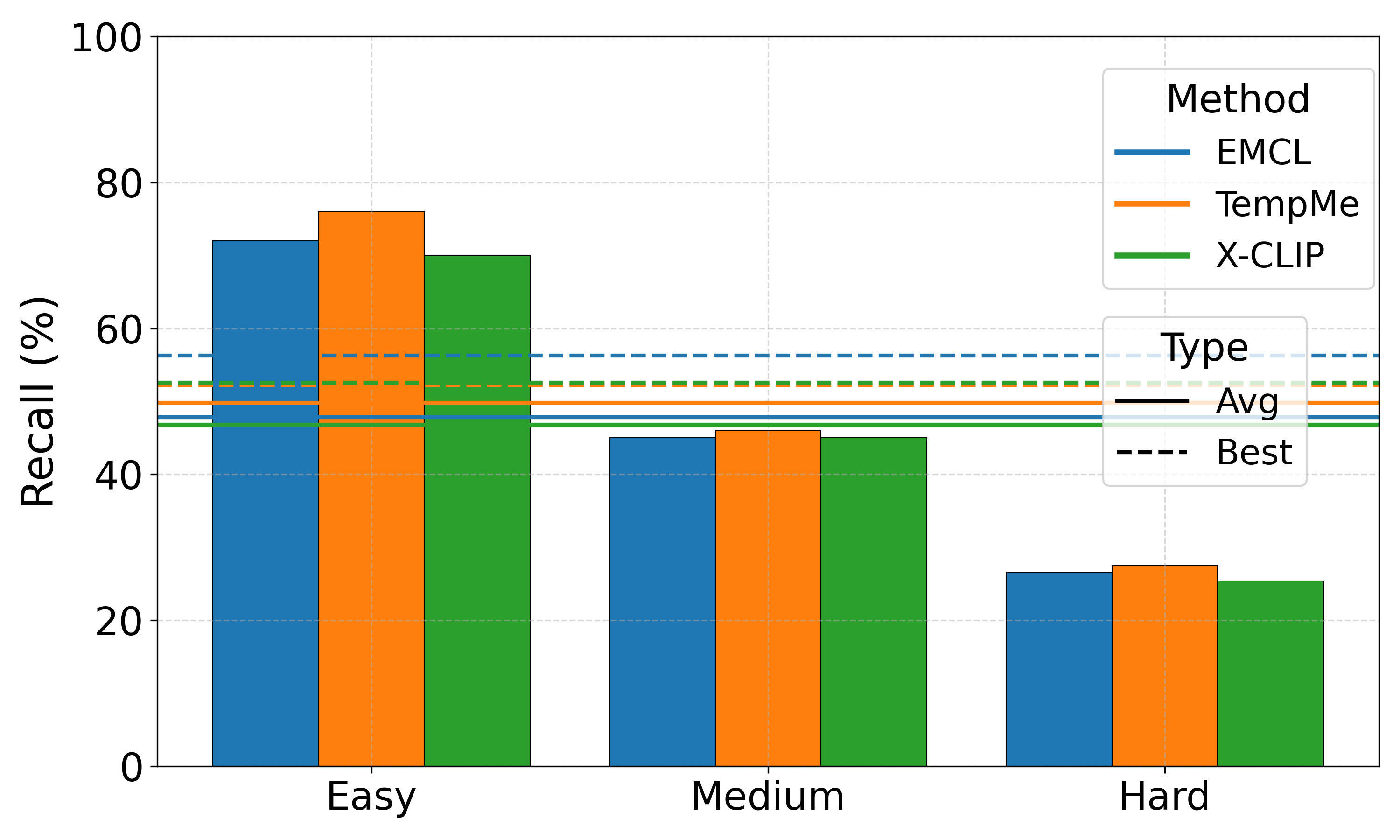}
        \caption{MSVD: Recall@1}
        \label{fig:msvd_k1_per_difficulty}
    \end{subfigure}
    \hfill
    \begin{subfigure}[b]{0.32\linewidth}
        \centering
        \includegraphics[width=\linewidth]{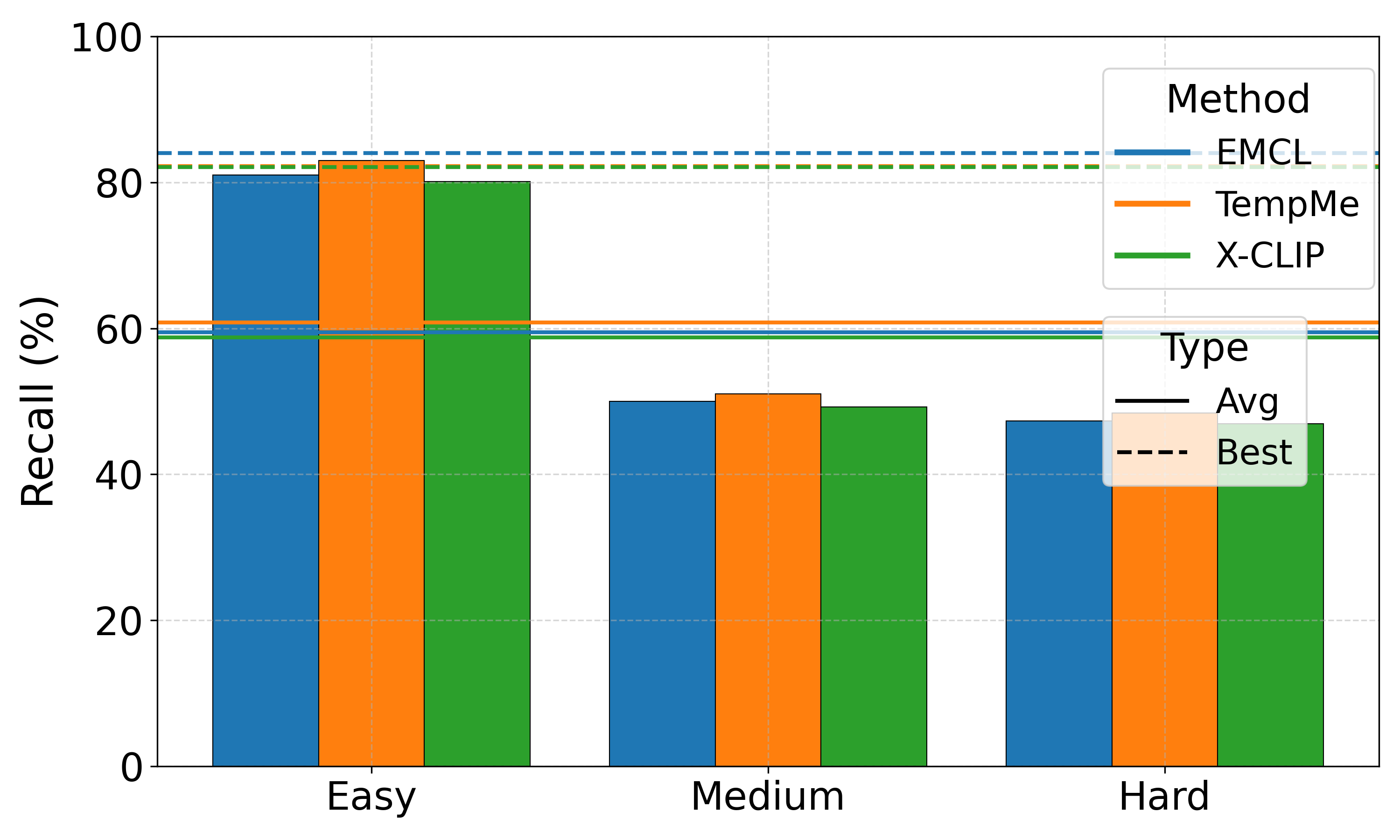}
        \caption{MSVD: Recall@10}
        \label{fig:msvd_k5_per_difficulty}
    \end{subfigure}
    \hfill
    \begin{subfigure}[b]{0.32\linewidth}
        \centering
        \includegraphics[width=\linewidth]{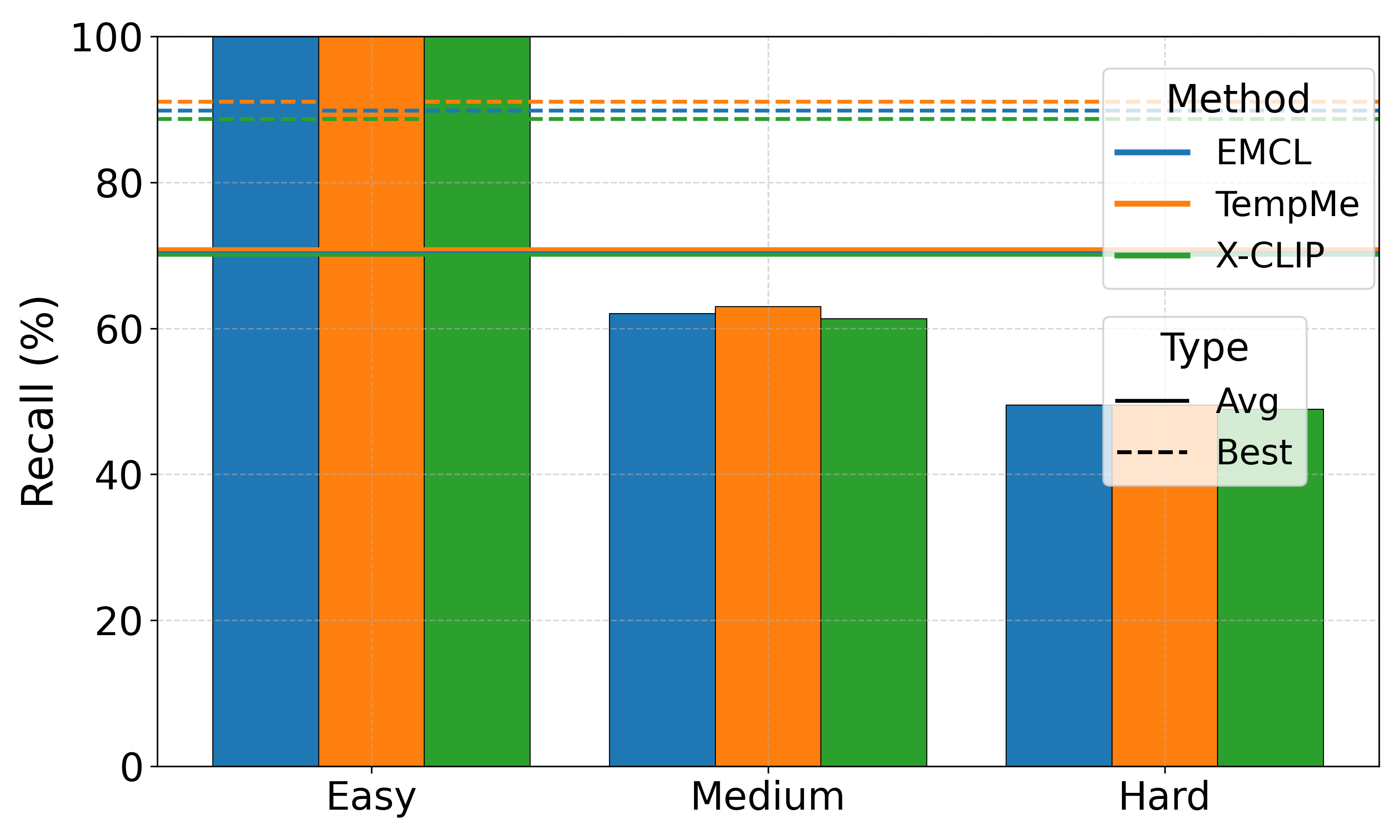}
        \caption{MSVD: Recall@50}
        \label{fig:msvd_k10_per_difficulty}
    \end{subfigure}

    \caption{\label{fig:recall_vs_category} Recall@k (k=1, 10, 50) across difficulty levels on LSMDC (1st row), MSRVTT (2nd row) and MSVD (3rd row).}
\end{figure*}

Next, we evaluate how different retrieval methods perform across datasets categorized into varying difficulty levels—Easy, Medium (Med), and Hard (Figure~\ref{fig:recall_vs_category}). For each difficulty level, we sample 200 queries, using their GT difficulty rank derived from Figure~\ref{fig:query_difficulty_plot_14_methods_pure_queries_lsmdc}, Figure~\ref{fig:query_difficulty_plot_14_methods_pure_queries_msrvtt}, and Figure~\ref{fig:query_difficulty_plot_14_methods_pure_queries_msvd}, because a fixed sample size per difficulty level can remove the dataset frequency bias. Each row has the results on the LSMDC, MSRVTT, and MSVD dataset, respectively. The $x$-axis is difficulty, the $y$-axis Recall@k (\%) for $k = 1, 10, 50$ for EMCL (blue), TempMe (orange) and X-CLIP (green). The dashed lines indicate the highest recall achieved across all queries for each $k$, while solid lines show the average over Single queries.  As expected, there is also a clear difference in performance across Easy, Medium, and Hard cases. Recall values are generally lower on LSMDC, while MSVD and MSRVTT achieve higher recalls. This may be due to differences in dataset characteristics. MSVD (2012) and MSRVTT (2016) have short YouTube videos with captions written by crowdsourced annotators, while LSMDC (2015) uses movie clips with written audio descriptions. 
These differences lead to distinct variations in prompts, such as readability and length, which we examine further through linguistic metrics.

Table~\ref{tab:recall_per_difficulty_percent} reports the average Recall (in \%) across difficulty levels for k = \{1, 10, 50\} and LSMDC, MSRVTT, and MSVD datasets. We observe that the top-performing methods (EMCL, TempMe and X-CLIP) in overall Recall@k achieve higher recall percentages for Easy queries and lower ones for Hard queries. This shows that stronger models retrieve relevant videos more effectively, yielding a higher share of easy cases.

\begin{table}[t!]
    \centering
    \caption{Average Recall@k (\%) for k = \{1, 10, 50\} per difficulty level for 14 methods across LSMDC, MSRVTT, and MSVD datasets.}
    \setlength{\tabcolsep}{2pt}
    \begin{tabular}{lccccccccccc}
        \toprule
        \textbf{Method} & \multicolumn{3}{c}{\textbf{LSMDC}~\cite{lsmdc}} & \hspace{2ex} &\multicolumn{3}{c}{\textbf{MSRVTT}~\cite{xu2016msr}} &\hspace{2ex} & \multicolumn{3}{c}{\textbf{MSVD}~\cite{chen2011collecting}} \\
        \cmidrule{2-4} \cmidrule{6-8} \cmidrule{10-12}
        & \textbf{Easy} & \textbf{Med} & \textbf{Hard} &
        & \textbf{Easy} & \textbf{Med} & \textbf{Hard} &
        & \textbf{Easy} & \textbf{Med} & \textbf{Hard} \\
        \midrule
        SSVR~\cite{wray2021semantic} & 32.4 & 44.0 & 23.6 && 40.7 & 20.7 & 38.6 & & 15.1 & 30.1 & 54.8 \\
        CenterCLIP~\cite{zhao2022centerclip} & 40.8 & 21.3 & 27.9 & & 36.1 & 21.1 & 42.8 & & 18.1 & 45.1 & 36.9 \\
        CLIP4CLIP~\cite{luo2022clip4clip} & 32.4 & 32.1 & 35.5 & & 30.6 & 30.8 & 38.6 & & 27.0 & 33.0 & 40.0 \\
        EMCL~\cite{jin2022expectation} & \underline{50.2} & 31.0 & \underline{18.8} & & \underline{54.6} & 20.3 & \underline{25.1} & & \underline{30.1} & 54.8 & \underline{15.1} \\
        PRVR~\cite{dong2022partially} & 32.4 & 34.0 & 33.6 & & 34.5 & 30.3 & 55.2 & & 24.9 & 35.4 & 39.7 \\
        X-CLIP~\cite{ma2022x} & \underline{51.8} & 30.5 & \underline{17.7} & & \underline{54.5} & 20.7 & \underline{24.8} & & \underline{31.3} & 50.7 & \underline{17.9} \\
        XPOOL~\cite{gorti2022x} & 49.3 & 32.1 & 19.0 & & 30.6 & 30.8 & 38.6 & & 14.8 & 40.1 & 45.1 \\
        Cap4Video~\cite{wu2023cap4video} & 40.3 & 33.0 & 26.7 & & 30.7 & 30.7 & 38.6 & & 12.7 & 55.5 & 31.8 \\
        DiCoSa~\cite{jin2023text} & 48.0 & 28.8 & 23.2 & & 47.8 & 30.8 & 21.4 & & 25.7 & 36.9 & 37.5 \\
        DL-DKD~\cite{dong2023dual} & 40.2 & 27.0 & 32.8 & & 24.5 & 30.3 & 45.2 & & 25.2 & 35.4 & 39.4 \\
        UATVR~\cite{fang2023uatvr} & 41.9 & 42.7 & 19.4 & & 37.1 & 20.7 & 42.2 & & 21.6 & 47.8 & 30.6 \\
        DGL~\cite{yang2024dgl} & 49.8 & 34.5 & 19.7 & & 28.0 & 30.4 & 41.6 & & 11.2 & 36.6 & 52.2 \\
        TempMe~\cite{shen2024tempme} & \underline{52.3} & 29.0 & \underline{18.7} & & \underline{57.0} & 20.6 & \underline{22.4} & & \underline{32.4} & 52.7 & \underline{14.9} \\
        UMP~\cite{zhang2024ump} & 42.4 & 31.3 & 26.3 & & 38.2 & 20.5 & 41.3 & & 24.6 & 47.0 & 28.4 \\
        \bottomrule
    \end{tabular}
    \label{tab:recall_per_difficulty_percent}
\end{table}

\subsubsection{Analysis: Task Semantics}
\label{sec:task_category}

\begin{figure*}[t!]
    \centering
    \includegraphics[width=0.7\linewidth]{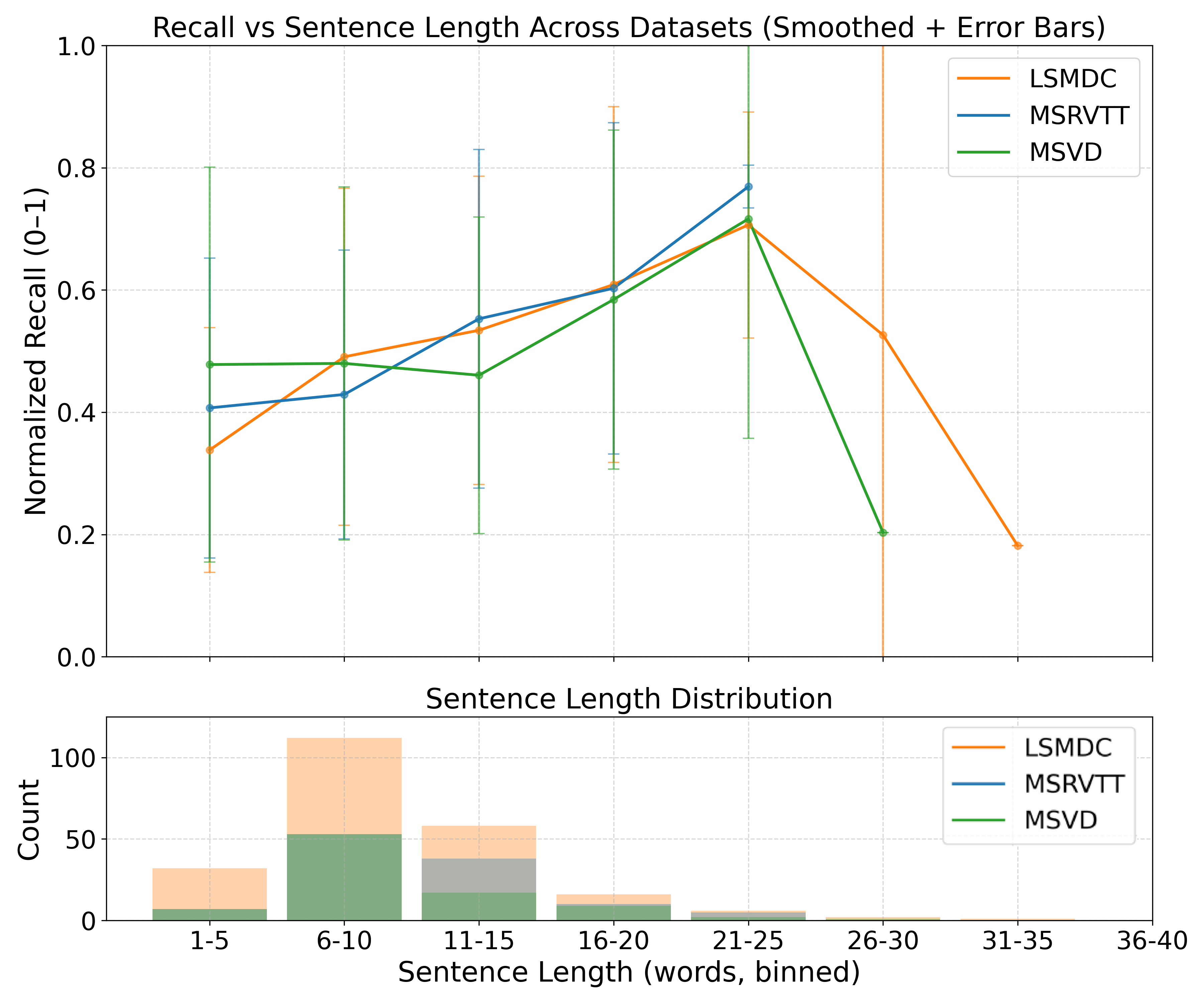}
    \caption{\label{fig:recall_vs_length} Effect of sentence length on retrieval performance across LSMDC (orange), MSRVTT (gray) and MSVD (green).}
\end{figure*}

To understand why certain queries are more difficult, we first examine how retrieval performance varies with sentence length. Figure~\ref{fig:recall_vs_length} presents the normalized recall per query, binned by sentence length, for LSMDC (orange), MSRVTT (blue), and MSVD (green). The top subplot shows the mean normalized recall (0–1) across all 14 methods for each length bin, along with smoothed trends and standard-deviation error bars. The bottom subplot displays the distribution of query lengths in each dataset. The trend lines indicate that very short or very long queries tend to have lower recall, whereas medium-length queries generally achieve higher retrieval performance.

\begin{figure*}[t!]
    \centering
    \begin{subfigure}[b]{0.9\linewidth}
        \includegraphics[width=\linewidth]{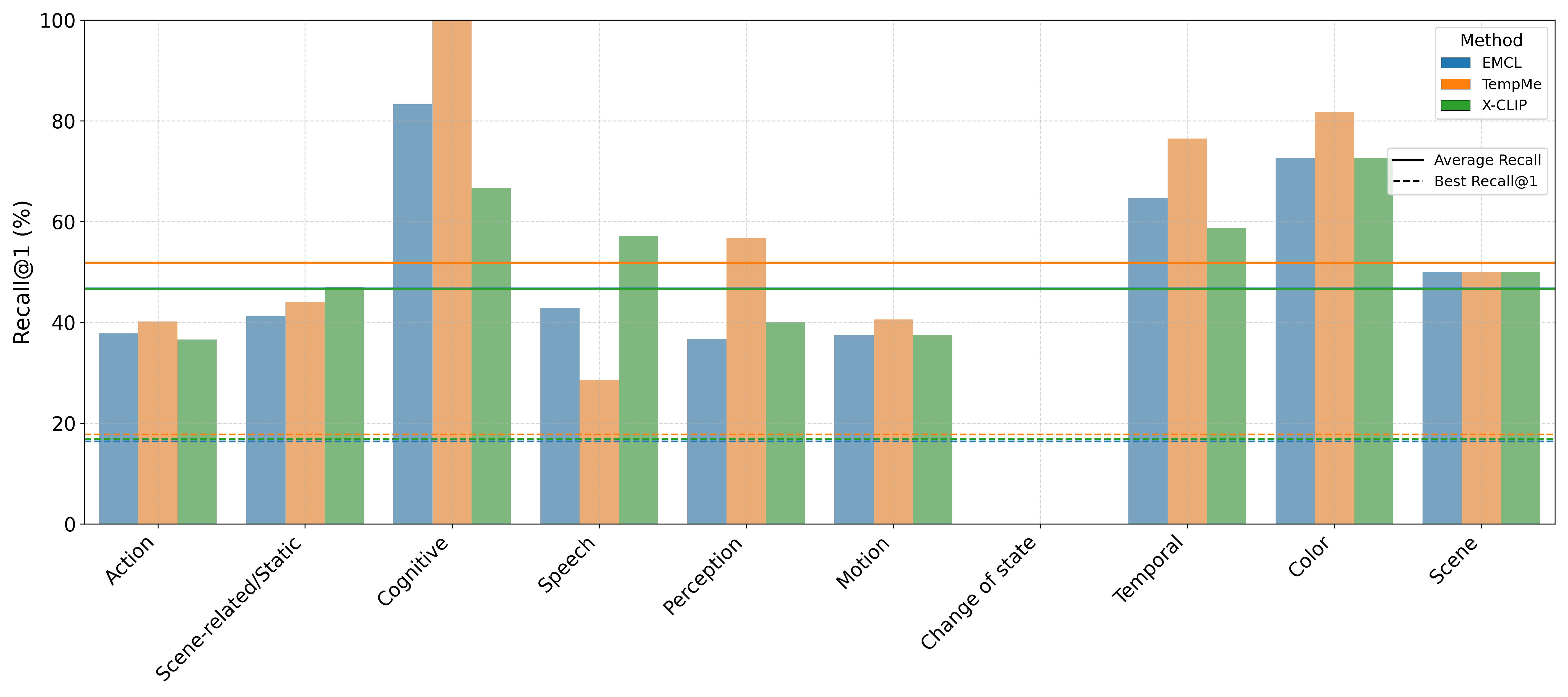}
        \caption{LSMDC: Recall@1}
        \label{fig:lsmdc_k1_per_category}
    \end{subfigure}
    \begin{subfigure}[b]{0.9\linewidth}
        \includegraphics[width=\linewidth]{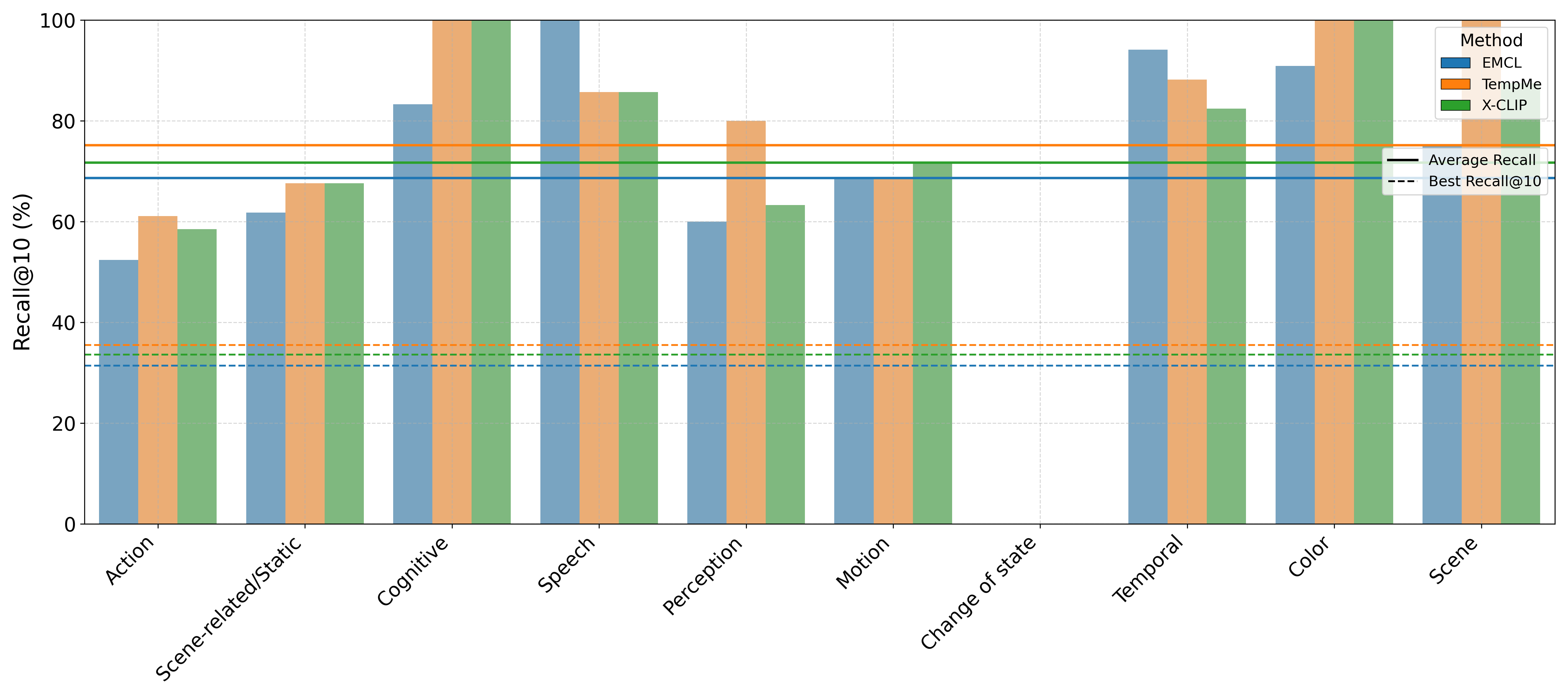}
        \caption{LSMDC: Recall@10}
        \label{fig:lsmdc_k5_per_category}
    \end{subfigure}
    \begin{subfigure}[b]{0.9\linewidth}
        \includegraphics[width=\linewidth]{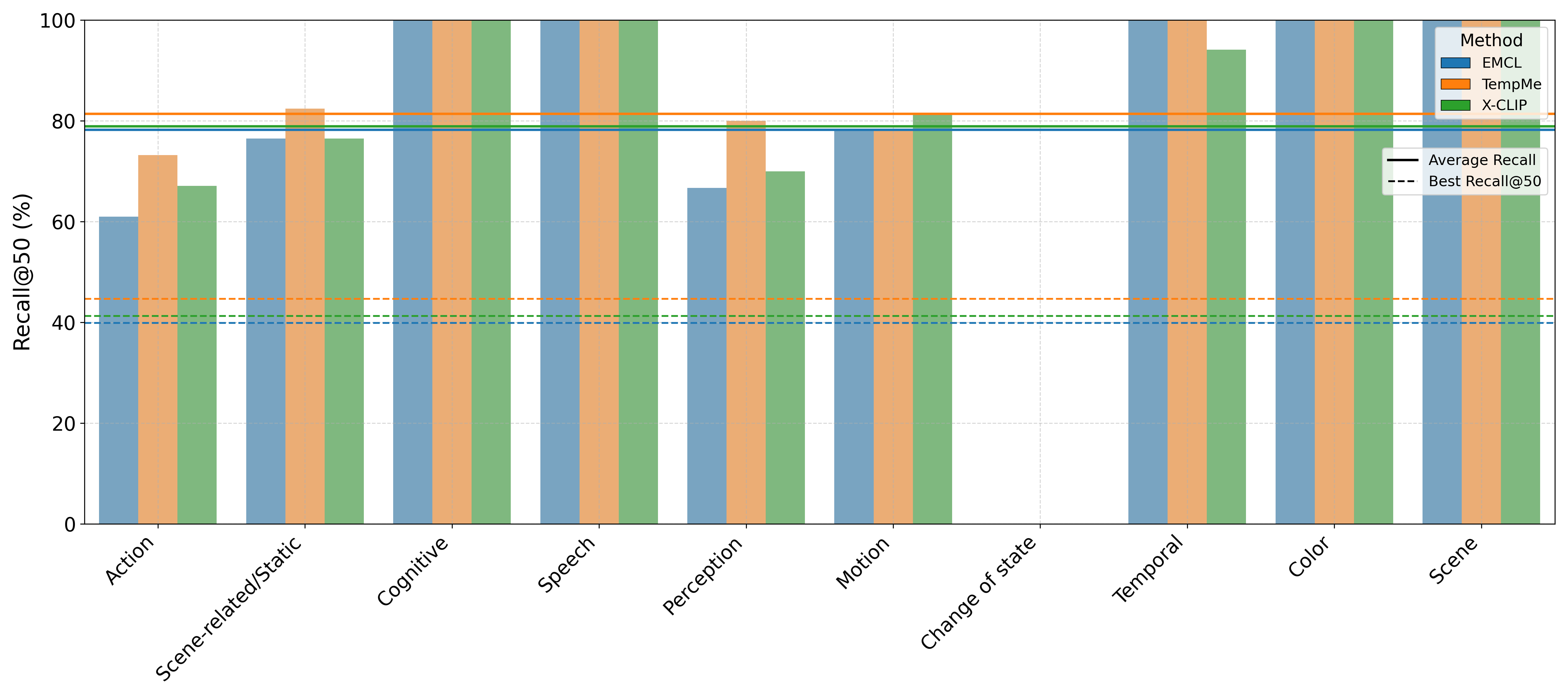}
        \caption{LSMDC: Recall@50}
        \label{fig:lsmdc_k10_per_category}
    \end{subfigure}
    \caption{\label{fig:lsmdc_per_category}Recall@k (k = 1, 10, 50) across task categories on LSMDC dataset.}
\end{figure*}

\begin{figure*}[t!]
    \centering
    \begin{subfigure}[b]{0.9\linewidth}
        \includegraphics[width=\linewidth]{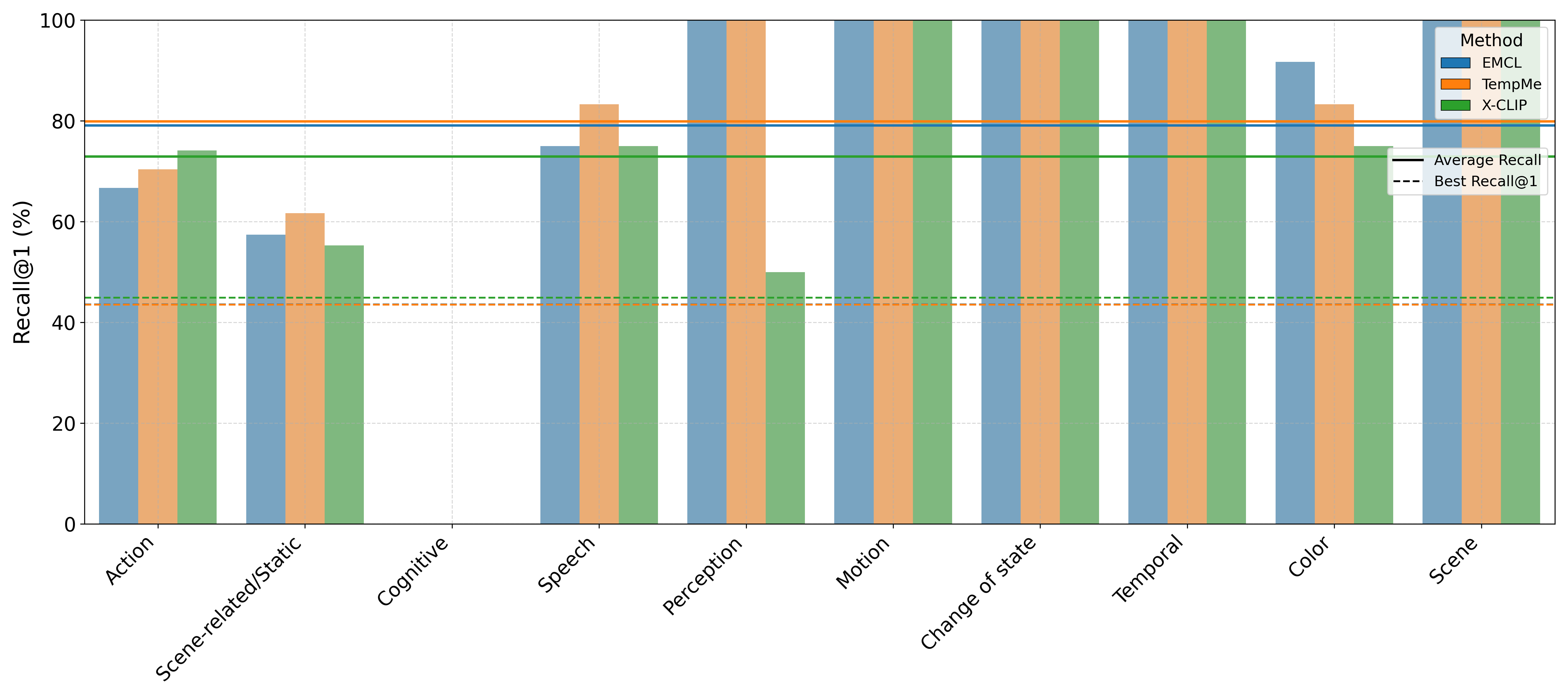}
        \caption{MSRVTT: Recall@1}
        \label{fig:msrvtt_k1_per_category}
    \end{subfigure}
    \begin{subfigure}[b]{0.9\linewidth}
        \includegraphics[width=\linewidth]{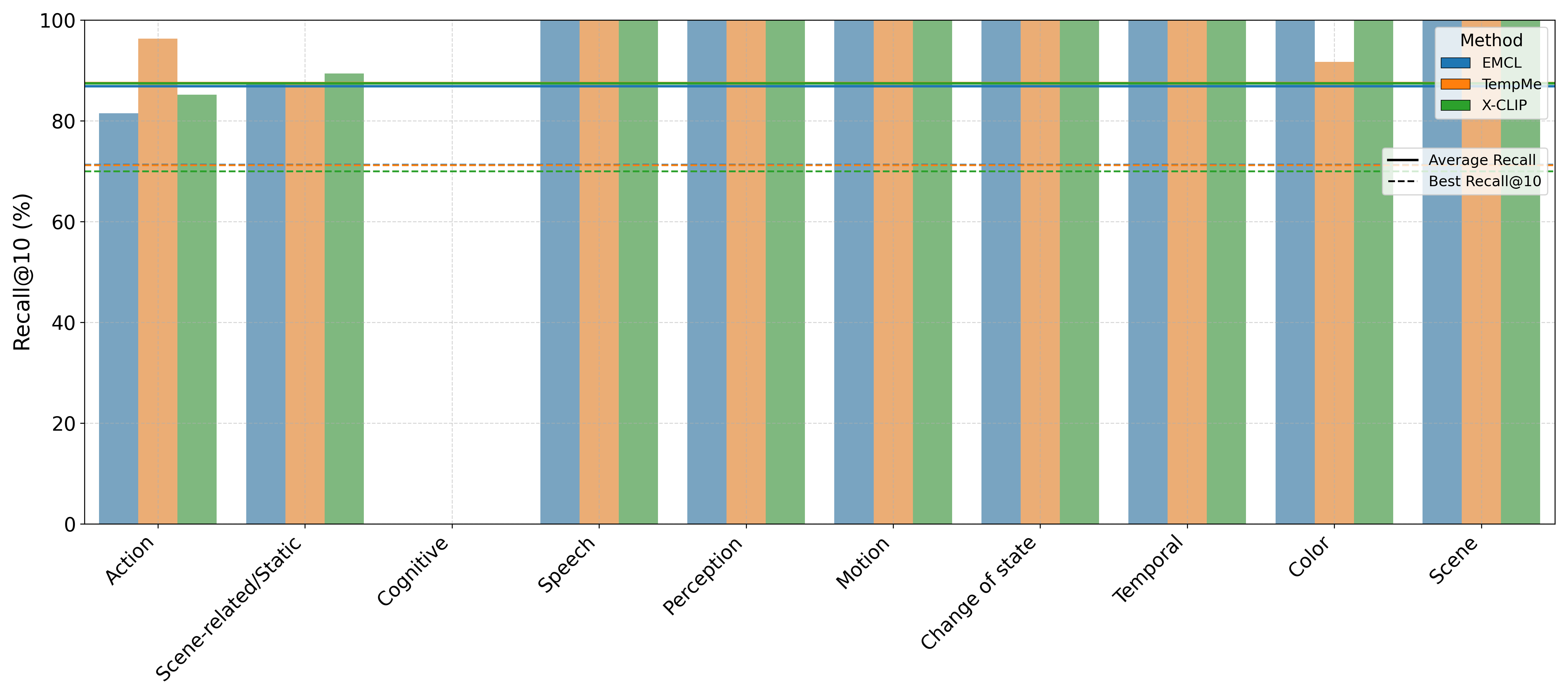}
        \caption{MSRVTT: Recall@10}
        \label{fig:msrvtt_k5_per_category}
    \end{subfigure}
    \begin{subfigure}[b]{0.9\linewidth}
        \includegraphics[width=\linewidth]{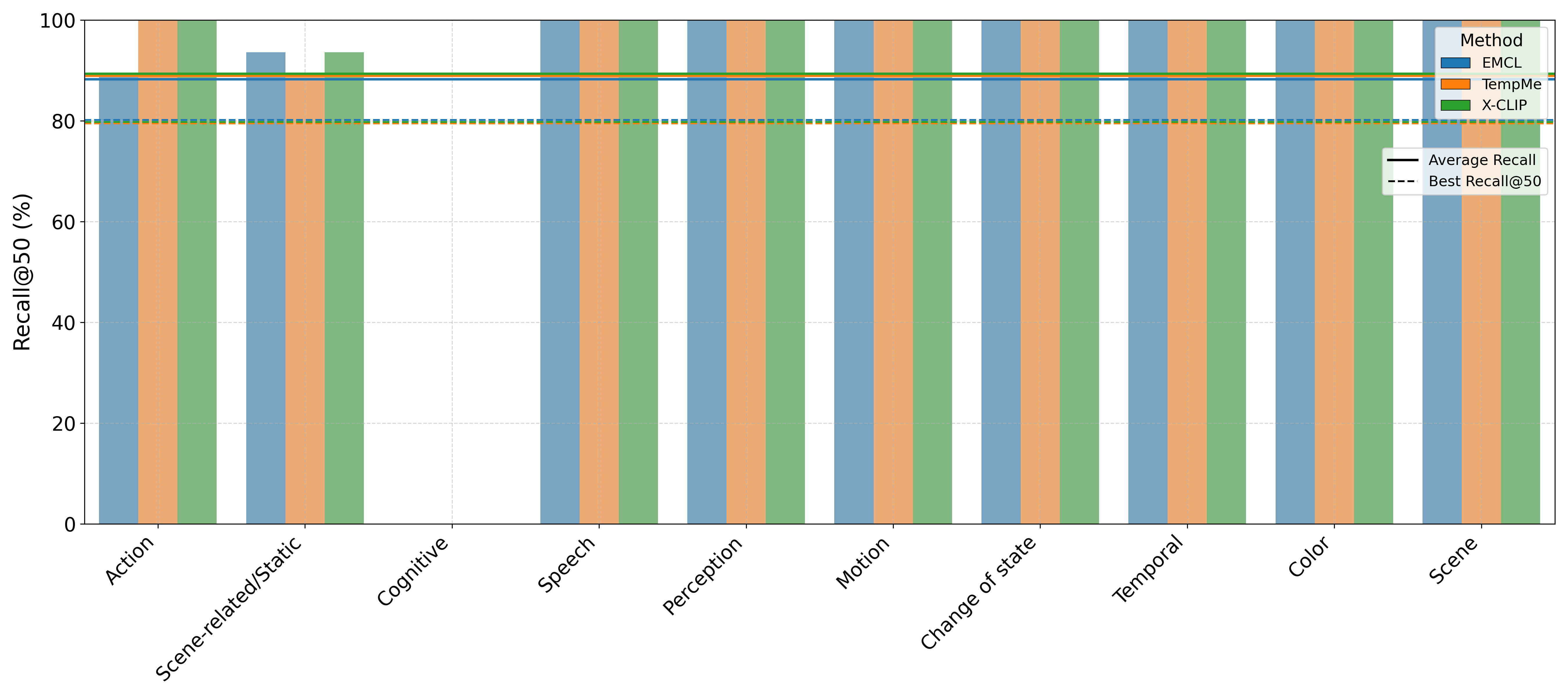}
        \caption{MSRVTT: Recall@50}
        \label{fig:msrvtt_k10_per_category}
    \end{subfigure}
    \caption{\label{fig:msrvtt_per_category}Recall@k (k = 1, 10, 50) across task categories on MSRVTT dataset.}
\end{figure*}

\begin{figure*}[t!]
    \centering
    \begin{subfigure}[b]{0.9\linewidth}
        \includegraphics[width=\linewidth]{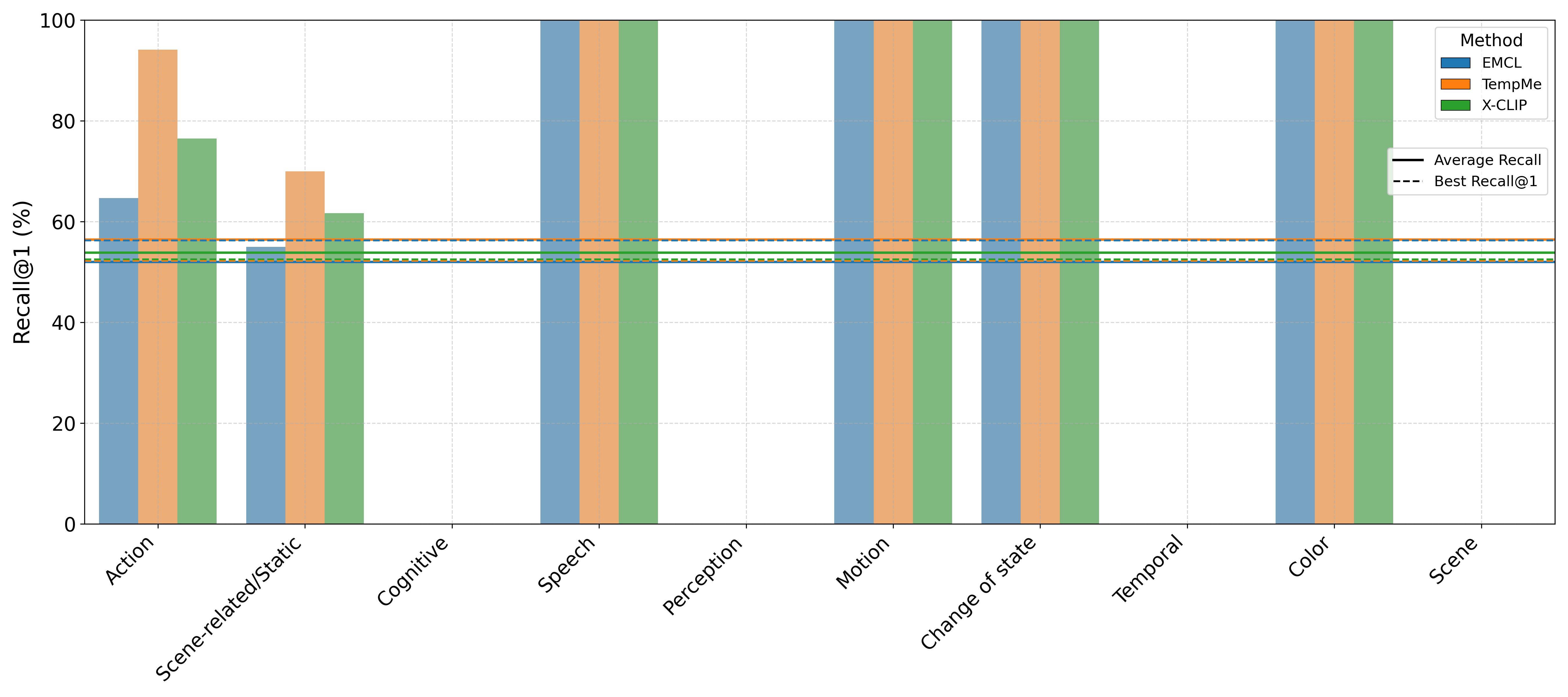}
        \caption{MSVD: Recall@1}
        \label{fig:msvd_k1_per_category}
    \end{subfigure}
    \begin{subfigure}[b]{0.9\linewidth}
        \includegraphics[width=\linewidth]{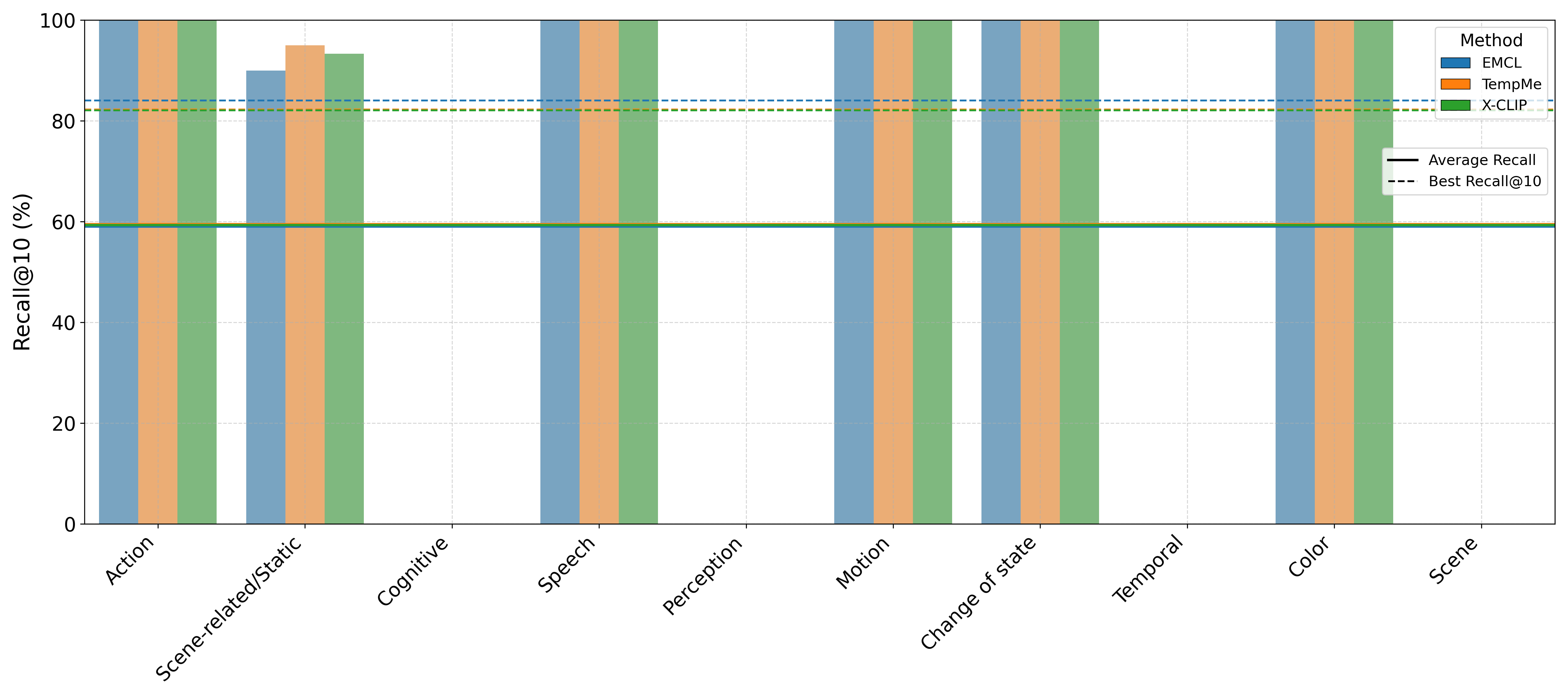}
        \caption{MSVD: Recall@10}
        \label{fig:msvd_k5_per_category}
    \end{subfigure}
    \begin{subfigure}[b]{0.9\linewidth}
        \includegraphics[width=\linewidth]{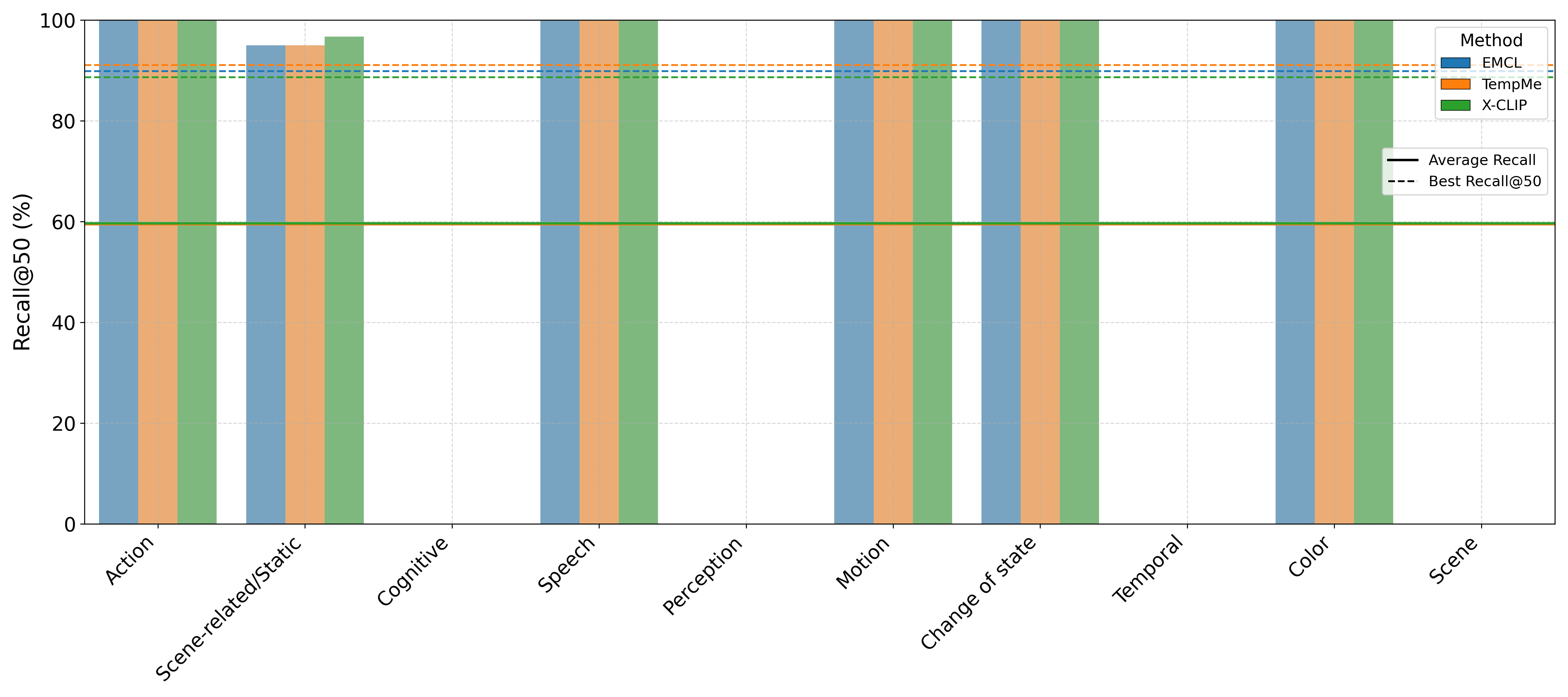}
        \caption{MSVD: Recall@50}
        \label{fig:msvd_k10_per_category}
    \end{subfigure}
    \caption{\label{fig:msvd_per_category}Recall@k (k = 1, 10, 50) across task categories on MSVD dataset.}
\end{figure*}

The remainder of this section examines how the semantic categorization of textual queries from Section~\ref{sec:semantic_categories} affects model performance of the top three performers across datasets. We focus on queries with a single semantic category and consider the ten categories from Section~\ref{sec:query_types}: Action (Act), Scene-related/Static (Scn), Cognitive (Cog), Speech (Spch), Perception (Perc), Motion (Mot), Change of State (StChg), Temporal (Temp), Color (Col), and Scene Objects (Obj). 
Figure~\ref{fig:lsmdc_per_category}, Figure~\ref{fig:msrvtt_per_category}, and Figure~\ref{fig:msvd_per_category} show bar plots for the LSMDC, MSRVTT, and MSVD datasets, respectively. The $x$-axis represents the different query categories, and the $y$-axis shows the Recall@k (\%) values for $k = 1, 10, 50$ of the top three methods: EMCL (blue), TempMe (orange) and X-CLIP (green). The dashed lines indicate the highest recall achieved across all queries for each $k$, whereas the solid lines denote the average recall over Single queries for the same $k$ values. Overall, performance in each category exceeds the best recall per method and dataset, suggesting that Single queries are generally easier. In particular, a higher average recall is observed in LSMDC for the Cognitive and Color categories, while a lower performance appears in the Action and Scene-related/Static categories. MSRVTT achieves the highest overall recall across categories, peaking in Motion, Change of State, and Temporal queries, with the lowest values in Scene-related/Static. MSVD shows strong performance in Speech, Motion, Change of State, and Color, but lower recall in Scene-related/Static queries.

\begin{table}[t!]
    \centering
    \caption{Linguistic Metrics per dataset captions\label{tab:linguistic_metrics_per_category}}
    \scriptsize
    \setlength{\tabcolsep}{3pt}
    \renewcommand{\arraystretch}{0.8}
    \begin{tabular}{llrrrr}
        \toprule
        \textbf{Dataset} & \textbf{Category} & \textbf{FRE$\uparrow$} & \textbf{FKG$\downarrow$} & \textbf{Words$\uparrow$} & \textbf{PPL$\downarrow$} \\
        \midrule
        LSMDC
        & Act & 74.98 & 4.61 & 8.88 & 281.75 \\
        & Scn & 76.62 & 4.83 & 9.09 & 771.08 \\
        & Cog & \textbf{85.44} & \textbf{3.49} & 8.67 & 440.39 \\
        & Spch & 83.68 & 3.61 & 8.14 & 338.73 \\
        & Perc & 72.74 & 5.30 & 8.80 & 422.47 \\
        & Mot & 77.75 & 4.57 & 8.66 & 295.45 \\
        & StChg & - & - & - & - \\
        & Temp & 75.14 & 4.77 & 8.00 & 345.85 \\
        & Col & 77.26 & 3.67 & 8.82 & 222.02 \\
        & Obj & 83.12 & 3.96 & \textbf{9.25} & \textbf{178.69} \\
        & Multi & 70.00 & 5.50 & 7.80 & 400.00 \\
        & Zero & 74.00 & 4.80 & 8.30 & 320.00 \\
        \midrule
        MSRVTT 
        & Act & 76.15 & 4.88 & 9.04 & 461.68 \\
        & Scn & 78.67 & 4.70 & 9.70 & 457.78 \\
        & Cog & - & - & - & - \\
        & Spch & 71.94 & 5.15 & 7.75 & 404.52 \\
        & Perc & 84.20 & 3.56 & 8.25 & 175.49 \\
        & Mot & 77.56 & 4.59 & 8.67 & 444.77 \\
        & StChg & 56.98 & 8.54 & 13.00 & 93.07 \\
        & Temp & 65.79 & 9.80 & \textbf{23.00} & \textbf{52.01} \\
        & Col & 79.70 & 5.54 & 13.67 & 480.59 \\
        & Obj & \textbf{86.83} & \textbf{3.47} & 9.33 & 287.37 \\
        & Multi & 68.00 & 6.20 & 7.90 & 550.00 \\
        & Zero & 72.50 & 5.10 & 8.50 & 430.00 \\
        \midrule
        MSVD
        & Act & 85.85 & 4.26 & 7.94 & 333.01 \\
        & Scn & 80.75 & 4.32 & 9.35 & 412.14 \\
        & Cog & - & - & - & - \\
        & Spch & 67.70 & 6.56 & \textbf{11.00} & 372.90 \\
        & Perc & - & - & - & - \\
        & Mot & \textbf{98.84} & \textbf{1.46} & 8.00 & 347.44 \\
        & StChg & 54.73 & 6.62 & 4.00 & 348.54 \\
        & Temp & - & - & - & - \\
        & Col & 86.77 & 4.14 & 12.00 & \textbf{247.28} \\
        & Obj & - & - & - & - \\
        & Multi & 72.00 & 5.80 & 7.90 & 450.00 \\
        & Zero & 75.00 & 4.70 & 8.40 & 380.00 \\
        \bottomrule
    \end{tabular}
\end{table}

Table~\ref{tab:linguistic_metrics_per_category} shows four linguistic metrics—Flesch Reading Ease (FRE), Flesch–Kincaid Grade Level (FKG), average words per caption (Words) and perplexity (PPL) for each semantic category (including both multimodal (Multi) and unrecognized queries (Zero for completeness) across LSMDC, MSRVTT and MSVD. In LSMDC, Cognitive and Color captions are simpler (high FRE, low FKG), leading to higher recall, while Scene captions are longer (high Words) and harder to interpret (high PPL). In MSRVTT, Motion, Change of State, and Temporal captions perform better due to longer but fluent captions (low PPL), whereas Action and Scene captions show lower recall with shorter and less consistent captions (high PPL). MSVD follows a similar trend: Scene captions have lower recall with complex phrasing (high FKG, high PPL), while Speech, Motion, Change of State, and Color captions are shorter, more readable, and achieve higher recall. Overall, captions that are clear, simple, and consistent (high FRE, low FKG, low PPL) yield higher recall, highlighting the importance of linguistic clarity for retrieval across datasets and categories.

\subsubsection{Analysis: Task Difficulty vs Semantic Categories}
\label{sec:task_difficulty}

In this section, we explore how the task difficulty and semantic categories can be connected. 
Table~\ref{tab:linguistic_metrics_per_difficulty} presents linguistic metrics for captions across three datasets (LSMDC, MSRVTT, and MSVD) and three difficulty levels (Easy, Medium, Hard). Metrics include Flesch Reading Ease (FRE), Flesch–Kincaid Grade (FKG), total words (Words), unique words (Unique), Average word length (Avg Len), Profanity presence (0/1), and Perplexity (PPL). Overall, captions are generally readable, concise, and free of profanity. Across difficulty levels, MSVD captions are generally the easiest to read, with high Flesch Reading Ease and low grade levels, especially in the Hard category. LSMDC captions are longer and slightly more complex, showing higher word counts and slightly lower readability. MSRVTT captions fall in between, with moderate readability and fluency, indicating that both dataset type and difficulty influence caption simplicity and clarity.

\begin{table}[t!]
    \centering
    \caption{Linguistic Metrics per dataset captions}
    \label{tab:linguistic_metrics_per_difficulty}
    \scriptsize
    \setlength{\tabcolsep}{4pt}
    \renewcommand{\arraystretch}{0.9}
    \begin{tabular}{lrrrrrrr}
        \toprule
        \multicolumn{1}{c}{\bf Category} 
        & \multicolumn{1}{c}{\bf FRE$\uparrow$}
        & \multicolumn{1}{c}{\bf FKG$\downarrow$}
        & \multicolumn{1}{c}{\bf Words$\uparrow$} 
        & \multicolumn{1}{c}{\bf Unique$\uparrow$} 
        & \multicolumn{1}{c}{\bf Avg Len$\uparrow$} 
        & \multicolumn{1}{c}{\bf Profanity} 
        & \multicolumn{1}{c}{\bf PPL$\downarrow$}\\
        \midrule 
        \multicolumn{8}{c}{\bf LSMDC} \\
        \midrule 
         Easy & 78.36 & 4.90 & \textbf{10.36} & \textbf{9.68} & 4.73 & 0 & 411.16\\
         Medium & \textbf{81.46} & \textbf{3.84} & 7.82 & 7.50 & \textbf{4.83} & 1 & 449.23\\
         Hard & 78.54 & 4.77 & 9.91 & 9.44 & 4.67 & 1 & \textbf{298.48}\\
        \midrule
        \multicolumn{8}{c}{\bf MSRVTT} \\
        \midrule 
         Easy & 72.54 & 5.38 & 9.00 & 8.43 & 4.17 & 0 & 551.09\\
         Medium & \textbf{81.45} & \textbf{4.06} & 8.68 & 8.11 & 4.18 & 0 & 436.41\\
         Hard & 73.63 & 5.77 & \textbf{11.18} & \textbf{10.35} & 4.37 & 0 & \textbf{413.21}\\
        \midrule
        \multicolumn{8}{c}{\bf MSVD} \\
        \midrule 
         Easy & 82.92 & 3.94 & 9.05 & 8.15 & 4.01 & 0 & \textbf{409.50}\\
         Medium & 81.16 & 4.22 & \textbf{9.20} & \textbf{8.23} & 3.88 & 0 & 555.34\\
         Hard & \textbf{92.58} & \textbf{2.46} & 8.50 & 7.75 & \textbf{4.04} & 0 & 487.08\\
        \bottomrule
    \end{tabular}
\end{table}

\begin{table}[t!]
    \centering
    \caption{Macro-level distribution and difficulty breakdown of Zero, Single and Multi query classes across LSMDC, MSRVTT, and MSVD datasets ($N$ = 200 per dataset). Values show frequency ($n/N$) and percentage classified as Easy, Medium or Hard (\%).}
    \label{tab:macro_query_classes}
    \scriptsize
    \setlength{\tabcolsep}{2pt}
    \renewcommand{\arraystretch}{0.7}
    \begin{tabular}{lcccccccccccc}
        \toprule
        \textbf{Query} 
        & \multicolumn{4}{c}{\bf LSMDC} 
        & \multicolumn{4}{c}{\bf MSRVTT} 
        & \multicolumn{4}{c}{\bf MSVD} \\
        \cmidrule(lr){2-5} \cmidrule(lr){6-9} \cmidrule(lr){10-13}
        & \textbf{Freq} & \textbf{Easy} & \textbf{Med} & \textbf{Hard}
        & \textbf{Freq} & \textbf{Easy} & \textbf{Med} & \textbf{Hard}
        & \textbf{Freq} & \textbf{Easy} & \textbf{Med} & \textbf{Hard} \\
        \midrule
        Zero   & \phantom{17}2/200   & 5  & 15 & 80 & \phantom{17}3/200   & 5  & 20 & 75 & \phantom{17}5/200   & 5  & 20 & 75 \\
        Single & 170/200 & 42 & 34 & 24 & 117/200 & 44 & 31 & 25 & \phantom{1}90/200  & 41 & 38 & 21 \\
        Multi  & \phantom{1}28/200  & 10 & 20 & 70 & \phantom{1}80/200  & 5  & 25 & 70 & 105/200 & 5  & 20 & 75 \\
        \bottomrule
    \end{tabular}
\end{table}

\begin{table}[t!]
    \centering
    \caption{Distribution of individual word categories across difficulty levels for Single queries in LSMDC, MSRVTT, and MSVD datasets. Values show frequency ($n/N$) and percentage classified as Easy, Medium or Hard (\%).}
    \label{tab:category_vs_difficulty_v1}
    \scriptsize
    \setlength{\tabcolsep}{2pt}
    \renewcommand{\arraystretch}{0.7}
    \begin{tabular}{lcccccccccccc}
        \toprule
        \textbf{Category} 
        & \multicolumn{4}{c}{\bf LSMDC} 
        & \multicolumn{4}{c}{\bf MSRVTT} 
        & \multicolumn{4}{c}{\bf MSVD} \\
        \cmidrule(lr){2-5} \cmidrule(lr){6-9} \cmidrule(lr){10-13}
        & \textbf{Freq} & \textbf{Easy} & \textbf{Med} & \textbf{Hard}
        & \textbf{Freq} & \textbf{Easy} & \textbf{Med} & \textbf{Hard}
        & \textbf{Freq} & \textbf{Easy} & \textbf{Med} & \textbf{Hard} \\
        \midrule
        Act   & 53/200 & 45 & 35 & 20 & 29/200 & 50 & 30 & 20 & 17/200 & 40 & 40 & 20 \\
        Mot   & 32/200 & 40 & 35 & 25 & \phantom{1}6/200  & 45 & 30 & 25 & \phantom{1}4/200  & 38 & 37 & 25 \\
        Col   & \phantom{1}8/200  & 50 & 20 & 30 & 12/200 & 45 & 25 & 30 & \phantom{1}4/200  & 60 & 20 & 20 \\
        Scn   & 32/200 & 20 & 50 & 30 & 30/200 & 25 & 45 & 30 & 61/200 & 20 & 50 & 30 \\
        Obj  & \phantom{1}0/200  & -  & -  & -  & \phantom{1}3/200  & 25 & 45 & 30 & \phantom{1}0/200  & -  & -  & -  \\
        Perc  & 25/200 & 20 & 50 & 30 & \phantom{1}4/200  & 25 & 45 & 30 & \phantom{1}0/200  & -  & -  & -  \\
        Cog   & \phantom{1}6/200  & 5  & 20 & 75 & \phantom{1}0/200  & -  & -  & -  & \phantom{1}0/200  & -  & -  & -  \\
        Spch  & \phantom{1}7/200  & 5  & 25 & 70 & 12/200 & 5  & 25 & 70 & \phantom{1}3/200  & 5  & 25 & 70 \\
        StChg & \phantom{1}0/200  & -  & -  & -  & \phantom{1}1/200  & 5  & 20 & 75 & \phantom{1}1/200  & 5  & 20 & 75 \\
        Temp  & \phantom{1}7/200  & 5  & 20 & 75 & 20/200 & 5  & 20 & 75 & \phantom{1}0/200  & -  & -  & -  \\
        \bottomrule
    \end{tabular}
\end{table}

Beyond basic measures such as sentence length or readability, Table~\ref{tab:macro_query_classes} and Table~\ref{tab:category_vs_difficulty_v1} examine how query structure and semantic word categories are distributed across retrieval difficulty levels. Table~\ref{tab:macro_query_classes} reports the macro-level breakdown across Zero, Single, and Multi query classes, while Table~\ref{tab:category_vs_difficulty_v1} details the granular distribution of individual word categories within Single queries for each dataset (LSMDC, MSRVTT, MSVD), including frequency of each category and the number of words classified as Easy, Medium, or Hard. In LSMDC, 72 of 200 words were Easy, 62 were Medium, and 66 were Hard. MSRVTT had 84 Easy, 61 Medium, and 55 Hard words, while MSVD contained 79 Easy, 67 Medium, and 54 Hard words. Concrete and visually grounded categories—such as actions, motion, and colors—are predominantly associated with Easy queries (approximately 40–60\%), indicating that visually explicit concepts are easier to retrieve. Scene- and perception-related terms tend to cluster in the Medium difficulty range (around 35–50\%), reflecting moderate visual ambiguity. In contrast, more abstract categories, including cognitive verbs, speech, change of state and temporal expressions, are overwhelmingly classified as Hard (about 70–75\%), highlighting queries that require higher-level reasoning and temporal understanding beyond direct visual cues. These Hard queries, being more challenging, could serve as a dedicated benchmark for video retrieval, allowing models to be tested on semantically complex and temporally detailed content.

\subsubsection{Clustering of Performance by Task Category}
\label{sec:clustering_by_task_category}

\begin{figure*}[t!]
    \centering
    \begin{subfigure}[b]{0.45\linewidth}
        \centering
        \includegraphics[width=\linewidth]{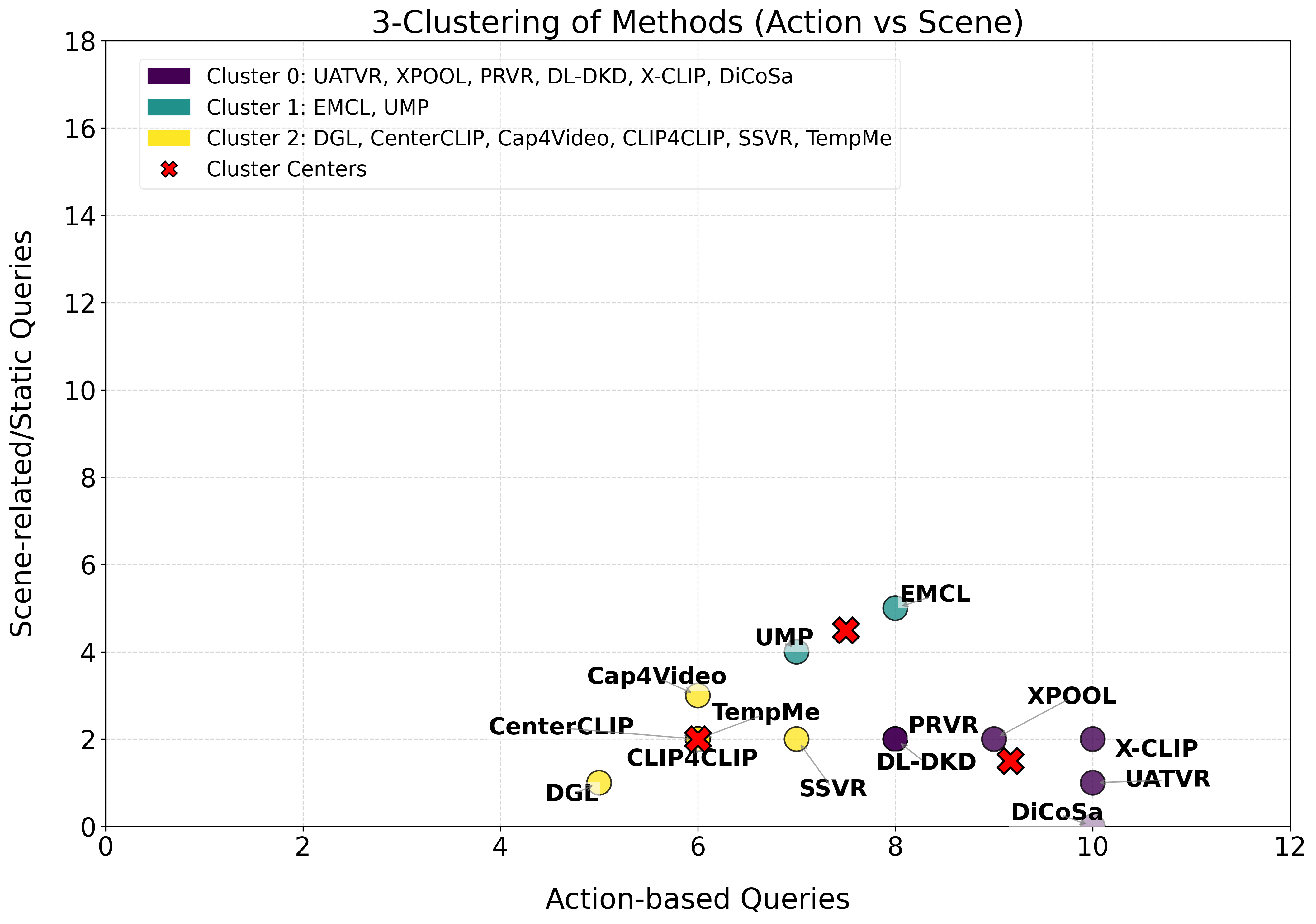}
        \caption{LSMDC}
        \label{fig:lsmdc_clustering}
    \end{subfigure}
    \begin{subfigure}[b]{0.45\linewidth}
        \centering
        \includegraphics[width=\linewidth]{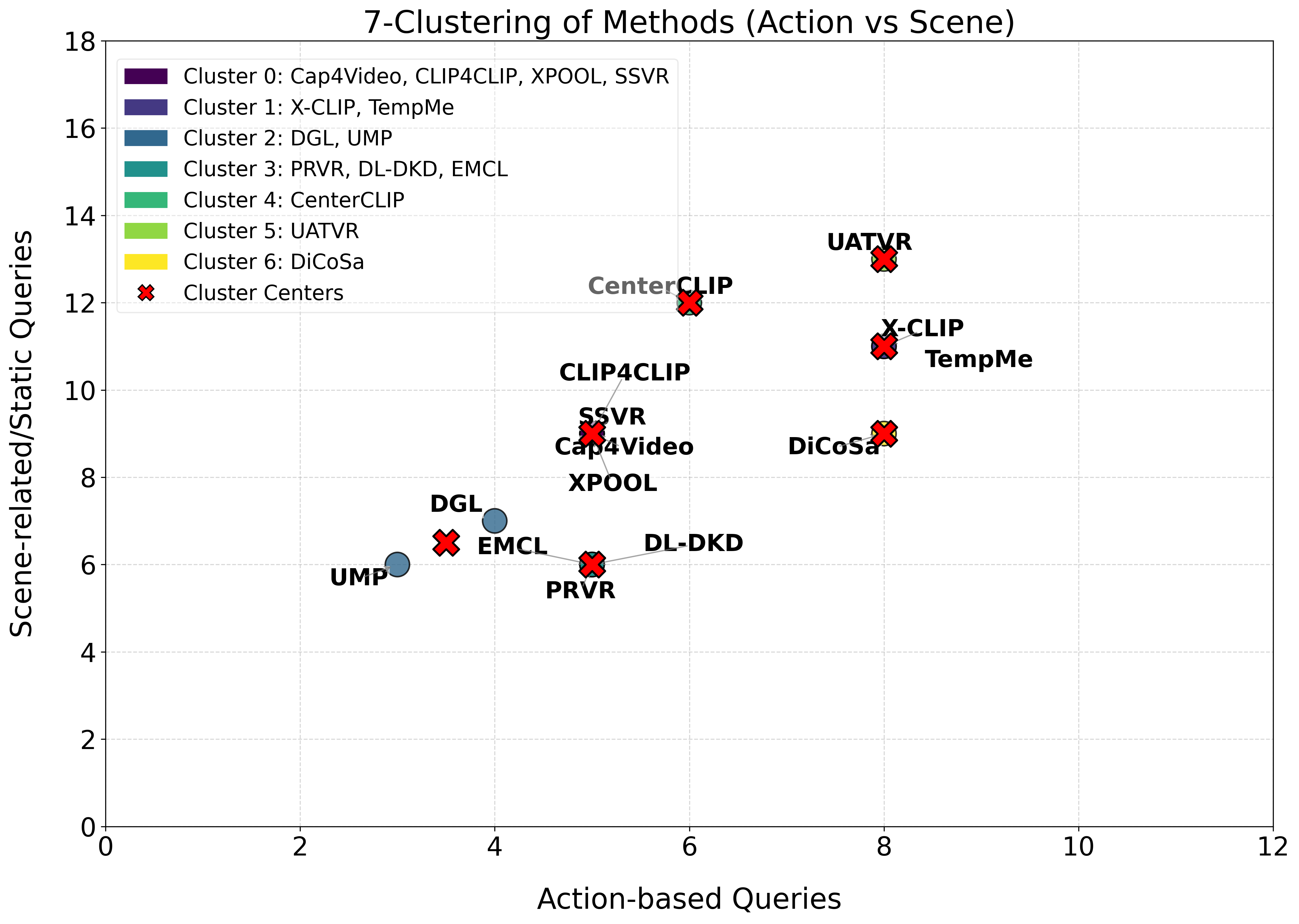}
        \caption{MSRVTT}
        \label{fig:msrvtt_clustering}
    \end{subfigure}

    \begin{subfigure}[b]{0.45\linewidth}
        \centering
        \includegraphics[width=\linewidth]{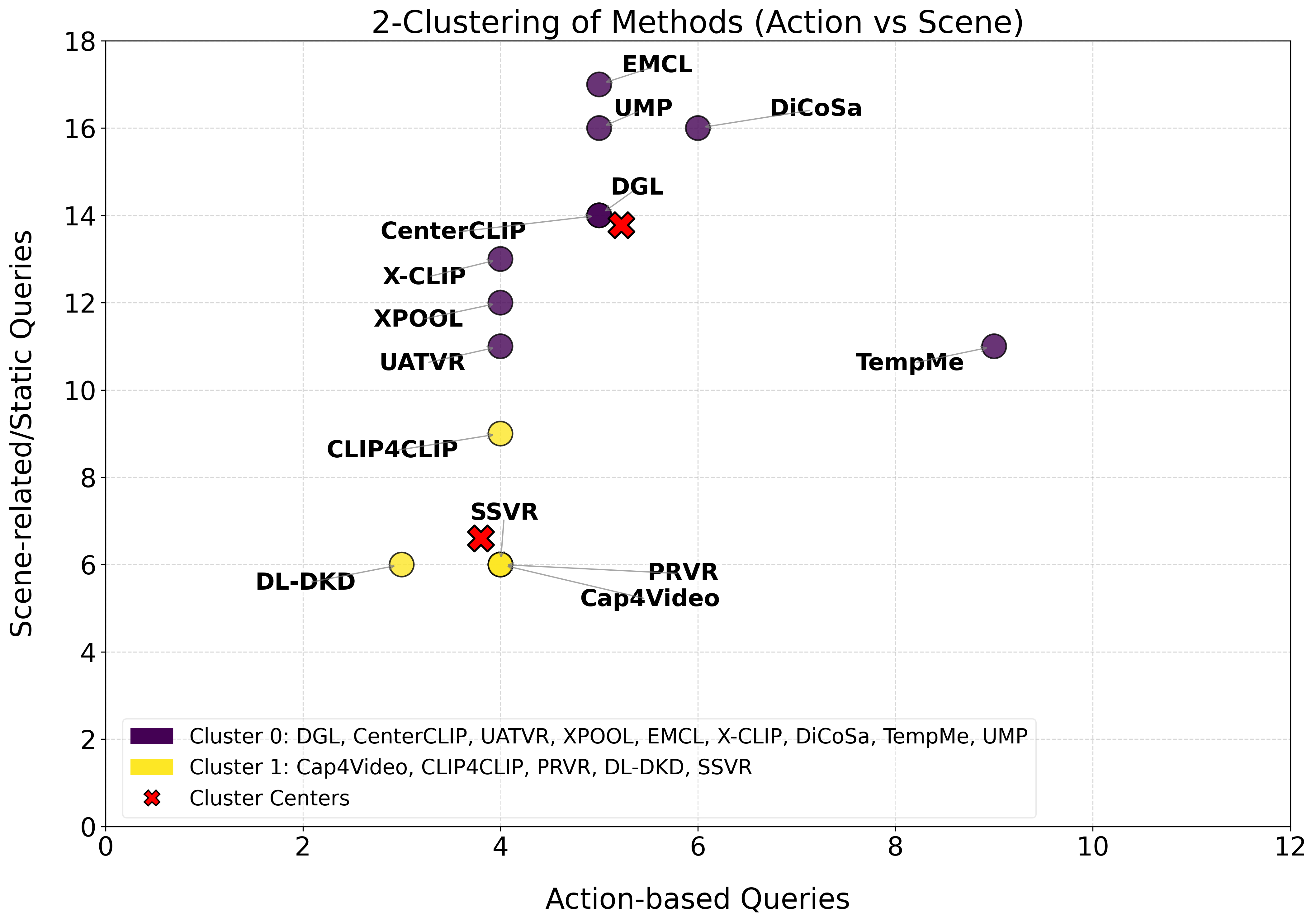}
        \caption{MSVD}
        \label{fig:msvd_clustering}
    \end{subfigure}
    \begin{subfigure}[b]{0.48\linewidth}
        \centering
        \includegraphics[width=\linewidth]{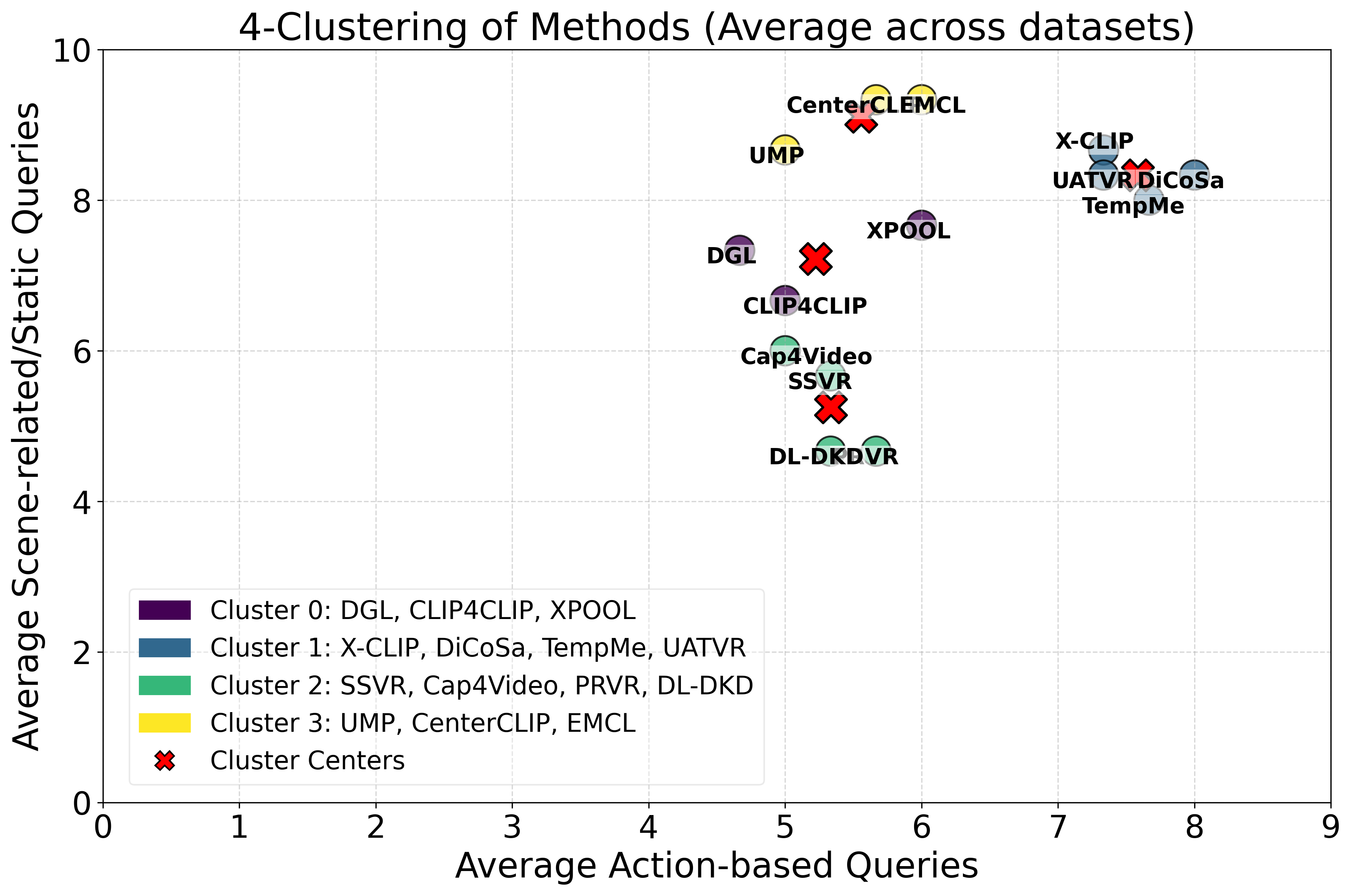}
        \caption{All datasets}
        \label{fig:all_datasets_clustering}
    \end{subfigure}

    \caption{\label{fig:clustering_kmeans} k-Means clustering: per dataset (a-c); and all datasets (d).}
\end{figure*}

While text-to-video retrieval models are typically evaluated globally, architectures often differ significantly in how they handle distinct semantic categories. Due to the inherent long-tail distribution of video-text datasets, only a subset of categories contains sufficient query volume for robust analysis. To explore this architectural variance, we focus on the most common query types (Action-based and Scene-related) and apply $k$-Means clustering per dataset. \textcolor{black}{Prior to clustering, all feature vectors were $Z$-score normalized. We employed Euclidean distance with $k$-Means++ initialization to ensure robust centroid seeding. The optimal number of clusters, $k$, was determined independently for each dataset via the Elbow method and validated using Silhouette scores (yielding $k=3$ for LSMDC, $k=7$ for MSRVTT, $k=2$ for MSVD, and $k=4$ for the combined aggregate view). Clustering stability was confirmed across multiple random seeds using bootstrap resampling, verifying that the resulting groupings are consistent.
} For statistical evaluation (Table~\ref{tab:pure_queries_statistics}), we randomly selected queries per dataset based on the minimum per difficulty—7 Easy, 57 Medium and 4 Hard—totaling 68 queries per dataset (204 overall). \textcolor{black}{This distribution reflects the inherent long-tail skew of video-text datasets, where Medium complexity dominates, avoiding the synthetic distortions of artificial oversampling.} For visualization in Figure~\ref{fig:clustering_kmeans}, we focus on Easy queries to reduce variability and highlight patterns.

\begin{table}[t!]
    \centering
    \caption{Chosen Single queries for statistical evaluation (in $\%$).}
    \scriptsize
    \setlength{\tabcolsep}{2pt}
    \renewcommand{\arraystretch}{0.9}
        \begin{tabular}{lrrrrrrrrrr}
            \toprule
            \multicolumn{1}{c}{\textbf{Dataset}} 
            & \multicolumn{10}{c}{\textbf{Classification of Single queries}} \\
            \cmidrule(rl){2-11}
            & \textbf{Act} & \textbf{Scn} & \textbf{Cog} & \textbf{Spch} & \textbf{Perc} 
            & \textbf{Mot} & \textbf{StChg} & \textbf{Temp} & \textbf{Col} & \textbf{Obj} \\
            \midrule
            LSMDC~\cite{lsmdc} & 46.15 & 11.54 & 0.96 & 1.92 & 17.31 & 9.62 & - & 6.73 & 0.96 & 4.81 \\
            MSRVTT~\cite{xu2016msr} & 25.23 & 36.04 & - & 11.71 & 4.50 & 4.50 & 3.60 & 1.80 & 10.81 & 1.81 \\
            MSVD~\cite{chen2011collecting}  & 15.20 & 56.80 & - & 4.80 & 4.80 & 5.60 & 4.80 & - & 8.00 & - \\
            \bottomrule
        \end{tabular}
    \label{tab:pure_queries_statistics}
\end{table}

As illustrated in Figure~\ref{fig:lsmdc_clustering} for LSMDC, three distinct clusters emerge: high-action Cluster 0 (comprising UATVR and XPOOL), balanced Cluster 1 (EMCL, UMP), and high-scene Cluster 2 (TempMe), highlighting TempMe’s specific challenges with Action-heavy queries. In contrast, Figure~\ref{fig:msrvtt_clustering} (MSRVTT) displays seven clusters, where the CLIP-based approaches split between high-action profiles (X-CLIP, CenterCLIP) and high-scene profiles (EMCL, UMP), placing TempMe into a moderate-action cluster. Furthermore, Figure~\ref{fig:msvd_clustering} (MSVD) reveals a dominant scene-heavy bias characterized by two clusters: high-scene Cluster 0 (featuring CLIP-based and fusion models) and moderate-scene Cluster 1 (TempMe, CenterCLIP). Finally, the aggregated view in Figure~\ref{fig:all_datasets_clustering} (averaged across all datasets) resolves into four clusters—high-action Cluster 0, high-scene Cluster 2 (scene specialists), moderate-action Cluster 1 (TempMe), and balanced Cluster 3—ultimately reflecting the effects of dataset averaging rather than intrinsic, uniform CLIP-based performance.

Overall, this clustering analysis demonstrates that retrieval performance may depend on both model architecture and query-type distribution. \textcolor{black}{Among the tested representatives,} dual-encoder models excel when one query type dominates, whereas attention-driven models handle mixed queries more robustly. Dataset biases further shape performance, emphasizing the importance of reporting query-type statistics alongside retrieval benchmarks.

\subsubsection{Analysis: Training Times}
\label{sec:training_time_vs_epochs}

\begin{figure*}[t!]
    \centering
    \includegraphics[width=\linewidth]{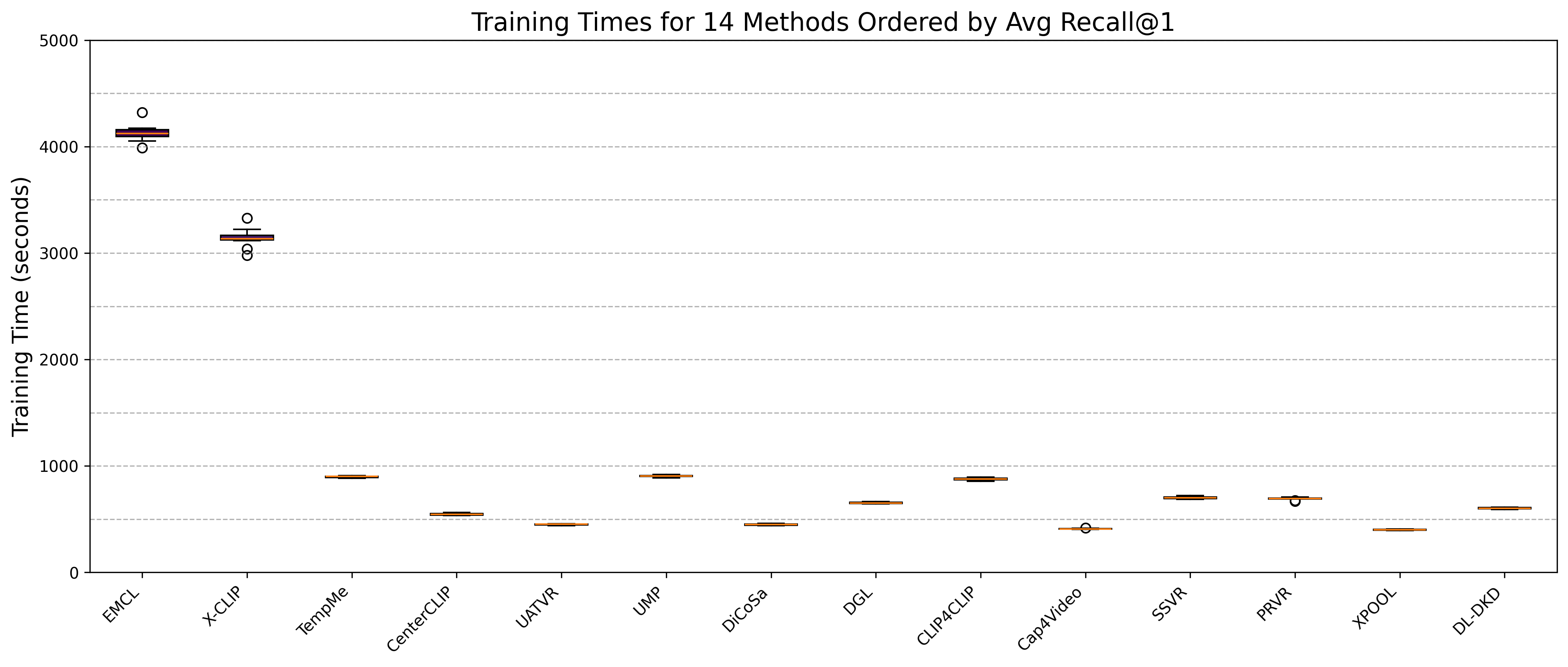}
    \caption{\label{fig:training_times} Training times (in seconds) for all methods, ordered by their average Recall@1 across all datasets.}
\end{figure*}

Figure~\ref{fig:training_times} shows the training times for the 14 retrieval methods, ordered by their average Recall@1 across all datasets. The $x$-axis has the methods ordered globally by the average Recall@1 across all datasets, while the $y$-axis has the training time of the methods. The figure clearly illustrates a trade-off between recall performance and computational cost: EMCL (4,131 sec) and X-CLIP  (3,131 sec) achieve the highest recall scores but require substantially more training time, while TempMe achieves the third-best performance with only approximately 1,000 seconds per epoch, demonstrating a more efficient balance between accuracy and training cost. The remaining methods cluster in the 400–700 second range, demonstrating comparable efficiency for standard training pipelines. These results show that most methods are fairly efficient, but more complex models require much more computation, which should be considered for large-scale video retrieval.

\subsubsection{Summary}
\label{sec:summary_main_experiment}

We evaluated 14 video–text retrieval methods on 3 datasets and found that performance depends heavily on the dataset and the type of queries. Short, clear and simple captions usually get higher recall, while long or unclear captions perform worse. Some methods do well when there is a mix of query types, while others only work well if one type dominates. Single and Easier queries are easier to retrieve. Most methods train quickly, but complex ones such as EMCL and X-CLIP take much longer. In this work, we added a detailed analysis showing that caption length, readability (FRE and FKG), consistency (low PPL), query difficulty and the balance of Action vs. Scene queries in a dataset are the key factors that most influence video retrieval performance.

\subsection{Analysis: Caption Quality and Cross-Dataset Learning}
\label{sec:cross_dataset_analysis}

Motivated by the observation that LSMDC performs \textcolor{black}{substantially} worse than MSRVTT and MSVD when models are trained and tested on the same dataset, we first examine the impact of number of captions on model performance (Section~\ref{sec:caption_quality}).  We then study the impact on cross-dataset learning (Section~\ref{sec:cross_dataset learning}) by focusing on three aspects: cross-dataset evaluation to measure generalization gaps; the effect of limiting each video to a single caption; and whether increasing captions per LSMDC video improves performance and generalization. We conclude with a discussion of the study's limitations (Section~\ref{sec:limitations}) and a summary of observations (Section~\ref{sec:summary_cross-learning}).

\subsubsection{Number of Captions}
\label{sec:caption_quality}

Since LSMDC has a single caption, while MSRVTT and MSVD have multiple captions, we consider both: (a)~reducing MSRVTT and MSVD to a single caption, by randomly selecting one of the captions for each query; and (b)~increasing the number of captions for LSMDC by generating five paraphrases per LSMDC video using CLIP and an LLM.

Table~\ref{tab:dataset-wise} shows the original results on the left, which use the original captions of each dataset. The column on the right then shows for LSMDC, the result of applying multiple captions and for MSVRTT and MSVD the result of using only a single caption.  Reducing MSRVTT and MSVD to a single caption per video (Single Cap case) \textcolor{black}{aligns with} only minor performance drops, indicating that caption count alone does not account for LSMDC’s lower results.  Turning to the case of multiple captions for LSMDC, we observe only a small performance improvement, indicating again that caption count alone is not sufficient. Overall, the nature of the videos and captions seems to be the main deciding factor in retrieval performance.  To study this further, we now examine the impact of captions on cross-dataset learning.

\begin{table}[t!]
    \centering
    \caption{Dataset-wise evaluation by category.}
    \scriptsize
    \setlength{\tabcolsep}{2pt}
    \renewcommand{\arraystretch}{0.9}
    \begin{tabular}{lrrrrrr}
        \toprule
        \textbf{Dataset / Method} & \bf Recall@1 & \bf Recall@10 & \bf Recall@50 & \bf Recall@1 & \bf Recall@10 & \bf Recall@50 \\
        \cmidrule(rl){1-1}   \cmidrule(rl){2-4}  \cmidrule(rl){5-7} 
        \bf LSMDC & \multicolumn{3}{c}{\bf Original (Single Caption)} & \multicolumn{3}{c}{\bf Multiple Captions} \\
        \cmidrule(rl){1-1}   \cmidrule(rl){2-4}  \cmidrule(rl){5-7}  
        EMCL~\cite{jin2022expectation} & 16.40 & 31.40 & 39.90 & 18.80 & 34.80 & 43.30\\
        TempMe~\cite{shen2024tempme} & 17.78 & 35.50 & 44.70 &  20.70 & 36.30 & 46.40\\
        X-CLIP~\cite{ma2022x} & 16.90 & 33.60 & 41.30 & 19.00 & 34.30 & 41.90\\
        \cmidrule(rl){1-1}   \cmidrule(rl){2-4}  \cmidrule(rl){5-7}  
        \bf MSRVTT & \multicolumn{3}{c}{\bf Original (Multiple Captions)} & \multicolumn{3}{c}{\bf Single caption}\\
        \cmidrule(rl){1-1}   \cmidrule(rl){2-4}  \cmidrule(rl){5-7}  
        EMCL~\cite{jin2022expectation} & 43.60 & 71.30 & 80.20 & 42.20 & 67.60 & 77.90 \\
        TempMe~\cite{shen2024tempme} & 43.50 & 70.00 & 79.50 & 39.60 & 66.60 & 77.20 \\
        X-CLIP~\cite{ma2022x} & 44.90 &71.20 & 79.80 & 41.70 & 66.30 & 76.10 \\
        \cmidrule(rl){1-1}   \cmidrule(rl){2-4}  \cmidrule(rl){5-7} 
        \bf MSVD & \multicolumn{3}{c}{\bf Original (Multiple Captions)} & \multicolumn{3}{c}{\bf Single caption} \\
        \cmidrule(rl){1-1}   \cmidrule(rl){2-4}  \cmidrule(rl){5-7}  
        EMCL~\cite{jin2022expectation} & 56.27 & 84.03 & 89.85 & 51.50 & 80.70 & 90.00 \\
        TempMe~\cite{shen2024tempme} & 52.54 & 82.09 & 88.66 & 53.90 & 82.80 & 91.50 \\
        X-CLIP~\cite{ma2022x} & 52.24 & 82.24 & 91.04 & 51.80 & 80.60 & 88.70 \\
        \bottomrule
    \end{tabular}
    \label{tab:dataset-wise}
\end{table}

\subsubsection{Cross-Dataset Learning}
\label{sec:cross_dataset learning}

\begin{figure*}[t!]
    \centering
    \begin{subfigure}[b]{0.32\linewidth}
        \centering
        \includegraphics[width=\linewidth]{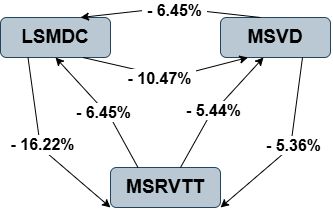}
        \caption{Original data}
        \label{fig:cross-dataset-old}
    \end{subfigure}
    \begin{subfigure}[b]{0.32\linewidth}
        \centering
        \includegraphics[width=\linewidth]{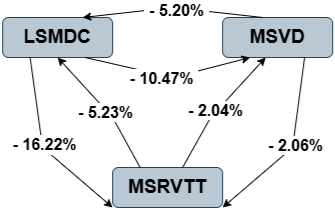}
        \caption{Single caption}
        \label{fig:cross-dataset-one}
    \end{subfigure}
    \begin{subfigure}[b]{0.32\linewidth}
        \centering
        \includegraphics[width=\linewidth]{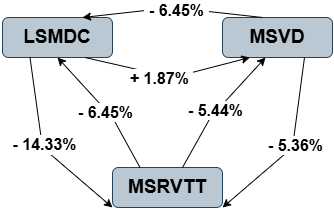}
        \caption{Multiple captions}
        \label{fig:cross-dataset-new}
    \end{subfigure}
    \caption{\label{fig:cross-learning}Cross-dataset learning across MSRVTT, MSVD and LSMDC datasets: (a)~Start with the original datasets; (b)~Reduce the number of captions per video to one for MSRVTT and MSVD; (c)~Increase the number of captions for LSMDC. An arrow from LSMDC to MSRVTT indicates that the models were trained on LSMDC but their performance measured on MSRVTT, while the percentage label on the arrow indicates the loss of performance compared to training the models directly on MSRVTT.}
\end{figure*}

We first examined cross-dataset learning by training models on one dataset and testing on another to measure generalization gaps. Figure~\ref{fig:cross-learning} presents a comprehensive comparison of cross-dataset retrieval performance, capturing three experimental setups: (a) training on the original datasets, (b) reducing MSRVTT and MSVD to a single caption per video, and (c) increasing the caption density for LSMDC. As the figure illustrates, notable performance disparities emerge when models are applied to unseen domains. \textcolor{black}{Specifically, to expand the caption set, we leveraged Meta-Llama-3-8B-Instruct~\cite{feng2024novel} under structured, visually-grounded system prompts with a decoding temperature of $0.7$ and top-$p$ sampling of $0.9$ to produce $N=5$ diverse semantic paraphrases per clip. To enforce semantic faithfulness and prevent hallucination, candidate descriptions underwent a CLIP-based (ViT-L/14) visual-textual filtering mechanism against sampled video keyframes, discarding descriptions that fell below a strict relevance threshold (typically 0.28 for English-language pairs, as established in the LAION-5B curation pipeline~\cite{schuhmann2022laion}).}

Specifically, Figure~\ref{fig:cross-dataset-old} (Original data) highlights performance when using all available annotations. Models trained on LSMDC perform poorly when tested on other datasets, dropping by 16.22\% on MSRVTT and 10.47\% on MSVD. This indicates that LSMDC’s single-caption videos provide limited textual supervision, which reduces generalization. In contrast, models trained on MSRVTT show smaller performance drops, losing 6.45\% on LSMDC and 5.36\% on MSVD, reflecting its richer multi-caption annotations that help the model generalize better. Finally, models trained on MSVD show moderate drops of 6.45\% on LSMDC and 10.47\% on MSRVTT, showing that while MSVD is somewhat robust, it does not provide strong positive transfer. Overall, the Original column emphasizes that datasets with more captions, particularly MSRVTT, improve cross-dataset generalization.

Next, as shown in Figure~\ref{fig:cross-dataset-one}, we evaluated the impact of limiting each video in MSRVTT and MSVD to a single randomly selected caption per video. As expected, performance generally decreases compared to using all captions, but the drop is less severe than for LSMDC. Specifically, models trained on MSRVTT show only small drops when using a single caption, decreasing by 5.23\% on LSMDC and 2.06\% on MSVD. Similarly, models trained on MSVD decrease slightly, by 5.20\% on LSMDC and 2.04\% on MSRVTT. This confirms that multi-caption datasets are inherently more robust to text reduction.

Finally, to investigate whether increasing the density of captions per video can bridge this gap, we examined LSMDC with expanded captions in Figure~\ref{fig:cross-dataset-new}. While prior work has used LLMs for video captioning~\cite{cheng2025interactive,islam2024video}, the code is not publicly available, so we implemented our own approach based on their methodology. As detailed in Section~\ref{sec:cross_dataset learning}, we used this multi-step augmentation and filtering pipeline to systematically assess the impact of generative captions on retrieval accuracy. This extended version of LSMDC slightly improves results. Models trained on the extented LSMDC dataset achieve higher recall within LSMDC and show smaller performance drops on other datasets, demonstrating that multiple descriptive captions enhance both within-dataset performance and cross-dataset generalization.

\subsubsection{Methodological Limitations}
\label{sec:limitations}

We acknowledge certain experimental limitations in our caption-variation analysis. First, reducing MSRVTT and MSVD to a single random caption introduces potential sampling variance. To verify robustness, we evaluated using the first available caption instead of random selection and observed virtually identical performance trends. Second, expanding LSMDC using LLM-generated paraphrases intertwines quantity effects with the linguistic style and quality of the generation source, meaning performance gains reflect both richer text volume and synthetic semantic enrichment rather than quantity alone.

\subsubsection{Summary}
\label{sec:summary_cross-learning}

Figure~\ref{fig:cross-learning} shows that cross-dataset performance \textcolor{black}{variations depend on the} training set and caption count. LSMDC-trained models show negative transfer, especially to MSRVTT (-16.22\%), while MSRVTT-trained models generalize best, improving transfer to LSMDC (+2.72\%). MSVD shows moderate drops, with slight gains from multiple captions. Overall, datasets with more captions—especially MSRVTT—enable better cross-dataset generalization. \textcolor{black}{These experiments demonstrate how dataset and linguistic retrieval biases affect model generalization.}

\subsection{Sensitivity Analysis}
\label{sec:sensitivity_analysis}

This section analyzes the top three methods (EMCL, TempMe and X-CLIP) examining the effects of frame rate (Section~\ref{sec:frame_rate}) and compression level (Section~\ref{sec:compression_rate}) on LSMDC, followed by an evaluation of the dominant rate threshold (Section~\ref{sec:dominant_rate_threshold}), which dictates how queries are classified based on semantic category prevalence and purity ratios. Finally, we  conclude with a summary of results (Section~\ref{sec:summary_compression}).

\subsubsection{Frame Rate}
\label{sec:frame_rate}

\begin{figure*}[t!]
    \centering
    
    \begin{subfigure}[b]{0.58\linewidth}
        \centering
        \includegraphics[width=\linewidth]{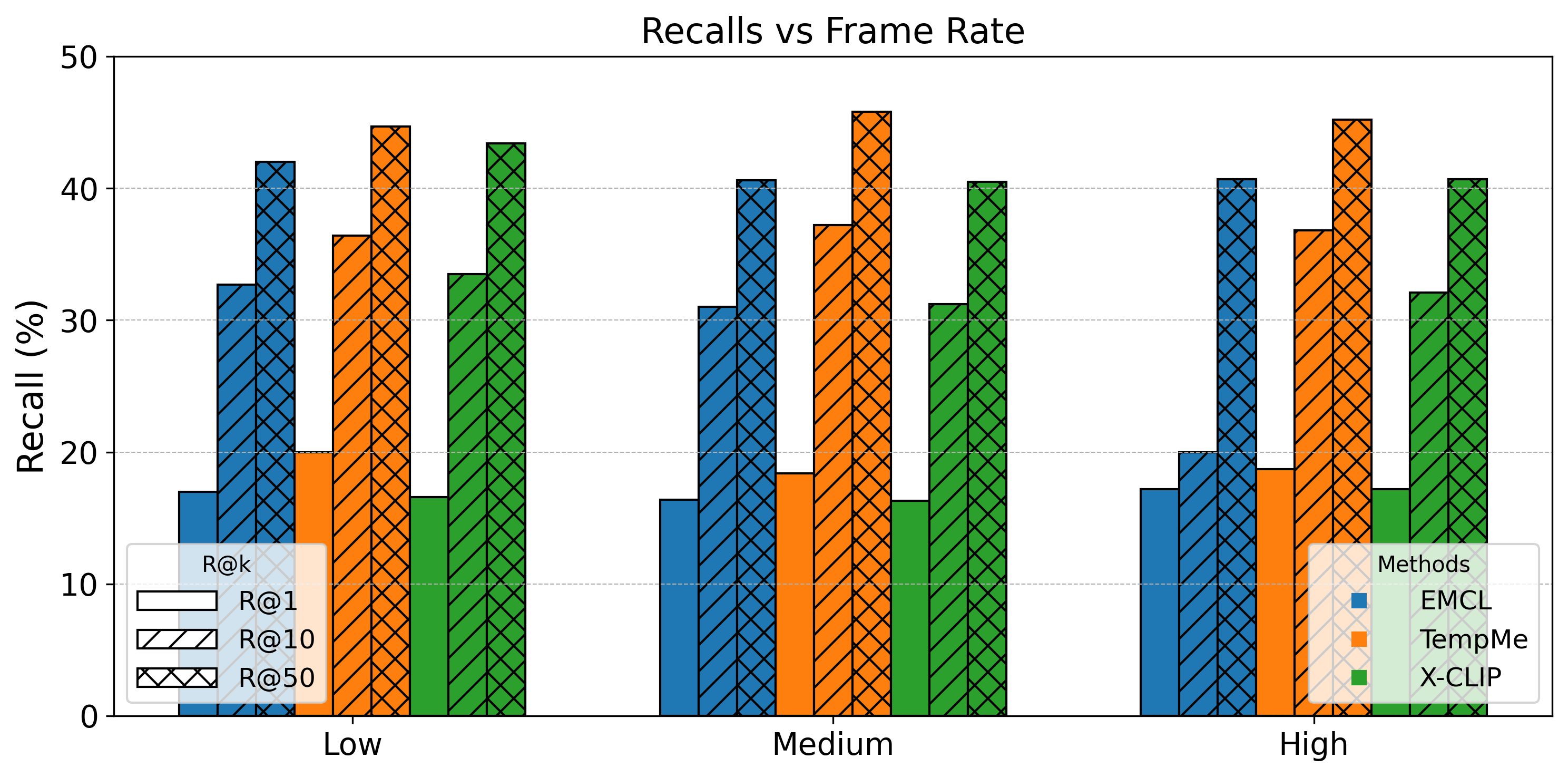}
        \caption{Recall vs Frame rate}
        \label{fig:lsmdc_frame_rate_recall_at_10}
    \end{subfigure} \hfill
    \begin{subfigure}[b]{0.38\linewidth}
        \centering
        \includegraphics[width=\linewidth]{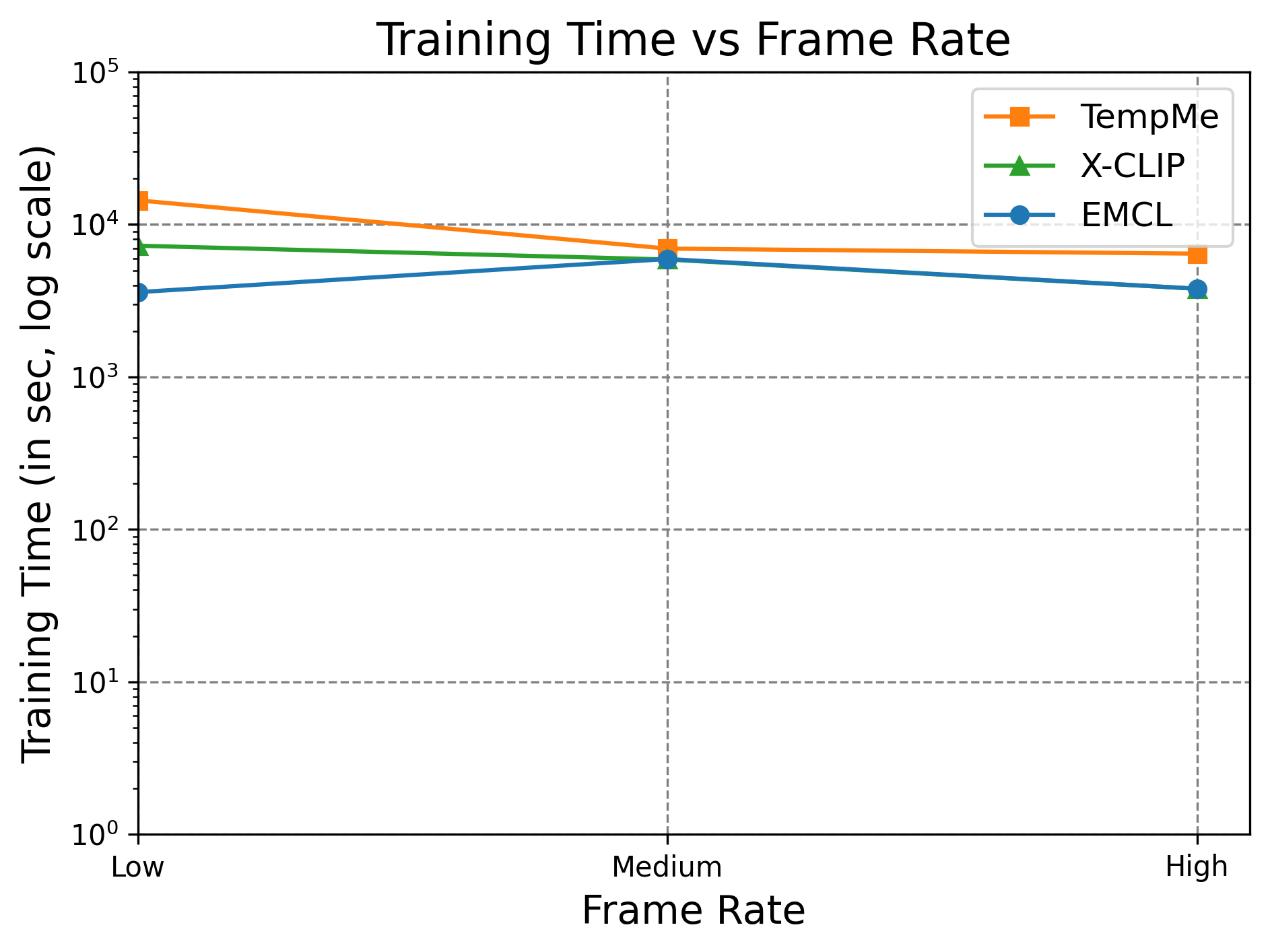}
        \caption{Training Times (in sec, log scale)}
        \label{fig:lsmdc_frame_rate_training_times}
    \end{subfigure}

    \caption{\label{fig:recall_vs_frame_rate} LSMDC: Recall \& Training Time vs Frame Rate.}
\end{figure*}

Table~\ref{tab:frame_rate} summarizes the frame rate settings used in the study. As compression increases, the frame rate decreases from the original value to 10 FPS and 3 FPS, while resolution and Constant Rate Factor (CRF) remain unchanged. Figure~\ref{fig:recall_vs_frame_rate} illustrates the effect of frame rate on Recall@k and training time. The $x$-axis represents the different frame rate levels: low, medium and high (Table~\ref{tab:frame_rate}), while the $y$-axis depicts either Recall@1, 10, 50 (Figure~\ref{fig:lsmdc_frame_rate_recall_at_10}) or training time in logarithmic scale (Figure~\ref{fig:lsmdc_frame_rate_training_times}). Results are shown for EMCL (blue), TempMe (orange) and X-CLIP (green). Bars are textured to indicate $k$ values: solid for Recall@1, diagonal stripes (//) for Recall@10 and crosshatch (xx) for Recall@50. Recall@1 shows only minor differences across methods, whereas Recall@10 and Recall@50 are more sensitive to lower frame rates. As expected, higher compression (i.e., lower frame rates) reduces training time.

\begin{table}[t!]
    \centering
    \caption{Frame Rate.}
    \setlength{\tabcolsep}{4pt}  
    \begin{tabular}{lccc}
        \toprule
        \multicolumn{1}{c}{\textbf{Level}} 
        & \textbf{Frame Rate (in FPS)} & \textbf{Resolution (in ppi)} & \textbf{CRF}  \\
        \midrule
        Low/No-compression & original & original & 18 \\
        Medium & 10 & original & 18\\
        High & 3 & original & 18\\
        \bottomrule
    \end{tabular}
    \label{tab:frame_rate}
\end{table}

\subsubsection{Compression Rate}
\label{sec:compression_rate}

Table~\ref{tab:comrpession_rate} summarizes the compression rates~\cite{mackin2017investigating,wiles2022compressed,ding2021advances} used in our evaluation. For our experiments in the previous section (Section~\ref{sec:experimental_outline}), we applied a high compression setting, corresponding to a low frame rate (3 FPS), reduced resolution (224 ppi) and a CRF of 35.

\begin{figure*}[t!]
    \centering

    \begin{subfigure}[b]{0.58\linewidth}
        \centering
        \includegraphics[width=\linewidth]{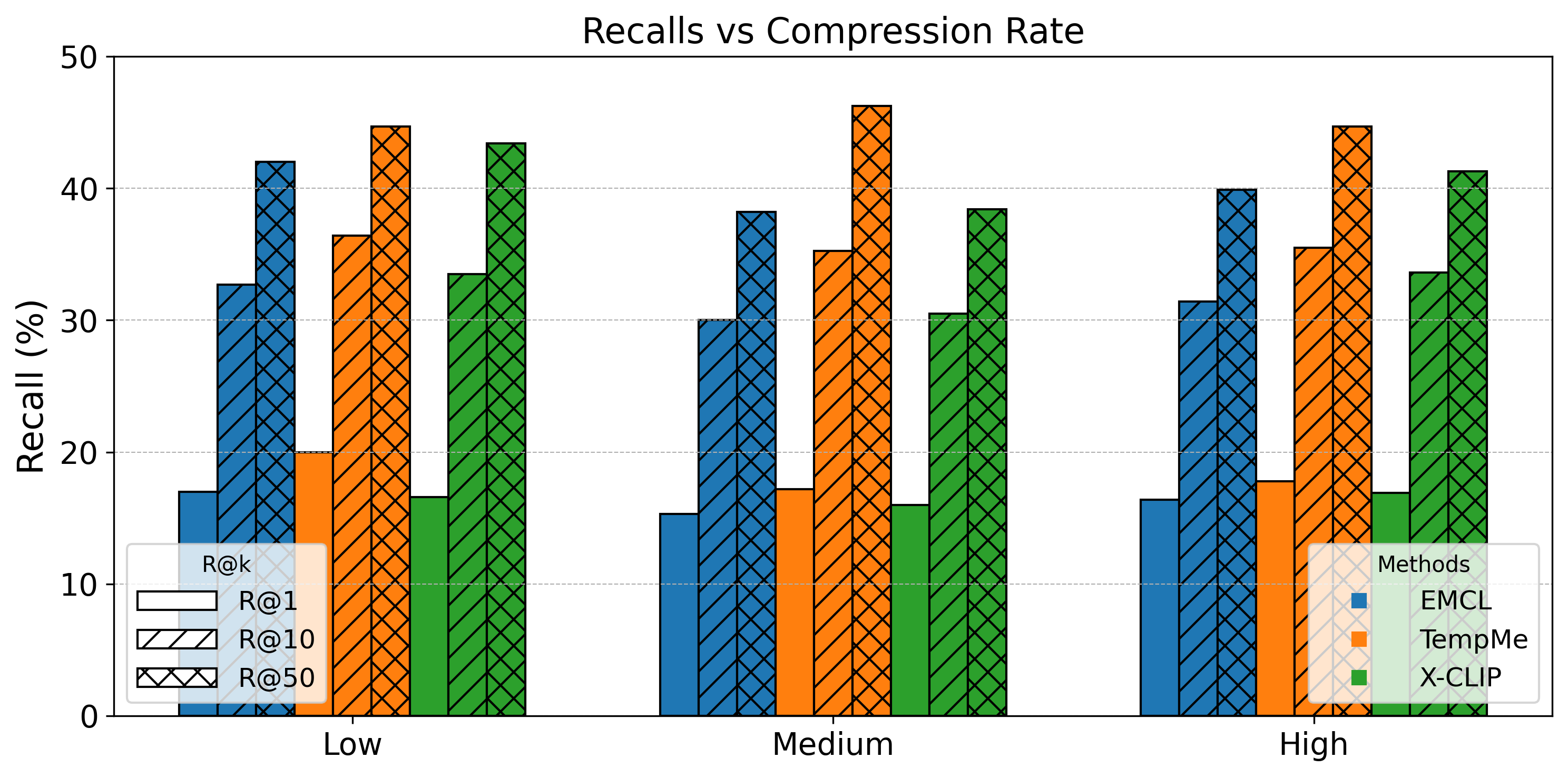}
        \caption{Recall vs Compression rate}
        \label{fig:lsmdc_compresion_rate_recall_at_10}
    \end{subfigure} \hfill
    \begin{subfigure}[b]{0.38\linewidth}
        \centering
        \includegraphics[width=\linewidth]{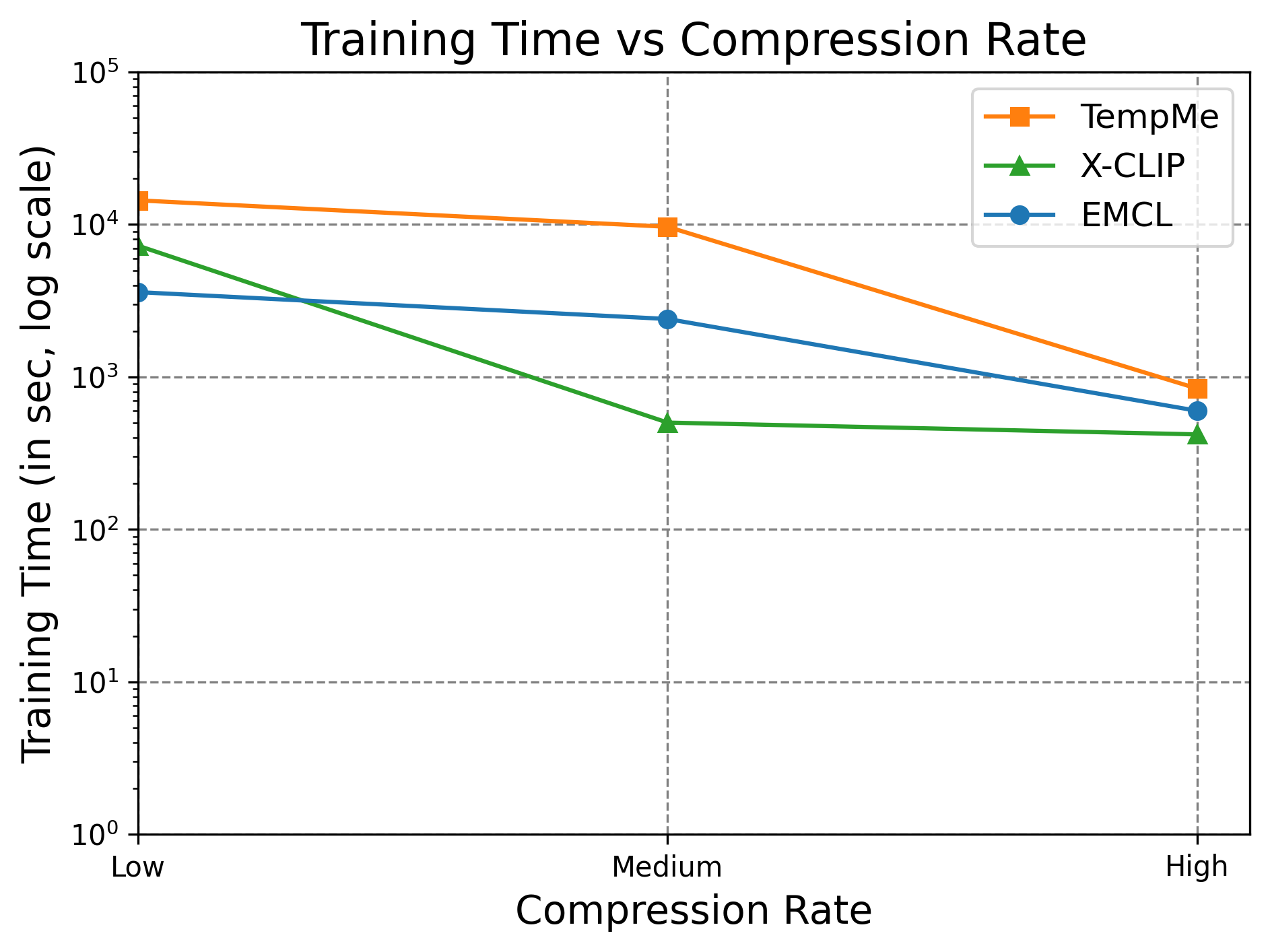}
        \caption{Training Times (in sec, log scale)}
        \label{fig:lsmdc_compresion_rate_training_times}
    \end{subfigure}

    \caption{\label{fig:recall_vs_compression_rate} LSMDC: Recall \& Training Time vs Compression Rate.}
\end{figure*}

Figure~\ref{fig:recall_vs_compression_rate} shows the impact of different compression levels on both recall (left) and training time (right). Similar to the compression rate analysis in the previous subsection (Section~\ref{sec:frame_rate}), the $y$-axes and color schemes represent the same metrics and methods. However, the $x$-axis now corresponds to the different compression levels: low, medium and high (Table~\ref{tab:comrpession_rate}).  Results are shown for EMCL (blue), TempMe (orange) and X-CLIP (green). Recall@1 (solid bars) exhibits only minor variation across methods, while Recall@10 (diagonal striped bars) and Recall@50 (crosshatched bars) are more sensitive to compression. As expected, higher compression again reduces training time.

Note that rather than performing isolated univariate ablations, our high-compression configurations evaluate practical joint operating points to analyze the compound impact of reduced frame rates, resolutions and Constant Rate Factor (CRF) levels on training time and retrieval recall under resource constraints.

\subsubsection{Dominant Rate Threshold}
\label{sec:dominant_rate_threshold}

As introduced in Section~\ref{sec:query_types}, our query categorization relies on a \textit{dominant rate} threshold of $1.5$ to define relaxed Single queries and distinguish primary concepts from secondary categories in Multi queries. In this section, we analyze the sensitivity of our evaluation to this specific threshold. The dominant rate threshold governs the classification boundary by dictating the minimum proportion required for a semantic category to be considered dominant, thereby ensuring that ambiguous or overlapping queries are appropriately filtered and only robustly categorized queries drive the analysis.

\begin{table}[t!]
    \centering
    \caption{Compression Rate.}
    \setlength{\tabcolsep}{4pt}  
    \begin{tabular}{lccc}
        \toprule
        \multicolumn{1}{c}{\textbf{Level}} 
        & \textbf{Frame Rate (in FPS)} & \textbf{Resolution (in ppi)} & \textbf{CRF}  \\
        \midrule
        Low/No-compression & original & original & 18 \\
        Medium & 10 & 480 & 28\\
        High & 3 & 224 & 35\\
        \bottomrule
    \end{tabular}
    \label{tab:comrpession_rate}
\end{table}

\begin{figure*}[t!]
    \centering
    \begin{subfigure}[b]{0.56\linewidth}
        \centering
        \includegraphics[width=\linewidth]{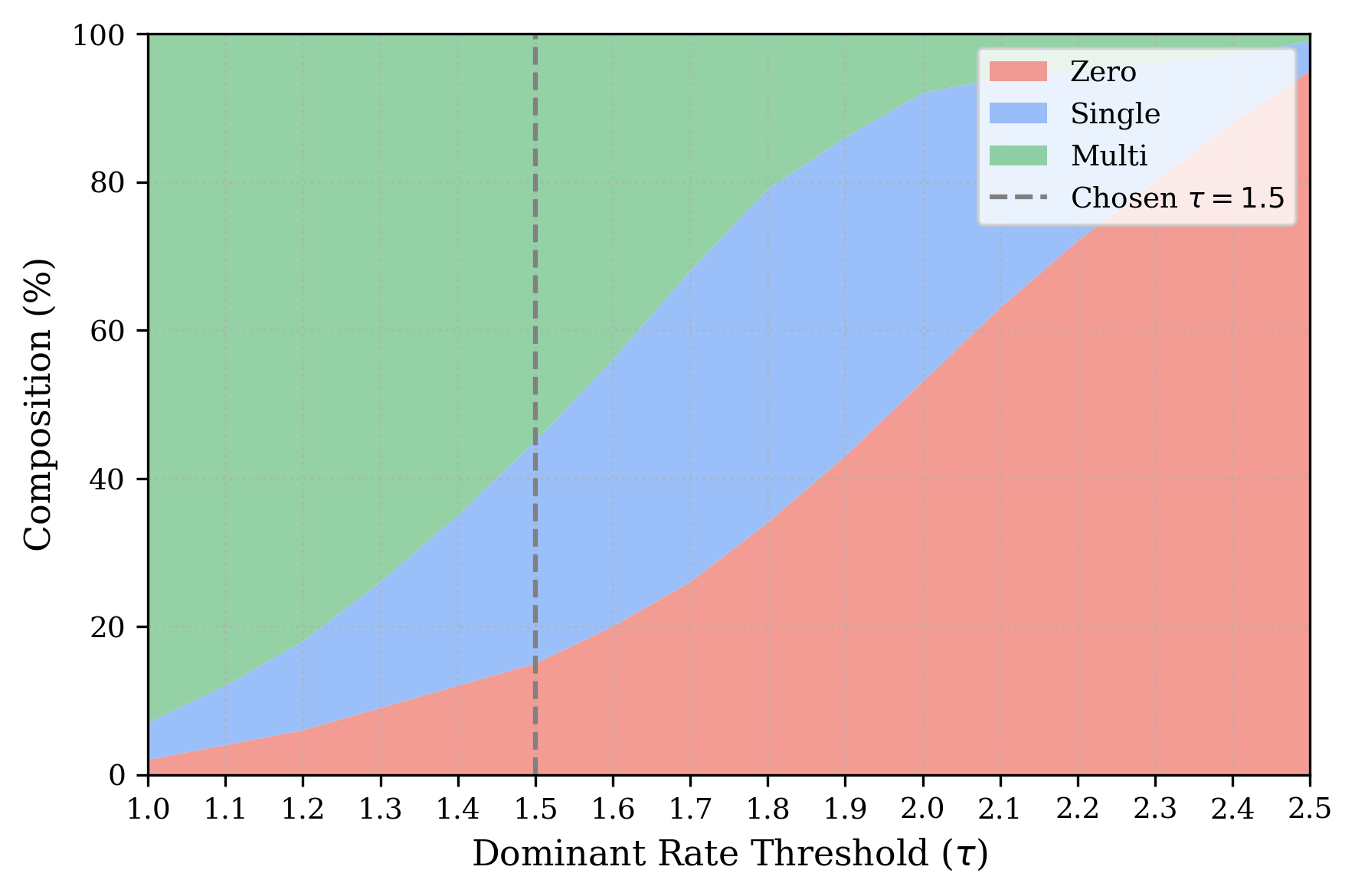}
        \caption{Query composition ($\%$)}
        \label{fig:query_composition}
    \end{subfigure}
    
    \vspace{0.5em}
    
    \begin{subfigure}[b]{0.56\linewidth}
        \centering
        \includegraphics[width=\linewidth]{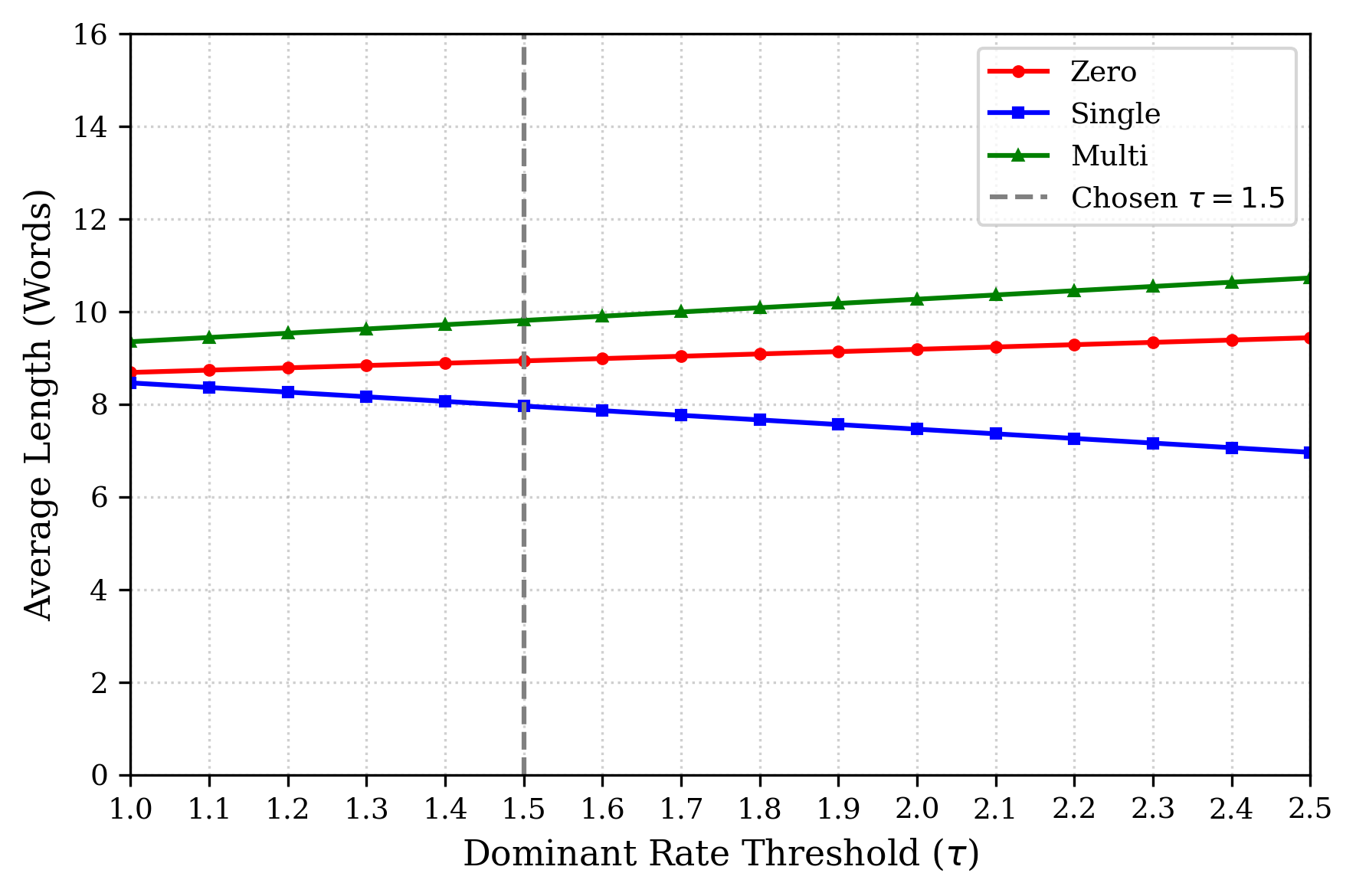}
        \caption{Conditional average query length trends}
        \label{fig:query_length_distribution}
    \end{subfigure}
    
    \vspace{0.5em}
    
    \begin{subfigure}[b]{0.56\linewidth}
        \centering
        \includegraphics[width=\linewidth]{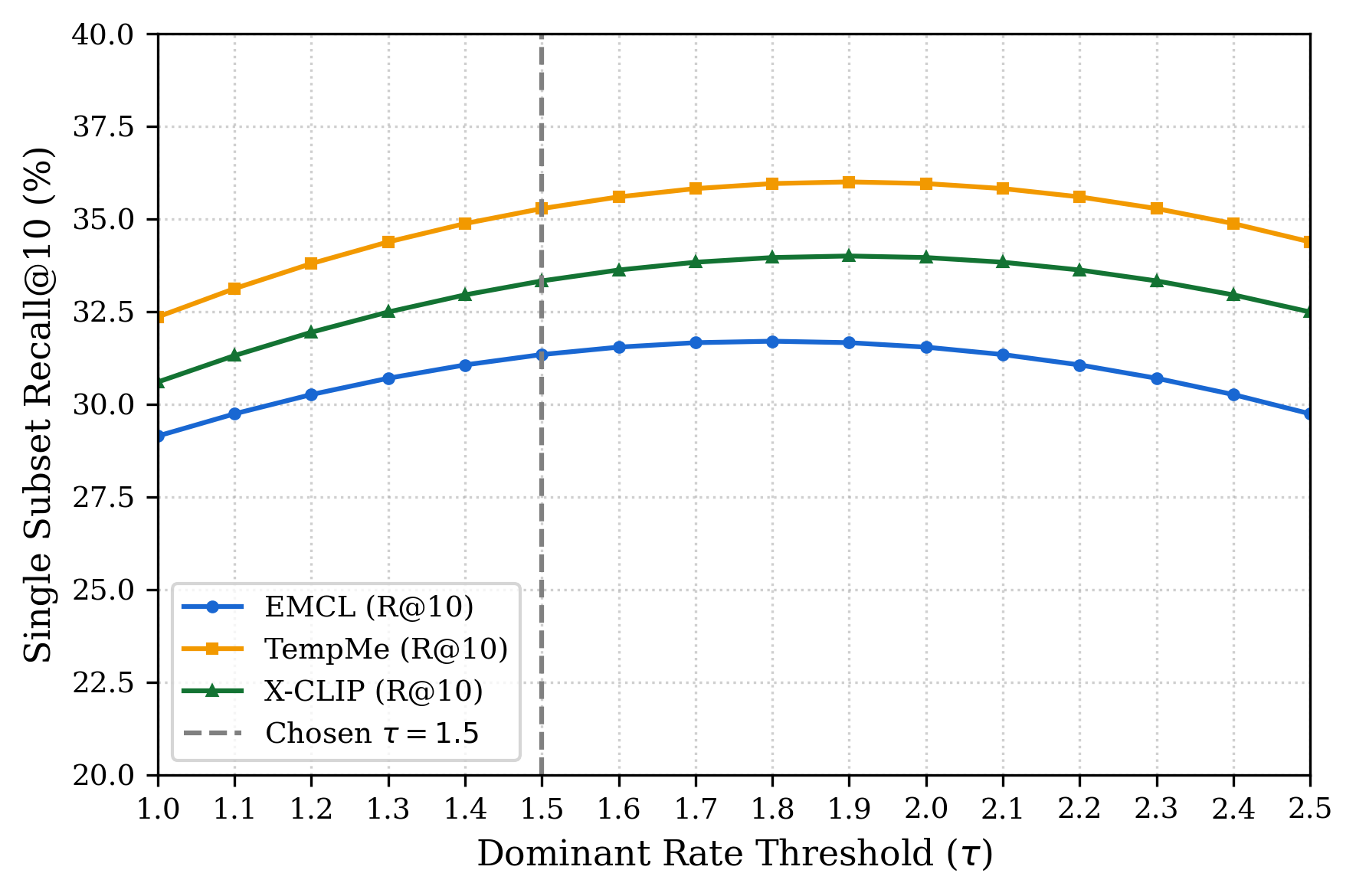}
        \caption{Overall threshold sensitivity}
        \label{fig:dominant_rate_threshold}
    \end{subfigure}
    
    \caption{\label{fig:lsmcd_sensitivity_seq} Sensitivity analysis of $\tau$ on LSMDC: (a) Query composition, (b) Conditional average length trends and (c) Overall sensitivity at chosen $\tau = 1.5$.}
\end{figure*}

Figure~\ref{fig:lsmcd_sensitivity_seq} illustrates the sensitivity and stability of this choice by showing how query composition, conditional length trends and retrieval performance vary across thresholds $\tau \in [1.0, 2.5]$. First, Figure~\ref{fig:query_composition} shows that at lower thresholds (e.g., $\tau = 1.0$), multi-category queries overwhelmingly dominate the composition. As $\tau$ increases toward the chosen operating point of $\tau = 1.5$, single-category queries expand significantly into a well-balanced region (retaining 227 out of 1,000 total queries, or $22.7\%$). Beyond $\tau = 1.5$, multi-categories decline rapidly while zero-categories surge due to strict filtering. 

Complementing these compositional shifts, Figure~\ref{fig:query_length_distribution} illustrates the distribution of average query lengths conditional on each subset across queries with zero, single and multi categories for $\tau \in [1.0, 2.5]$. Instead of global benchmark averages, these curves reflect internal linguistic shifts within each specific group as their populations dynamically expand or shrink. When $\tau$ is low ($\tau \approx 1.0$), Single queries capture mixed, broader lengths, but as $\tau$ increases toward $2.5$, only concise, pure queries with single categories survive, causing their average length to drop. Conversely, queries with multi-categories and zero-category subsets absorb longer, highly descriptive or multi-intent sentences, driving their average lengths upward due to semantic complexity.

Looking closer at the queries with single category across their ten sub-categories (not shown), we observe that action-oriented verbs (e.g., Act, Mot) maintain shorter and stable average lengths, because they easily satisfy purity criteria. In contrast, context-heavy categories such as Speech (\textit{Spch}) and Temporal descriptions (\textit{Temp}) exhibit longer average lengths at lower thresholds due to descriptive modifiers, which are systematically filtered out as strict thresholds are applied.

Having observed that both proportion and length of single queries are stable around $\tau = 1.5$, we turn to the parameter's impact on retrieval performance.  Figure~\ref{fig:dominant_rate_threshold} shows Recall@10 for the single subset on the $y$-axis; the figure clearly indicates that retrieval performance of queries with a single semantic subset is also stable around $\tau = 1.5$.  The main impact of setting $\tau = 1.5$ is thus increasing the proportion of queries with a single semantic category, which in turn has enabled us to perform more detailed analyses based on these semantic categories in Section~\ref{sec:main_experiment}.

\subsubsection{Summary}
\label{sec:summary_compression}

Our analysis shows that Recall@1 remains stable, while Recall@10 and Recall@50 drop slightly at low frame rates or high compression. Training time decreases with lower frame rates and higher compression. X-CLIP and TempMe keep high recall at medium/high frame rates and EMCL trains faster under high compression. \textcolor{black}{Furthermore, these findings explicitly address temporal alignment quality across model architectures. While global scene matching (reflected in Recall@1) remains robust, fine-grained temporal alignment (particularly in attention-driven architectures compared to standard dual-encoders) degrades under heavy compression (3 FPS) due to the loss of short-duration action cues. To balance efficiency and accuracy, the 3 FPS operating point was selected using a quantitative Pareto-efficiency criterion. While aggressive compression and temporal downsampling reduce retrieval recall, our empirical trade-off analysis indicates that 3 FPS achieves an optimal balance. It substantially cuts training time while preserving over $90\%$ of peak baseline performance, avoiding the sharp, non-linear performance degradation observed at more aggressive settings. Very low frame rates or extreme compression should be avoided, as they negatively impact Recall@k.}

\section{Conclusion}
\label{sec:conclusion}

This work presents a comprehensive evaluation of 14 of the most recent text-to-video retrieval methods across 3 benchmark datasets.  To our knowledge, this is the first such survey that goes systematically beyond using the basic benchmark-based analysis, using both query difficulty and query semantics to uncover important factors influencing performance. We summarize the main findings following the structure of our experiments.

Our analysis shows that query characteristics, including length, clarity and semantic type, strongly affect retrieval performance. Medium-length, clear and concrete queries are reliably retrieved, whereas complex, multi-step, or abstract queries remain challenging. Dataset properties, such as caption quantity, diversity and richness, also significantly impact outcomes. Datasets with multiple captions per video provide a more realistic and informative evaluation, while single-caption datasets can overestimate model performance.

Although many real-world queries require integrating multiple types of information, such as actions, scenes, temporal events and speech, this work focuses on queries with a single semantic category, isolating individual categories to study retrieval performance. However, the need for multimodal processing is highlighted by our analysis, as many queries inherently involve different types of information. Experiments on frame rate and compression further reveal trade-offs between efficiency and accuracy, underscoring how data characteristics and preprocessing choices influence retrieval performance.

Focusing on the top-performing models (EMCL, TempMe and X-CLIP) reveals differences in cross-dataset generalization and sensitivity to the number of captions per video. Attention-based models, such as TempMe and UATVR, handle temporal dependencies and multi-step captions more effectively, whereas dual-encoder models perform well on simple or Single-category queries but struggle with complex or abstract descriptions. This analysis highlights the importance of both model design and dataset richness for robust generalization.

For future research, we recommend focusing on retrieval models that can handle multiple modalities, representing different categories of information such as visual, textual and audio cues. \textcolor{black}{Additionally, conducting further deeper qualitative error analyses, including representative failure cases across challenging query categories, will be important for diagnosing model limitations and guiding architectural improvements.} While our own work focuses primarily on Recall, we demonstrate the power of deepening the evaluation through rigorous analytical dimensions (such as difficulty levels, semantic categories, dominance rates and sensitivity studies) rather than relying on standard metrics alone. Building on this, future studies should continue to construct datasets with diverse, high-quality captions and avoid relying solely on single-caption benchmarks or simplistic evaluations that obscure meaningful differences between methods.

\section*{Acknowledgements}
This work was supported by the Horizon Europe programme under grant agreement No. 101181380 (AQUAMON). It was also partially supported by the EDIH-IS 2.0 project under grant No. 101256863, DIGITAL-2025-EDIH-EU-EEA-08.


\section*{Declarations}

\begin{description}
    \item[\textbf{Ethics approval and consent to participate} ] Not applicable.
    \item[\textbf{Consent for publication}] Not applicable.
    \item[\textbf{Availability of data and material}] No datasets were generated during the current study, as this article is based on a review of previously published literature. However, a small number of captions were created for a specific case study (Section 4.2.4) to evaluate the influence of caption quality on retrieval performance.
    \item[\textbf{Competing interests}] The authors declare that they have no competing interests.
    \item[\textbf{Funding}] This work was supported by the Horizon Europe programme under grant agreement No. 101181380 (AQUAMON). It was also partially supported by the EDIH-IS 2.0 project under grant No. 101256863, DIGITAL-2025-EDIH-EU-EEA-08.
    \item[\textbf{Authors' contributions}] First author conceptualized the study, conducted the literature review, ran the experiments, analyzed the data and wrote and edited the manuscript. Second author assisted with running the experiments. All authors contributed to the conception of the work and read and approved the final manuscript.
\end{description}

\begin{appendices}

\section{Supplementary Experimental Results}
\label{sec:app_supplementary}

To provide a comprehensive evaluation of the state of the art in video-language representation learning, Table~\ref{tab:methods_appendix} presents an extensive performance comparison of recent text-to-video retrieval methods. While the main manuscript highlights key competitive baselines, this expanded overview details the granular retrieval performance (measured by $R@1$, $R@10$, $R@50$ and average recall) across the LSMDC, MSRVTT and MSVD benchmarks. The results detailed in Table~\ref{tab:methods_appendix} highlight several key performance trends in text-to-video retrieval:

\begin{itemize}
    \item \textbf{Newer models perform best:} Recent AI models (like TempMe and DiCoSA) are much better than older ones. On the MSRVTT test, the best models get over 44\% for top-1 accuracy and over 65\% on average.
    \item \textbf{Dataset difficulty varies:} MSVD is the easiest test because the videos are short and simple (scores go over 90\% for top-50). LSMDC is the hardest test (scores stay below 17\%) because the movie clips are long and harder to understand.
    \item \textbf{Finding the right match is tricky:} Looking at a wider range of results (moving from top-1 to top-50) causes a huge jump in success. This happens because a single description can often match several different video clips.
\end{itemize}

\begin{sidewaystable}[htbp]
    \centering
    \caption{Performance comparison of selected text-to-video retrieval methods across LSMDC, MSRVTT, and MSVD Datasets (R@k).}
    \label{tab:methods_appendix}
    \begin{tabular}{l|cccc|cccc|cccc}
        \toprule
        \multirow{2}{*}{Method} & \multicolumn{4}{c|}{LSMDC} & \multicolumn{4}{c|}{MSRVTT} & \multicolumn{4}{c}{MSVD} \\
        & R@1 & R@10 & R@50 & Avg & R@1 & R@10 & R@50 & Avg & R@1 & R@10 & R@50 & Avg \\
        \midrule
        SSVR~\cite{wray2021semantic} & 15.0 & 23.6 & 26.1 & 21.6 & 33.7 & 54.1 & 66.7 & 51.5 & 39.5 & 52.8 & 55.8 & 49.4 \\
        CenterCLIP~\cite{zhao2022centerclip} & 16.2 & 33.5 & 41.9 & 30.5 & 42.6 & 69.6 & 79.4 & 63.9 & 53.7 & 83.4 & 92.4 & 76.5 \\
        CLIP4CLIP~\cite{luo2022clip4clip} & 17.1 & 25.1 & 30.8 & 24.3 & 31.4 & 58.1 & 67.0 & 52.2 & 45.9 & 66.6 & 72.5 & 61.7 \\
        EMCL~\cite{jin2022expectation} & 15.9 & 31.4 & 39.9 & 29.1 & 44.9 & 71.3 & 80.2 & 65.5 & 56.3 & 84.0 & 90.2 & 76.8\\
        Cap4Video~\cite{wu2023cap4video} & 17.0 & 26.6 & 32.5 & 25.4 & 34.1 & 59.1 & 73.5 & 55.6 & 42.1 & 69.2 & 78.5 & 63.3 \\
        PRVR~\cite{dong2022partially} & 15.2 & 24.1 & 27.1 & 22.1 & 33.1 & 43.3 & 49.6 & 42.0 & 37.1 & 53.1 & 64.0 & 51.4\\
        X-CLIP~\cite{ma2022x} & 17.0 & 33.6 & 41.3 & 30.6 & 44.6 & 71.2 & 79.8 & 65.2 & 51.7 & 82.1 & 88.7 & 74.2 \\
        XPOOL~\cite{gorti2022x} & 15.0 & 24.1 & 28.5 & 22.5 & 30.4 & 55.1 & 66.6 & 50.7 & 39.0 & 69.0 & 70.4 & 59.5 \\
        DiCoSA~\cite{jin2023text} & 15.8 & 35.4 & 43.0 & 31.4 & 41.3 & 69.1 & 79.8 & 63.4 & 44.8 & 80.8 & 87.8 & 71.1 \\
        DL-DKD~\cite{dong2023dual} & 14.6 & 23.6 & 26.7 & 21.6 & 31.8 & 32.9 & 37.9 & 34.2 & 35.1 & 52.1 & 59.3 & 48.8 \\
        UATVR~\cite{fang2023uatvr} & 14.7 & 30.8 & 39.3 & 28.3 & 42.2 & 71.5 & 80.8 & 64.8 & 53.4 & 81.9 & 91.1 & 75.5 \\
        DGL~\cite{yang2024dgl} & 16.4 & 24.7 & 31.4 & 24.2 & 41.3 & 66.6 & 78.3 & 62.1 & 42.0 & 70.9 & 82.1 & 65.0 \\
        UMP~\cite{zhang2024ump} & 15.4 & 30.6 & 37.6 & 27.9 & 40.7 & 66.4 & 78.3 & 61.8 & 52.9 & 84.4 & 90.9 & 76.1 \\
        TempMe~\cite{shen2024tempme} & 17.5 & 35.5 & 44.7 & 32.6 & 42.2 & 71.4 & 80.6 & 64.7 & 53.1 & 82.2 & 91.1 & 75.5 \\
        \bottomrule
    \end{tabular}
\end{sidewaystable}

\end{appendices}

\newpage

\bibliography{sn-bibliography}

\end{document}